\documentclass[twocolumn, aps, superscriptaddress, showpacs, floatfix]{revtex4-1}
\usepackage{graphicx}
\usepackage{dcolumn}
\usepackage{bm}
\usepackage[colorlinks, linkcolor=blue, anchorcolor=blue, citecolor=blue, urlcolor=blue]{hyperref}
\usepackage{amsmath}
\usepackage{multirow}
\usepackage{longtable}
\usepackage{supertabular}
\usepackage{CJK, upgreek, fancyhdr}
\usepackage{lastpage}
\usepackage{longtable}
\usepackage{changes}
\usepackage{etoolbox}
\usepackage{setspace}
\usepackage{tabularx}
\usepackage{lineno, hyperref}
\usepackage{supertabular}
\usepackage[normalem]{ulem}
\usepackage{eqnarray}
\usepackage{booktabs} 
\usepackage{subfigure}
\usepackage{amsmath}
\usepackage{graphicx}
\usepackage{float} 
\usepackage{subfigure}
\usepackage[thicklines]{cancel}
\usepackage{ulem}
\begin{document}

\preprint{APS/123-QED}

\title{Laser-assisted deformed $\alpha$ decay of the ground state even-even nuclei}

\author{Jun-Hao Cheng}
\affiliation{Department of Physics, National University of Defense Technology, 410073 Changsha, People's Republic of China}
\author{Wen-Yu Zhang}
\affiliation{Department of Physics, National University of Defense Technology, 410073 Changsha, People's Republic of China}
\author{Qiong Xiao}
\affiliation{Department of Physics, National University of Defense Technology, 410073 Changsha, People's Republic of China}
\author{Jun-Gang Deng}
\affiliation{College of Science, China Three Gorges University, 443002 Yichang, People's Republic of China}
\author{Tong-Pu Yu}
\email{tongpu@nudt.edu.cn}
\affiliation{Department of Physics, National University of Defense Technology, 410073 Changsha, People's Republic of China}

\begin{abstract}

In the present work, the influence of ultra-intense laser fields on the $\alpha$ decay half-life of the deformed ground state even-even nucleus with the mass number $52 \leq Z \leq 118$ is systematically studied. The calculations show that the laser field changes the $\alpha$ decay half-life by varying the $\alpha$ decay penetration probability in a small range. Moreover, the analytical formulas for the rate of change of the $\alpha$ decay penetration probability in the ultra-intense laser fields have been derived by the spherical approximation, which agrees well with the numerical solutions for nuclei with more significant proton numbers. This provides a fast way to estimate the rate of change of the $\alpha$ decay penetration probability for superheavy nuclei. Furthermore, the relationship between laser properties and the average rate of change of the $\alpha$ decay penetration probability is investigated. The calculations indicate that the shorter the wavelength of the laser pulse is, the larger the average rate of change of the penetration probability.

\end{abstract}

\maketitle

\section{Introduction}

During the past twenty decades, many decay modes and exotic nuclei have been discovered with the advent of radioactive ion beam facilities around the world, e.g., Dubna, Rikagaku Kenkyusho (RIKEN), Heavy Ion Research Facility in Lanzhou (HIRFL), Berkeley, GSI, and Grand Accelerateur National d’Ions Lourds (GANIL) \cite{doi:10.1146/annurev.nucl.55.090704.151604, RevModPhys.72.733, RevModPhys.84.567, PhysRevLett.110.242502, PhysRevC.87.044335, PhysRevC.91.051302, 10.1140/epja/i2015-15088-9}. As one of the main decay modes of superheavy nuclei, $\alpha$ decay has always attracted much attention in synthesizing and researching superheavy nuclei \cite{PhysRevLett.112.092501}. Theoretically, $\alpha$ decay was one of the early successes of quantum mechanics. Gamow \cite{gamow1928quantentheorie}, Condon, and Gurney \cite{gurney1928wave} independently used barrier tunneling theory based on quantum mechanics to calculate $\alpha$ decay lifetimes. Experimentally, $\alpha$ decay spectra of neutron-deficient nuclei and heavy and superheavy nuclei provide important nuclear structural information, which is an irreplaceable means for researchers to understand the structure and stability of heavy and superheavy nuclei \cite{PhysRevLett.104.042701}. Meanwhile, $\alpha$ decay process is also essential for crucial issues such as understanding the nuclear cluster structure in superheavy nuclei \cite{PhysRevLett.87.192501, PhysRevC.69.044318, PhysRevC.73.064304}, studying the chronology of the solar system \cite{doi:10.1126/science.1215510} and finding stable superheavy element islands \cite{RevModPhys.72.733}.

The advent of laser fields with a wide range of frequencies, intensities, and durations provides a unique opportunity to study nuclear physics in the laboratory. Studying laser-nucleus interactions has been driven by the rapid development of intense laser technologies over the past few decades, e.g. the chirped pulse amplification technique \cite{9894358, STRICKLAND1985219}. Recently, it took only ten months for the peak laser field intensity to be increased from $5.5 \times 10^{22} {\ } \rm{W}/cm^{2}$ to the current $10^{23} {\ } \rm{W}/cm^{2}$ \cite{Yoon:19, Yoon:21}. Furthermore, the Shanghai Ultra Intensive Ultrafast Laser Facility (SULF) \cite{Li:18, Yu:18} or the Extreme Light Infrastructure for Nuclear Physics (ELI-NP) \cite{Mi_icu_2019, doi:10.1063/1.5093535} is expected to further increase the peak laser intensity by one to two orders of magnitude in the short term from the existing intensity. The rapid rise in peak intensity and energy of delivered lasers has made direct laser-nuclear interactions one of the hottest topics in nuclear physics \cite{PhysRevLett.127.052501, PhysRevC.100.064610, PhysRevC.101.044304, V2020, PhysRevC.106.034612, PhysRevC.106.024606, PhysRevC.105.054001, PhysRevC.99.044610, PhysRevC.100.041601, Li_2021, PhysRevC.104.044614, Lv2020}. Recent works have noted that the extreme laser fields can directly increase the probability of light nuclear fusion and heavy nuclear fission \cite{PhysRevC.102.011601, PhysRevC.102.064629}. Excitingly, J. Feng {\it{et al.}} experimentally presented femtosecond pumping of isomeric nuclear states by the Coulomb excitation of ions with the quivering electrons induced by laser fields \cite{PhysRevLett.128.052501}. This adds to the study of direct laser-nucleus interactions. For decay, many efforts have been dedicated to discussing how high-intensity lasers can interfere with the half-life of the natural decay of nuclei \cite{Mi_icu_2013, PhysRevLett.119.202501, Kis_2018, BAI201823, PhysRevC.99.044610, Mi_icu_2019, PhysRevLett.124.212505}. From an energy point of view, the cycle-averaged kinetic energy of emitted particles in the laser field with an intensity of $10^{23} {\ } \rm{W}/cm^{2}$ can exceed 3 MeV, which is already on the order of decay energy \cite{PhysRevC.102.064629}. However, the theoretical calculations can not be verified due to the lack of experimental data. It is essential to find a reasonable and adequate experimental scheme for studying the effect of laser light on nuclear decay. A feasible experimental protocol requires an evaluation of effects of the laser and properties of the nucleus on the laser-nucleus interaction and a way to select the right nucleus and adjust the laser parameters to obtain significant experimental results, which has seldom been investigated in detail.

Even-even nuclei capable of $\alpha$ decay are characterised by a large half-life time span and a wide variety of parent nuclei, which has the potential to become the object of future experimental studies of direct laser-nuclear interactions. To date, many approaches have been used to study nucleus $\alpha$ decay, such as the deformed Cosh potentials \cite{SOYLU201559}, the deformed Woods-Saxon (WS) potential \cite{PhysRevC.83.067302, PhysRevC.85.044324}, the Gamow-like model \cite{PhysRevC.87.024308, CHENG2019350}, the liquid drop model \cite{GUO2015110, PhysRevC.74.017304, PhysRevC.48.2409}, the cluster model \cite{PhysRevLett.65.2975, PhysRevC.74.014304, XU2005303}, the coupled-channels method \cite{Dzyublik2017, PhysRevC.101.044304, PhysRevC.92.051301, PhysRevC.73.014315, PhysRevC.78.034608}, the deformed version of the density-dependent cluster model (DDCM) with microscopic double-folding potentials \cite{PhysRevC.73.041301, ISMAIL2017202, doi:10.1142/S0217732308029885} and others \cite{PhysRevLett.59.262, PhysRevC.81.064318, PhysRevC.78.057302, PhysRevC.81.064318, QI2014203, PhysRevC.101.034307, PhysRevC.102.044314, Deng_2021, DENG2021136247}. These models reproduce, to varying degrees, the potential of the emitting particles in the parent nucleus. In the present work, consideration of the effect of the deformation of the nucleus on the $\alpha$ decay half-life is necessary since the laser-nucleus interaction introduces a new electric dipole term in the nuclear Hamiltonian, which is closely related to the angle between the vector $E(t)$ and the vector $r$. Taking into account of the deformation of the parent nucleus, we systematically study the rate of change of the $\alpha$ decay half-life of the deformed ground state even-even nucleus with the mass number $52 \leq Z \leq 118$ by using the state-of-the-art laser. The Coulomb potential of the emitted $\alpha$ particle-daughter nucleus is calculated by the double-folding model \cite{PhysRevC.73.041301}. Moreover, we chose the deformed Woods-Saxon nuclear potential in the calculation \cite{PhysRevC.83.067302}, which has been shown in our previous work to be competent for calculating the total potential energy between the nucleus-emitted particle \cite{PhysRevC.105.024312}. Furthermore, we give an analytical formula for calculating the rate of change of penetration probability $\Delta P$ and investigate the relationship between the rate of change of penetration probability and the properties of the parent nucleus itself. Finally, we investigate the relationship between the laser properties and the average rate of change of the $\alpha$ decay penetration probability $\Delta P_{avg}$. The calculations show that the shorter the wavelength of the laser pulse is, the larger the average rate of change of the $\alpha$ decay penetration probability.

This article is organised as follows. In the next section, the theoretical framework for calculating the $\alpha$ decay half-life in ultra-intense laser fields is described in detail. In Section \ref{section 3}, the detailed calculation results and discussions are provided. In Section \ref{section 4}, a brief summary is given.

\section{THEORETICAL FRAMEWORK}
\label{section 2}
\subsection{The theoretical method}

$\alpha$ decay half-life $T_{1/2}$, an important indicator of nuclear stability, can be written as
\begin{equation}
\label{eq1}
T_{1/2}=\frac{\hbar \rm{ln}2}{\Gamma}, 
\end{equation}
where $\hbar$ represents the reduced Planck constant, and $\Gamma$ is the $\alpha$ decay width depending on the $\alpha$ particle formation probability $S_{\alpha}$, the normalized factor $F$ and penetration probability $P$. In the density-dependent cluster model (DDCM), the $\alpha$ decay width can be written as \cite{PhysRevC.73.041301}
\begin{equation}
\label{eq2}
\Gamma=\frac{\hbar^{2}}{4\mu}S_{\alpha}FP, 
\end{equation}
where $\mu=\frac{M_d M_{\alpha}}{M_d+M_{\alpha}}$ is the reduced mass of the daughter nucleus and the $\alpha$ particle in the center-of-mass coordinate with $M_\alpha$ and $M_{d}$ being masses of the $\alpha$ particle and the daughter nucleus, respectively.

Considering the influence of the nucleus deformation, we obtain the total penetration probability $P$ by averaging $P_{\varphi}$ in all directions. This is widely used in both $\alpha$ decay and fusion reaction calculations \cite{PhysRevC.85.044324, PhysRevC.86.044317, NI2015108, PhysRevC.83.044317, STEWART1996332}, which can be written as \cite{PhysRevC.73.041301}
\begin{equation}
P=\frac{1}{2}\int_{0}^{\pi}P_{\varphi}\, \rm{sin}\varphi d\varphi, 
\label{eq4}
\end{equation}
\begin{equation}
P_{\varphi}=\exp\! [- 2 \int_{R_{2}}^{R_{3}} k(r, t, \varphi, \theta)\, dr].
\label{eq5}
\end{equation}
Similarly, the total normalised factor $F$ can be obtained by averaging $F_{\varphi}$ in all directions. It is given by \cite{PhysRevLett.59.262}
\begin{equation}
F=\frac{1}{2}\int_{0}^{\pi} F_{\varphi} \rm{sin}\varphi d\varphi, 
\end{equation}

\begin{equation}
F_{\varphi} \int_{R_1}^{R_2} dr \frac{1}{k(r, t, \varphi, \theta)} cos^2(\int_{R_1}^{r} dr' k(r', t, \varphi, \theta)-\frac{\pi}{4})=1
\end{equation}

where the classical turning points $R_{1}$, $R_{2}$ and $R_{3}$ can be determined by the equation $V(r, t, \varphi, \theta) = Q_{\alpha}$. $\varphi$ represents the orientation angle of the symmetry axis of the daughter nucleus with respect to the emitted $\alpha$ particle. $\theta$ is related to the laser-nucleus interaction, which will be provided in more detail in the following subsection. $k(r, t, \varphi, \theta)$ is the wave number, which can be written as
\begin{equation}
k(r, t, \varphi, \theta)= \sqrt{\frac{2\mu}{\hbar^{2}} \mid V(r, t, \varphi, \theta)-Q_{\alpha}\mid }, 
\end{equation}
where $r$ is the separation between the mass center of $\alpha$ particle and the mass center of core and $Q_{\alpha}$ is the $\alpha$ decay energy.

In this work, the total interaction potential $V(r, t, \varphi, \theta)$ between the daughter nucleus and the emitted $\alpha$ particle can be given by
\begin{equation}
V(r, t, \varphi, \theta)=\lambda(\varphi) V_N(r, \varphi)+V_l(r)+V_C(r, \varphi)+V_i(r, t, \varphi, \theta).
\label{6}
\end{equation}
where $V_l(r)$, $V_C(r, \varphi)$ and $V_N(r, \varphi)$ are the centrifugal, Coulomb, and nuclear potentials, respectively. $V_i(r, t, \varphi, \theta)$ describes the interaction of the electromagnetic field with the decay system \cite{Mi_icu_2013}, which will be provided in more detail in the following subsection. Meanwhile, $\lambda(\varphi)$ can be obtained by u\rm{sin}g the Bohr-Sommerfeld quantization condition.

In the present work, the emitted $\alpha$-daughter nucleus nuclear potential $V_N(r, \varphi)$ was chosen as the classic Woods-Saxon (WS) \cite{PhysRevC.83.067302} nuclear potential. For the Woods-Saxon form, the nuclear potential is approximated as the axial deformation \cite{PhysRevC.83.067302}, which can be written as
\begin{equation}
V_N(r, \varphi)= \frac{V'}{1+exp[(r-R_d(\varphi))/s]}, 
\end{equation}
with $R_d(\varphi)=r_0A_d^{1/3}[1+\beta_2 Y_{20}(\varphi)+\beta_4 Y_{40}(\varphi)+\beta_6 Y_{60}(\varphi)]$. Here, $Y_{ml}(\varphi)$ represents shperical harmonics function, $\beta_2$, $\beta_4$ and $\beta_6$ respectively denote the calculated quadrupole, hexadecapole and hexacontatetrapole deformation of the nuclear ground-state.  $A_d$ represents the mass number of the daughter nucleus. By systematically searching for the radius $r_0$, the diffuseness $s$, and the depth of the nuclear potential $V'$, we find that the most convenient value for the attenuation calculation is $r_0=1.06$, $s = 0.88 {\ } \rm{fm}$ and $V'=161.95 {\ } \rm{MeV}$ in the case of $S_{\alpha}=0.43$ \cite{XU2005303}.

In this work, we obtain the deformed Coulomb potential by the double-folding mode. It can be written as
\begin{equation}
V_C(\mathop{r}^{\rightarrow}, \varphi)=\int_{}^{}\int_{}^{}\frac{\rho_{d}(\vec{r}_d)\rho_{\alpha}(\vec{r}_\alpha)}{\mid \vec{r}+\vec{r}_d-\vec{r}_\alpha \mid } d\vec{r}_\alpha d\vec{r}_d, 
\end{equation}
where $\rho_{\alpha}$ and $\rho_{d}$ are the density distributions of the emitted $\alpha$ particle and the daughter nucleus, respectively. $\vec{r}_\alpha$ and $\vec{r}_d$ are the radius vectors in the charge distributions of the emitted $\alpha$ particle and daughter nuclei. Simplified appropriately by the Fourier transform \cite{PhysRevC.61.044607, ISMAIL200353, Gao_Long_2008}, the Coulomb potential can be approximated as
\begin{equation}
V_C(\mathop{r}^{\rightarrow}, \varphi)=V_C^{(0)}(\vec{r}, \varphi)+V_C^{(1)}(\vec{r}, \varphi)+V_C^{(2)}(\vec{r}, \varphi), 
\end{equation}
where $V_C^{(0)}(\vec{r}, \varphi)$, $V_C^{(1)}(\vec{r}, \varphi)$ and $V_C^{(2)}(\vec{r}, \varphi)$ are the bare Coulomb interaction, linear Coulomb coupling and second-order Coulomb coupling, respectively \cite{PhysRevC.61.044607}.

The Langer modified $V_l(r)$ is chosen in the form \cite{doi:10.1063/1.531270}. It can be written as
\begin{equation}
V_{l}(r)=\frac{\hbar^2(l+\frac{1}{2})^2}{2{\mu}r^2}, 
\end{equation}
where $l$ is the orbital angular momentum carried by the $\alpha$ particle. In the present work, we have only focused on the $\alpha$ decay of the even-even nuclei,  thus $l=0$ is taken in the calculations.

\subsection{Laser-nucleus interaction}
\subsubsection{ The quasistatic approximation}
The full width at half maximum (FWHM) of laser pulses with peak intensities exceeding $10^{23} {\ } \rm{W}/cm^{2}$ currently available in the laboratory is approximately $19.6$ $\rm{fs} {\ } (=1.96 \times 10^{-14}$ s). The laser cycles produced by a near-infrared laser with a wavelength of approximately 800 nm and an X-ray free-electron laser \cite{9760631} with a photon energy of 10 keV are approximately $10^{-15}$ s and $10^{-19}$ s, respectively. For $\alpha$ decay, the emitted $\alpha$ particle oscillates back and forth at high frequencies within the parent nuclei, with a small probability of tunneling out whenever the preformed $\alpha$ particle hits the potential wall. The time scale of the emitted $\alpha$ particle pas\rm{sin}g through the potential wall can be estimated.

Since the typical decay energy for $\alpha$ decay is approximately several MeV, the velocity of the preformed $\alpha$ particles is approximately $10^7$ m/s and the size of the parent nucleus is approximately 1 fm, the frequency of the oscillations can be roughly estimated to be $10^{22}$ Hz. The length of the tunnel path is less than 100 fm, and the time for the emitted $\alpha$ particle to pass through the tunnel is under $10^{-20}$ s. The highest peak intensity laser pulse currently achievable has an optical period much longer than this time. Therefore, the laser field does not change significantly during the passage of the emitted $\alpha$ particles through the potential barrier and the process can be considered quasistatic. A similar quasistatic approximation is usually used to describe the tunneling ionisation of atoms in strong-field atomic physics \cite{PhysRevA.54.R2551, PhysRevA.63.011404}. It is also shown that failure to consider this quasistatic conditions can lead to inaccurate theoretical calculations, e.g., Ref. \cite{PhysRevLett.119.202501}.

Finally, the kinetic energy of the emitted $\alpha$ particles is only a few MeV. They move much slower than the speed of light in vacuum. This means that the effect of the laser electric field on the emitted $\alpha$ particles is expected to be much larger than that of the laser magnetic field. Therefore, we can neglect the magnetic component of the laser field in the current work.

\subsubsection{The relative motion of the emitted $\alpha$ particle and the daughter nucleus in the center of mass coordinates}

In the quasistatic approximation, the time-dependent Schrödinger equation (TDSE) can be used to describe the interaction between the daughter nucleus and the emitted $\alpha$ particle \cite{PhysRevC.102.064629}, which can be written as
\begin{equation}
i \hbar \frac{\partial \Phi(\vec{r}_\alpha, \vec{r}_d, t)}{\partial t}=H(t)\Phi(\vec{r}_\alpha, \vec{r}_d, t), 
\end{equation}
where $H(t)$ is the time-dependent minimum-coupling Hamiltonian. Since the existing intense laser wavelengths are much larger than the spatial scale of $\alpha$ decay, the spatial dependency of the vector potential in a radiation gauge can be ignored. The time-dependent minimum-coupling Hamiltonian can be given by
\begin{equation}
H(t)=\sum_i\frac{1}{2m_i}[\vec{p_i}-\frac{q_i}{c}\vec{A}(t)]^2+V(r), 
\end{equation}
where $i$ represents the parameters related to the emitted $\alpha$ particle and the daughter nucleus, respectively.

For the center of mass coordinates ($\vec{R}, \vec{P}, \vec{r}, \vec{p}$):
\begin{equation}
\begin{aligned}
&\ \vec{r}_\alpha=\vec{R}+m_d\vec{r}/(m_\alpha+m_d)\\
&\ \vec{p}_\alpha=\vec{p}+m_\alpha\vec{P}/(m_\alpha+m_d)\\
&\ \vec{r}_d=\vec{R}-m_\alpha \vec{r}/(m_\alpha+m_d)\\
&\ \vec{p}_d=-\vec{p}+m_d\vec{P}/(m_\alpha+m_d).\\
\end{aligned}
\end{equation}
The time-dependent minimum-coupling Hamiltonian can be written as
\begin{equation}
H(t)=\frac{1}{2M}[\vec{P}-\frac{q}{c}\vec{A}(t)]^2+\frac{1}{2\mu}[\vec{p}-\frac{Q_{eff}}{c}\vec{A}(t)]^2+V(r), 
\end{equation}
where $M=m_\alpha+m_d$, $q=q_\alpha+q_d$. The effective charge for relative motion $Q_{eff}$ describes the tendency of the laser electric field to separate the emitted $\alpha$ particle from the daughter nuclei. It can be written as
\begin{equation}
Q_{eff}=\frac{q_\alpha m_d-q_d m_\alpha}{M}.
\end{equation}

By introducing unitary transformations, the wave function can be transformed into the center of mass coordinates
\begin{equation}
\phi(\vec{r}, \vec{R}, t)=\hat{U_r}\hat{U_R}\Phi(\vec{r}, \vec{R}, t), 
\end{equation}
where $\hat{U_r}={\rm{exp}}[-i\frac{Q_{eff}}{c}\vec{A}(t) \cdot \vec{r}/\hbar]$ and $\hat{U_R}={\rm{exp}}[-i\frac{q}{c}\vec{A}(t) \cdot \vec{R}/\hbar]$. The TDSE can be rewritten as
\begin{eqnarray}
i \hbar \frac{\partial \phi(\vec{r}, \vec{R}, t)}{\partial t}=&[-\frac{{\hbar}^2}{2\mu}\nabla^2_r+V(r)-Q_{eff}\vec{r}\vec{E}(t)\nonumber\\
&\ -\frac{{\hbar}^2}{2M}\nabla^2_R-q\vec{R}\vec{E}(t)]\phi(\vec{r}, \vec{R}, t), 
\end{eqnarray}
where $c\vec{E}(t)=-d\vec{A}(t)/dt$ is the time-dependent laser electric field. By factorizing the wave function $\phi(\vec{r}, \vec{R}, t)=\chi_1(\vec{R}, t)\chi_2(\vec{r} , t)$, we split the TDSE into two separate equations describing the center of mass coordinates and the relative motion between the daughter nucleus and the emitted $\alpha$ particle. They can be written as
\begin{equation}
i \hbar \frac{\partial\chi_1(\vec{R} , t)}{\partial t}=[-\frac{{\hbar}^2}{2M}\nabla^2_R-q\vec{R}\vec{E}(t)]\chi_1(\vec{R} , t), 
\end{equation}
\begin{equation}
i \hbar \frac{\partial\chi_2(\vec{r} , t)}{\partial t}=[-\frac{{\hbar}^2}{2\mu}\nabla^2_r+V(r)-Q_{eff}\vec{r}\vec{E}(t)]\chi_2(\vec{r} , t).
\end{equation}
The equation of relative motion is related to the laser electric field. The interaction potential energy between the relative motion particle and the laser field can be written as
\begin{equation}
V_i(\mathop{r}^{\rightarrow}, t, \theta)=-Q_{eff}\mathop{r}^{\rightarrow} \cdot \mathop{E}^{\rightarrow}(t) =-Q_{eff} r E(t) \rm{cos}\theta, 
\end{equation}
where $\theta$ is the angle between vector the $\vec{r}$ and vector $\vec{E}(t)$.

\subsubsection{Laser-nucleus interaction}

The laser electric field with a linearly polarized Gaussian plane wave form can be expressed as
\begin{equation}
E(t)=E_0 f(t) \rm{sin} (\omega \emph{t}), 
\end{equation}
where $\omega$ is the angular frequency. The peak of the laser electric field $E_0$ is related to the peak of the laser intensity $I_0$, which can be given by \cite{Mi_icu_2019}
\begin{equation}
E_0 [{\rm{V cm^{-1}}}]=(\frac{2I_0}{c\epsilon_0})^{1/2}=27.44(I_0[{\rm{W cm^{-2}}}])^{1/2}, 
\end{equation}
where $\epsilon_0$ and $c$ are the permittivity of free space and the speed of light in vacuum, respectively.

The sequence of Gaussian pulses with an envelope function of temporal profile $f(t)$ can be given by
\begin{equation}
f(t)=\rm{exp}(-\frac{\emph{t}^2}{\tau^2}), 
\end{equation}
where $\tau$ represents the pulse width of the envelope, which can be written in the form related to the pulse period $T_0$
\begin{equation}
\tau=x T_0.
\end{equation}
In the present work, we write $E(t)$ in the form related to the wavelength $\lambda$ for the discussion in the next section. It can be written as

\begin{equation}
\label{eq 29}
E(t)=E_0{\rm{exp}} (-\frac{t^2}{x^2 T_0^2}){\rm{sin}} (\omega t)=E_0{\rm{exp}} (-\frac{c^2 t^2}{\lambda^2 x^2}){\rm{sin}}(2 \pi \frac{c}{\lambda}t), 
\end{equation}
where the pulse period $T_0=1/\nu$ and $\nu=\omega/2\pi=c/\lambda$ is the laser frequency. 

The laser electric fields should also change the proton emission energy $Q_{\alpha}$. The change in the decay energy $\Delta Q_{\alpha}$ is equal to the energy of the emitted $\alpha$ particle accelerated by the laser electric field during the penetration of the potential barrier. It can be given by \\
\begin{equation}
\Delta Q_{\alpha}=eZ_\alpha E(t)R_d(\varphi)\rm{cos}\theta.
\end{equation}
The $\alpha$ decay energy with considering the laser electric field effect $Q^*_{\alpha}$ can thus be rewritten as 
\begin{equation}
Q^*_{\alpha}=Q_{\alpha}+\Delta Q_{\alpha}
\end{equation}

\begin{figure}
\includegraphics[width=9.5cm]{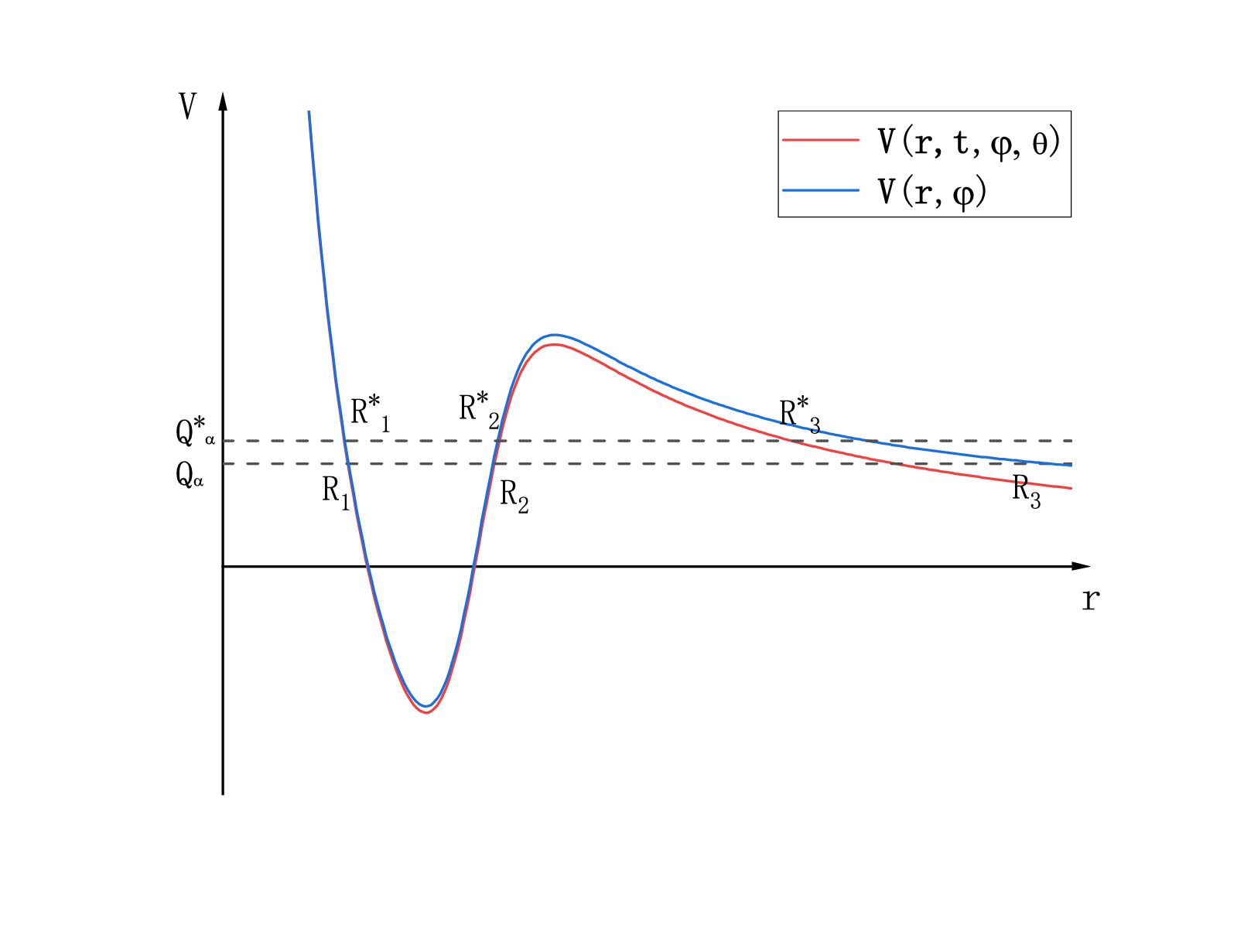}
\caption{(Color online) Schematic diagram of the total emitted $\alpha$ particle-daughter nucleus interaction potential with and without considering the laser electric field influence.}
\label{fig 1}
\end{figure}

In this framework, the total emitted $\alpha$ particle-daughter nucleus interaction potential with and without considering the laser electric field influence is shown in Fig. \ref{fig 1}. In this figure, the blue and red curves represent the total potential $V(r, \varphi)$ and the total potential $V(r, t, \varphi, \theta)$ in the laser electric field, respectively. $Q^*_{\alpha}$ and $Q_{\alpha}$ correspond to the $\alpha$ decay energy with and without considering the laser electric field effect, respectively. $R^*_{i}(i=1, 2, 3)$ and $R_{i}(i=1, 2, 3)$ refer to the classical turning points with and without considering the laser electric field effect, respectively. The schematic diagram shows that the laser electric field affects both the position of the classical turning points and the kinetic energy of the emitted $\alpha$ particle.

\section{RESULTS AND DISCUSSION}
\label{section 3}
In the present work, the $\alpha$ decay energy $Q_\alpha$, parity, spin, and the $\alpha$ decay half-lives experimental data for 190 ground state even-even nuclei from Z = 52 to Z = 118 are taken from the latest evaluated nuclear properties table NUBASE2020 \cite{NUBASE2020} and the latest evaluated atomic mass table AME2020 \cite{CPC-2021-0034, CPC-2020-0033}. $\beta_2$, $\beta_4$ and $\beta_6$ are taken from FRDM2012 \cite{MOLLER20161}. To describe the effect of the laser electric field on the $\alpha$ decay half-life, we define the rate of relative change of the $\alpha$ decay half-life $\Delta T$, 
\begin{equation}
\label{33}
\Delta T=\frac{T(E, \theta)-T(E=0, \theta)}{T(E=0, \theta)}.
\end{equation}

The normalized factor $F$ is determined by the principal quantum number $G$ \cite{PhysRevC.81.064309}, which is very insensitive to the external laser field because the integration is performed inside the nucleus from $R_1$ to $R_2$. One can safely treat the normalized factor $F$ as a laser-independent constant \cite{PhysRevC.102.064629}. Thus we assume that the external laser fields mainly affect the half-life of $\alpha$ decay by modifying the $\alpha$ decay penetration probability. The rate of the relative change of penetration probability $\Delta P$ is defined as
\begin{equation}
\Delta P=\frac{P(E, \theta)-P(E=0, \theta)}{P(E=0, \theta)}.
\label{eq34}
\end{equation}
As $\theta \ne 0$, we can consider laser field strength as a smaller laser field strength $E(t)\rm{cos}\theta$. In this work, we only consider the case of $\theta=0$, Eq. (\ref{33}) can be rewritten as
\begin{equation}
\label{eq35}
\Delta T=\frac{P(E=0)-P(E)}{P(E)}.
\end{equation}

\subsection{Gaussian laser-assisted $\alpha$ decay for the ground state even-even nuclei}

Based on the high-intensity laser pulses available in the current laboratory, we systematically investigate the effect of a peak intensity of $10^{23} {\ } \rm{W}/cm^{2}$ laser pulse on the $\alpha$ decay of the ground state even-even nuclei. The detailed results are listed in Table \ref{table 1}. In this table, the first four columns represent the parent nuclei, the minimum orbital angular momentum $l_{min}$, the $\alpha$ decay energy $Q_{\alpha}$, and the logarithmic form of the experimental $\alpha$ decay half-lives, respectively. The following three columns represent the logarithmic form of the theoretical $\alpha$ decay half-lives without considering the laser field, the rate of the relative change of penetration probability with $I_0=10^{23} {\ } \rm{W}/cm^{2}$, the rate of the relative change of $\alpha$ decay half-life with $I_0=10^{23} {\ } \rm{W}/cm^{2}$, respectively. The standard deviation $\sigma$ indicates the divergence between the theoretical $\alpha$ decay half-lives and the experimental data. It can be written as
\begin{equation}
\sigma=\sqrt{\sum [{\rm{lg}{T^{\rm{exp}}_{1/2}}(s)}-{\rm{lg}{T^{\rm{cal}}_{1/2}}(s)}]^2/n}.
\end{equation}
From Table \ref{table 1}, we can obtain the standard deviation $\sigma=0.325$. Moreover, it can be seen from Table \ref{table 1} that the difference between the theoretical half-life and the experimental data for the vast majority of nuclei is trivial. This means that the theoretical $\alpha$ decay half-lives can reproduce the experimental data well, and the model we use is trustworthy.

It can also be seen from Table \ref{table 1} that the $\mathcal{\alpha}$ decay penetration probability and the $\mathcal{\alpha}$ decay half-life of different parent nuclei have different rates of change under the influence of laser intensity of $10^{23} {\ } \rm{W}/cm^{2}$, and the rate of change ranges from $0.0009\%$ to $0.135\%$. As a particular case, the parent nucleus $^{108}$Xe corresponds to both $\Delta P$ and $\Delta T$ equal to 0. The reason is that $A_\alpha=2Z_\alpha=4$, if $A_d=2Z_d$, then $Q_{eff}=0$, so $\Delta P$ and $\Delta T$ are equal to 0. In other words, if the daughter nuclei and the emitted $\mathcal{\alpha}$ particles have the same charge-to-mass ratio, they will move cooperatively in the laser field, and the laser electric field will not have the effect of separating the two particles. As the angle between the vector $\vec{r}$ and the vector $\vec{E}(t)$ is equal to 0, the addition of the laser field increases the kinetic energy of the emitted $\mathcal{\alpha}$ particles and reduces the distance between the classical turning points $R_{2}$ and $R_{3}$, which leads to an increase in the penetration probability of $\alpha$ decay and a decrease in the half-life of $\mathcal{\alpha}$ decay. Furthermore, as seen from Table \ref{table 1}, the most sensitive parent nuclei to the high-intensity laser are $^{144}$Nd, which has the lowest decay energy among all parent nuclei.

To investigate which properties of the parent nucleus are related to $\Delta T$ and $\Delta P$, we rewrite Eq. (\ref{eq5}) as follows:

\begin{equation}
\label{eq37}
P_{\varphi}=\exp\! [- \frac{2(2\mu)^{1/2}}{\hbar} \int_{R_{2}}^{R_{3}} \sqrt{V_{\varphi lCN}(1+\frac{V_i(r, t, \varphi, \theta)}{V_{\varphi lCN}})}\, dr], 
\end{equation}
where $V_{\varphi lCN}=\lambda(\varphi) V_N(r, \varphi)+V_l(r)+V_C(r, \varphi)-Q_{\alpha}$ represents the integrand function without laser modification. Since $V_i(r, t, \varphi, \theta) \ll V_{\varphi lCN}$, the laser electric field can be regarded as a perturbation, and we take the Taylor expansion of Eq. (\ref{eq37})
\begin{equation}
\begin{split}
P_{\varphi}=&\exp\! [- \frac{2(2\mu)^{1/2}}{\hbar}\int_{R_{2}}^{R_{3}}\sqrt{V_{\varphi lCN}}\\
&\times (1+\frac{V_i(r, t, \varphi, \theta)}{2{V_{\varphi lCN}}}+\frac{V_i^2(r, t, \varphi, \theta)}{8 V^2_{\varphi lCN}}+...) \, dr]\\
\approx &\exp\! [- \frac{2(2\mu)^{1/2}}{\hbar}\int_{R_{2}}^{R_{3}}\sqrt{V_{\varphi lCN}}\\
&\times (1+\frac{V_i(r, t, \varphi, \theta)}{2{V_{\varphi lCN}}}+\frac{V_i^2(r, t, \varphi, \theta)}{8 V^2_{\varphi lCN}}) \, dr]\\
=&\exp\! [\chi_{\varphi}^{(0)}+\chi_{\varphi}^{(1)}+\chi_{\varphi}^{(2)}]\\
=&\exp\! [\chi_{\varphi}^{(0)}] \exp\! [\chi_{\varphi}^{(1)}+\chi_{\varphi}^{(2)}], 
\label{eq38}
\end{split}
\end{equation}
where $\chi_{\varphi}^{(0)}$, $\chi_{\varphi}^{(1)}$, and $\chi_{\varphi}^{(2)}$ can be expressed as
\begin{equation}
\chi_{\varphi}^{(0)}= - \frac{2(2\mu)^{1/2}}{\hbar} \int_{R_{2}}^{R_{3}} \sqrt{V_{\varphi lCN}}\, dr =P_{\varphi}(E=0), 
\end{equation}
\begin{equation}
\chi_{\varphi}^{(1)}= E(t)\times \frac{(2\mu)^{1/2}Q_{eff}\rm{cos}\theta}{\hbar} \int_{R_{2}}^{R_{3}} \frac{r}{\sqrt{V_{\varphi lCN}}}\, dr, 
\label{eq 40}
\end{equation}
\begin{equation}
\begin{split}
\chi_{\varphi}^{(2)}=&E^2(t)\times \frac{(2\mu)^{1/2}(Q_{eff}\rm{cos}\theta)^2}{4\hbar} \int_{R_{2}}^{R_{3}} \frac{r^2}{V_{\varphi lCN}^{3/2}}\, dr\\
=& \frac{1}{c\epsilon_0}I(t)\times \frac{(2\mu)^{1/2}(Q_{eff}\rm{cos}\theta)^2}{2\hbar} \int_{R_{2}}^{R_{3}} \frac{r^2}{V_{\varphi lCN}^{3/2}}\, dr, 
\end{split}
\end{equation}
where $I(t)$ is the laser field intensity proportional to the square of the laser electric field intensity $E^{2}(t)$. $P_{\varphi}(E=0)$  is the $\alpha$ penetration probability without laser modification. The rate of change of penetration probability can be written as
\begin{equation}
\begin{split}
\Delta P_{\varphi}\approx&\frac{\exp\! [\chi_{\varphi}^{(0)}] \exp\! [\chi_{\varphi}^{(1)}+\chi_{\varphi}^{(2)}]-\exp\! [\chi_{\varphi}^{(0)}]}{\exp\! [\chi_{\varphi}^{(0)}]}\\
=&\exp\! [\chi_{\varphi}^{(1)}+\chi_{\varphi}^{(2)}]-1.
\label{42}
\end{split}
\end{equation}
As $\chi_{\varphi}^{(1)}+\chi_{\varphi}^{(2)}$ approaches 0, $\exp [\chi_{\varphi}^{(1)}+\chi_{\varphi}^{(2)}]$ approaches 1+$\chi_{\varphi}^{(1)}+\chi_{\varphi}^{(2)}$, Eq. (\ref{42}) can be rewritten as:
\begin{equation}
\begin{split}
\Delta P_{\varphi}=\chi_{\varphi}^{(1)}+\chi_{\varphi}^{(2)}.
\label{eq41}
\end{split}
\end{equation}
Similarly, Eq. (\ref{eq35}) can be rewritten as
\begin{equation}
\Delta T_{\varphi}=-\frac{\chi_{\varphi}^{(1)}+\chi_{\varphi}^{(2)}}{1+\chi_{\varphi}^{(1)}+\chi_{\varphi}^{(2)}}.
\end{equation}

Due to the difficulty in integrating Eq. (\ref{eq4}), the analytical solution of the rate of the relative change of penetration probability $\Delta P$ can not be obtained precisely. To proceed, we used a spherical Gamow-like model that has been shown to be capable of reproducing the experimental data of $\alpha$ decay well instead of Eqs. (\ref{eq4}) and (\ref{eq5}) \cite{PhysRevC.87.024308}. In this model, the total potential $V(r)$ between the emitted $\alpha$ particle and the daughter nucleus is considered as a square situation well in the case of $r < R_{in}$, where $R_{in} =1.15(A_{d}^{1/3}+ A_{\alpha}^{1/3} )$ $\rm{fm}$ is the geometrical touching distance. For the case of $r > R_{in}$, the total potential $V(r)$ between the emitted $\alpha$ particle and the daughter nucleus is reduced to the Coulomb potential between the two particles.This means that the intersection point $R_{out}$ of $\alpha$ and $Q_{\alpha}$ is determined as $R_{out} = Z_d Z_\alpha e^2/Q_{\alpha}$. For the spherical approximation, Eq. (\ref{eq41}) can be rewritten as

\clearpage
\begingroup
\renewcommand*{\arraystretch}{1.535}
\setlength{\tabcolsep}{15pt}
\begin{longtable*}{ccccccc}
\caption{The effect of laser pulse with a peak intensity of $10^{23} {\ } \rm{W}/cm^{2}$ on the $\alpha$ decay of the ground state even-even nuclei. Here ${\text{lg}T_{1/2}^{\text{cal}}}$ is the logarithmic form of the $\mathcal{\alpha}$ decay half-life calculated in this work.}
\renewcommand\arraystretch{100}
\setlength{\tabcolsep}{20pt}
\setlength\LTleft{-20in}
\setlength\LTright{-1in plus 2 fill}
\label{table 1} \\
\hline Nucleus & $l_{min}$ & $Q_{\alpha}(\rm{MeV})$ &${\text{lg}{T_{1/2}^{\text{exp}}}}$ (s)&${\text{lg}{T_{1/2}^{\text{cal}}}}$ (s)&$\Delta P$& $\Delta T$ \\ \hline
\endfirsthead
\multicolumn{6}{c}%
{{\tablename\ \thetable{} -- continued from previous page}} \\
\hline Nucleus & $l_{min}$ & $Q_{\alpha}(\rm{MeV})$ &${\text{lg}{T_{1/2}^{\text{exp}}}}$ (s)&${\text{lg}{T_{1/2}^{\text{cal}}}}$ (s)&$\Delta P$& $\Delta T$ \\ \hline
\endhead
\hline \multicolumn{6}{r}{{Continued on next page}} \\
\endfoot
\hline \hline
\endlastfoot
$^{106}$Te&0&4.285&-4.108&-4.279&1.15$\times$$10^{-4}$&-1.15$\times$$10^{-4}$\\
$^{108}$Te&0&3.42&0.628&0.447&1.85$\times$$10^{-4}$&-1.85$\times$$10^{-4}$\\
$^{108}$Xe&0&4.575&-4.143&-4.391&0&0\\
$^{110}$Xe&0&3.875&-0.84&-1.031&1.45$\times$$10^{-4}$&-1.45$\times$$10^{-4}$\\
$^{112}$Xe&0&3.331&2.324&2.347&2.10$\times$$10^{-4}$&-2.10$\times$$10^{-4}$\\
$^{114}$Ba&0&3.585&1.694&1.884&1.75$\times$$10^{-4}$&-1.75$\times$$10^{-4}$\\
$^{144}$Nd&0&1.901&22.859&23.083&1.35$\times$$10^{-3}$&-1.35$\times$$10^{-3}$\\
$^{146}$Sm&0&2.529&15.332&15.41&7.86$\times$$10^{-4}$&-7.86$\times$$10^{-4}$\\
$^{148}$Sm&0&1.987&23.298&23.45&1.28$\times$$10^{-3}$&-1.28$\times$$10^{-3}$\\
$^{148}$Gd&0&3.271&9.352&9.186&4.33$\times$$10^{-4}$&-4.33$\times$$10^{-4}$\\
$^{150}$Gd&0&2.807&13.752&13.726&6.16$\times$$10^{-4}$&-6.16$\times$$10^{-4}$\\
$^{152}$Gd&0&2.204&21.533&21.558&1.09$\times$$10^{-3}$&-1.09$\times$$10^{-3}$\\
$^{150}$Dy&0&4.351&3.107&2.812&2.09$\times$$10^{-4}$&-2.09$\times$$10^{-4}$\\
$^{152}$Dy&0&3.726&6.93&6.885&3.47$\times$$10^{-4}$&-3.47$\times$$10^{-4}$\\
$^{154}$Dy&0&2.945&13.976&13.658&5.94$\times$$10^{-4}$&-5.94$\times$$10^{-4}$\\
$^{152}$Er&0&4.934&1.057&0.812&1.94$\times$$10^{-4}$&-1.94$\times$$10^{-4}$\\
$^{154}$Er&0&4.28&4.677&4.435&2.64$\times$$10^{-4}$&-2.64$\times$$10^{-4}$\\
$^{156}$Er&0&3.481&9.989&10.057&4.20$\times$$10^{-4}$&-4.20$\times$$10^{-4}$\\
$^{154}$Yb&0&5.474&-0.355&-0.642&1.59$\times$$10^{-4}$&-1.59$\times$$10^{-4}$\\
$^{156}$Yb&0&4.809&2.408&2.501&2.09$\times$$10^{-4}$&-2.09$\times$$10^{-4}$\\
$^{156}$Hf&0&6.025&-1.638&-1.942&1.34$\times$$10^{-4}$&-1.34$\times$$10^{-4}$\\
$^{158}$Hf&0&5.405&0.808&0.629&1.68$\times$$10^{-4}$&-1.68$\times$$10^{-4}$\\
$^{160}$Hf&0&4.902&3.276&3.048&2.15$\times$$10^{-4}$&-2.15$\times$$10^{-4}$\\
$^{162}$Hf&0&4.417&5.687&5.788&2.76$\times$$10^{-4}$&-2.76$\times$$10^{-4}$\\
$^{174}$Hf&0&2.494&22.8&23.9&1.11$\times$$10^{-3}$&-1.10$\times$$10^{-3}$\\
$^{158}$W&0&6.615&-2.845&-3.153&1.15$\times$$10^{-4}$&-1.15$\times$$10^{-4}$\\
$^{160}$W&0&6.065&-0.989&-1.185&1.38$\times$$10^{-4}$&-1.38$\times$$10^{-4}$\\
$^{162}$W&0&5.678&0.42&0.365&1.66$\times$$10^{-4}$&-1.66$\times$$10^{-4}$\\
$^{164}$W&0&5.278&2.218&2.173&1.93$\times$$10^{-4}$&-1.93$\times$$10^{-4}$\\
$^{166}$W&0&4.856&4.738&4.319&2.33$\times$$10^{-4}$&-2.33$\times$$10^{-4}$\\
$^{168}$W&0&4.5&6.2&6.368&2.85$\times$$10^{-4}$&-2.85$\times$$10^{-4}$\\
$^{180}$W&0&2.515&25.7&25.247&1.16$\times$$10^{-3}$&-1.16$\times$$10^{-3}$\\
$^{162}$Os&0&6.765&-2.678&-2.849&1.12$\times$$10^{-4}$&-1.12$\times$$10^{-4}$\\
$^{164}$Os&0&6.485&-1.662&-1.874&1.30$\times$$10^{-4}$&-1.30$\times$$10^{-4}$\\
$^{166}$Os&0&6.142&-0.593&-0.639&1.46$\times$$10^{-4}$&-1.46$\times$$10^{-4}$\\
$^{168}$Os&0&5.816&0.685&0.679&1.65$\times$$10^{-4}$&-1.65$\times$$10^{-4}$\\
$^{170}$Os&0&5.536&1.889&1.891&1.93$\times$$10^{-4}$&-1.93$\times$$10^{-4}$\\
$^{172}$Os&0&5.224&3.207&3.373&2.22$\times$$10^{-4}$&-2.22$\times$$10^{-4}$\\
$^{174}$Os&0&4.871&5.251&5.234&2.62$\times$$10^{-4}$&-2.62$\times$$10^{-4}$\\
$^{186}$Os&0&2.821&22.8&22.594&9.55$\times$$10^{-4}$&-9.54$\times$$10^{-4}$\\
$^{166}$Pt&0&7.295&-3.532&-3.746&1.03$\times$$10^{-4}$&-1.03$\times$$10^{-4}$\\
$^{168}$Pt&0&6.985&-2.695&-2.807&1.15$\times$$10^{-4}$&-1.15$\times$$10^{-4}$\\
$^{170}$Pt&0&6.708&-1.856&-1.874&1.33$\times$$10^{-4}$&-1.33$\times$$10^{-4}$\\
$^{172}$Pt&0&6.463&-0.994&-1.015&1.41$\times$$10^{-4}$&-1.41$\times$$10^{-4}$\\
$^{174}$Pt&0&6.363&0.061&0.039&1.62$\times$$10^{-4}$&-1.62$\times$$10^{-4}$\\
$^{176}$Pt&0&5.885&1.197&1.253&1.82$\times$$10^{-4}$&-1.82$\times$$10^{-4}$\\
$^{178}$Pt&0&5.573&2.428&2.626&2.10$\times$$10^{-4}$&-2.10$\times$$10^{-4}$\\
$^{180}$Pt&0&5.276&4.028&4.065&2.40$\times$$10^{-4}$&-2.40$\times$$10^{-4}$\\
$^{182}$Pt&0&4.951&5.623&5.793&2.81$\times$$10^{-4}$&-2.81$\times$$10^{-4}$\\
$^{184}$Pt&0&4.599&7.768&7.885&3.35$\times$$10^{-4}$&-3.35$\times$$10^{-4}$\\
$^{186}$Pt&0&4.32&9.728&9.718&3.83$\times$$10^{-4}$&-3.83$\times$$10^{-4}$\\
$^{188}$Pt&0&4.007&12.528&12.019&4.63$\times$$10^{-4}$&-4.63$\times$$10^{-4}$\\
$^{190}$Pt&0&3.269&19.183&18.796&7.17$\times$$10^{-4}$&-7.16$\times$$10^{-4}$\\
$^{170}$Hg&0&7.775&-3.509&-4.396&9.73$\times$$10^{-5}$&-9.73$\times$$10^{-5}$\\
$^{172}$Hg&0&7.525&-3.636&-3.7&1.07$\times$$10^{-4}$&-1.07$\times$$10^{-4}$\\
$^{174}$Hg&0&7.233&-2.699&-2.814&1.19$\times$$10^{-4}$&-1.19$\times$$10^{-4}$\\
$^{176}$Hg&0&6.897&-1.651&-1.718&1.27$\times$$10^{-4}$&-1.27$\times$$10^{-4}$\\
$^{178}$Hg&0&6.398&-0.526&-0.598&1.50$\times$$10^{-4}$&-1.50$\times$$10^{-4}$\\
$^{180}$Hg&0&6.258&0.73&0.554&1.70$\times$$10^{-4}$&-1.69$\times$$10^{-4}$\\
$^{182}$Hg&0&5.995&1.892&1.625&1.93$\times$$10^{-4}$&-1.93$\times$$10^{-4}$\\
$^{184}$Hg&0&5.66&3.442&3.125&2.18$\times$$10^{-4}$&-2.18$\times$$10^{-4}$\\
$^{186}$Hg&0&5.204&5.701&5.446&2.60$\times$$10^{-4}$&-2.60$\times$$10^{-4}$\\
$^{188}$Hg&0&4.709&8.722&8.312&3.28$\times$$10^{-4}$&-3.28$\times$$10^{-4}$\\
$^{178}$Pb&0&7.789&-3.602&-3.75&1.06$\times$$10^{-4}$&-1.06$\times$$10^{-4}$\\
$^{180}$Pb&0&7.419&-2.387&-2.642&1.18$\times$$10^{-4}$&-1.18$\times$$10^{-4}$\\
$^{182}$Pb&0&7.065&-1.26&-1.492&1.33$\times$$10^{-4}$&-1.33$\times$$10^{-4}$\\
$^{184}$Pb&0&6.774&-0.213&-0.478&1.47$\times$$10^{-4}$&-1.47$\times$$10^{-4}$\\
$^{186}$Pb&0&6.471&1.072&0.648&1.63$\times$$10^{-4}$&-1.63$\times$$10^{-4}$\\
$^{188}$Pb&0&6.109&2.468&2.123&1.88$\times$$10^{-4}$&-1.88$\times$$10^{-4}$\\
$^{190}$Pb&0&5.697&4.245&3.981&2.27$\times$$10^{-4}$&-2.27$\times$$10^{-4}$\\
$^{192}$Pb&0&5.221&6.546&6.419&2.74$\times$$10^{-4}$&-2.74$\times$$10^{-4}$\\
$^{194}$Pb&0&4.738&9.944&9.301&3.40$\times$$10^{-4}$&-3.40$\times$$10^{-4}$\\
$^{210}$Pb&0&3.793&16.567&16.082&6.46$\times$$10^{-4}$&-6.45$\times$$10^{-4}$\\
$^{186}$Po&0&8.502&-4.469&-5.035&9.50$\times$$10^{-5}$&-9.50$\times$$10^{-5}$\\
$^{188}$Po&0&8.083&-3.569&-3.911&1.11$\times$$10^{-4}$&-1.11$\times$$10^{-4}$\\
$^{190}$Po&0&7.693&-2.611&-2.778&1.23$\times$$10^{-4}$&-1.23$\times$$10^{-4}$\\
$^{192}$Po&0&7.32&-1.492&-1.594&1.39$\times$$10^{-4}$&-1.39$\times$$10^{-4}$\\
$^{194}$Po&0&6.987&-0.407&-0.473&1.90$\times$$10^{-4}$&-1.90$\times$$10^{-4}$\\
$^{196}$Po&0&6.658&0.775&0.764&1.74$\times$$10^{-4}$&-1.74$\times$$10^{-4}$\\
$^{198}$Po&0&6.31&2.266&2.14&2.36$\times$$10^{-4}$&-2.36$\times$$10^{-4}$\\
$^{200}$Po&0&5.981&3.793&3.577&2.25$\times$$10^{-4}$&-2.25$\times$$10^{-4}$\\
$^{202}$Po&0&5.7&5.143&4.87&2.53$\times$$10^{-4}$&-2.53$\times$$10^{-4}$\\
$^{204}$Po&0&5.485&6.275&6.069&2.80$\times$$10^{-4}$&-2.80$\times$$10^{-4}$\\
$^{206}$Po&0&5.327&7.144&6.78&3.04$\times$$10^{-4}$&-3.04$\times$$10^{-4}$\\
$^{208}$Po&0&5.216&7.961&7.391&3.23$\times$$10^{-4}$&-3.23$\times$$10^{-4}$\\
$^{210}$Po&0&5.407&7.078&6.302&3.07$\times$$10^{-4}$&-3.07$\times$$10^{-4}$\\
$^{212}$Po&0&8.954&-6.531&-6.805&1.41$\times$$10^{-4}$&-1.41$\times$$10^{-4}$\\
$^{214}$Po&0&7.833&-3.787&-3.785&1.22$\times$$10^{-4}$&-1.22$\times$$10^{-4}$\\
$^{216}$Po&0&6.906&-0.842&-0.736&1.98$\times$$10^{-4}$&-1.98$\times$$10^{-4}$\\
$^{218}$Po&0&6.115&2.269&2.452&2.95$\times$$10^{-4}$&-2.95$\times$$10^{-4}$\\
$^{194}$Rn&0&7.863&-3.108&-2.626&1.21$\times$$10^{-4}$&-1.21$\times$$10^{-4}$\\
$^{196}$Rn&0&7.616&-2.328&-1.879&1.35$\times$$10^{-4}$&-1.35$\times$$10^{-4}$\\
$^{198}$Rn&0&7.35&-1.163&-1.005&1.53$\times$$10^{-4}$&-1.53$\times$$10^{-4}$\\
$^{200}$Rn&0&7.044&0.07&0.135&1.65$\times$$10^{-4}$&-1.65$\times$$10^{-4}$\\
$^{202}$Rn&0&6.773&1.09&1.138&1.77$\times$$10^{-4}$&-1.77$\times$$10^{-4}$\\
$^{204}$Rn&0&6.547&2.012&2.019&1.94$\times$$10^{-4}$&-1.94$\times$$10^{-4}$\\
$^{206}$Rn&0&6.384&2.737&2.675&2.14$\times$$10^{-4}$&-2.14$\times$$10^{-4}$\\
$^{208}$Rn&0&6.261&3.367&3.189&2.22$\times$$10^{-4}$&-2.21$\times$$10^{-4}$\\
$^{210}$Rn&0&6.159&3.954&3.597&2.34$\times$$10^{-4}$&-2.34$\times$$10^{-4}$\\
$^{212}$Rn&0&6.385&3.157&2.619&2.23$\times$$10^{-4}$&-2.23$\times$$10^{-4}$\\
$^{214}$Rn&0&9.208&-6.587&-6.711&1.11$\times$$10^{-4}$&-1.11$\times$$10^{-4}$\\
$^{216}$Rn&0&8.197&-4.538&-4.094&1.41$\times$$10^{-4}$&-1.41$\times$$10^{-4}$\\
$^{218}$Rn&0&7.262&-1.472&-1.123&1.82$\times$$10^{-4}$&-1.82$\times$$10^{-4}$\\
$^{220}$Rn&0&6.405&1.745&2.143&2.38$\times$$10^{-4}$&-2.38$\times$$10^{-4}$\\
$^{222}$Rn&0&5.59&5.519&5.93&3.17$\times$$10^{-4}$&-3.17$\times$$10^{-4}$\\
$^{202}$Ra&0&7.88&-2.387&-1.979&1.32$\times$$10^{-4}$&-1.32$\times$$10^{-4}$\\
$^{204}$Ra&0&7.636&-1.222&-1.2&1.45$\times$$10^{-4}$&-1.45$\times$$10^{-4}$\\
$^{206}$Ra&0&7.416&-0.62&-0.408&1.54$\times$$10^{-4}$&-1.54$\times$$10^{-4}$\\
$^{208}$Ra&0&7.273&0.104&0.062&1.65$\times$$10^{-4}$&-1.65$\times$$10^{-4}$\\
$^{210}$Ra&0&7.151&0.602&0.486&1.73$\times$$10^{-4}$&-1.73$\times$$10^{-4}$\\
$^{214}$Ra&0&7.273&0.387&0.006&1.74$\times$$10^{-4}$&-1.74$\times$$10^{-4}$\\
$^{216}$Ra&0&9.525&-6.764&-6.616&1.05$\times$$10^{-4}$&-1.05$\times$$10^{-4}$\\
$^{218}$Ra&0&8.541&-4.587&-4.297&1.32$\times$$10^{-4}$&-1.32$\times$$10^{-4}$\\
$^{220}$Ra&0&7.594&-1.742&-1.442&1.69$\times$$10^{-4}$&-1.69$\times$$10^{-4}$\\
$^{222}$Ra&0&6.678&1.526&1.866&2.25$\times$$10^{-4}$&-2.25$\times$$10^{-4}$\\
$^{224}$Ra&0&5.789&5.497&5.851&3.04$\times$$10^{-4}$&-3.04$\times$$10^{-4}$\\
$^{226}$Ra&0&4.871&10.703&11.159&4.41$\times$$10^{-4}$&-4.41$\times$$10^{-4}$\\
$^{208}$Th&0&8.204&-2.62&-2.21&1.35$\times$$10^{-4}$&-1.35$\times$$10^{-4}$\\
$^{210}$Th&0&8.069&-1.796&-1.777&1.35$\times$$10^{-4}$&-1.35$\times$$10^{-4}$\\
$^{212}$Th&0&7.958&-1.499&-1.456&1.45$\times$$10^{-4}$&-1.45$\times$$10^{-4}$\\
$^{214}$Th&0&7.827&-1.06&-1.04&1.51$\times$$10^{-4}$&-1.51$\times$$10^{-4}$\\
$^{216}$Th&0&8.073&-1.58&-1.823&1.44$\times$$10^{-4}$&-1.44$\times$$10^{-4}$\\
$^{218}$Th&0&9.849&-6.914&-6.826&1.00$\times$$10^{-4}$&-1.00$\times$$10^{-4}$\\
$^{220}$Th&0&8.974&-4.991&-4.715&1.21$\times$$10^{-4}$&-1.21$\times$$10^{-4}$\\
$^{222}$Th&0&8.132&-2.65&-2.407&1.51$\times$$10^{-4}$&-1.50$\times$$10^{-4}$\\
$^{224}$Th&0&7.299&0.017&0.282&1.92$\times$$10^{-4}$&-1.92$\times$$10^{-4}$\\
$^{226}$Th&0&6.453&3.265&3.64&2.44$\times$$10^{-4}$&-2.44$\times$$10^{-4}$\\
$^{228}$Th&0&5.52&7.781&8.231&3.49$\times$$10^{-4}$&-3.49$\times$$10^{-4}$\\
$^{230}$Th&0&4.77&12.376&12.848&4.85$\times$$10^{-4}$&-4.85$\times$$10^{-4}$\\
$^{232}$Th&0&4.082&17.645&18.248&6.83$\times$$10^{-4}$&-6.82$\times$$10^{-4}$\\
$^{216}$U&0&8.53&-2.161&-2.432&1.29$\times$$10^{-4}$&-1.29$\times$$10^{-4}$\\
$^{218}$U&0&8.775&-3.451&-3.141&1.24$\times$$10^{-4}$&-1.24$\times$$10^{-4}$\\
$^{222}$U&0&9.478&-5.328&-5.326&1.11$\times$$10^{-4}$&-1.11$\times$$10^{-4}$\\
$^{224}$U&0&8.628&-3.402&-3.123&1.40$\times$$10^{-4}$&-1.40$\times$$10^{-4}$\\
$^{226}$U&0&7.701&-0.57&-0.274&1.79$\times$$10^{-4}$&-1.79$\times$$10^{-4}$\\
$^{228}$U&0&6.799&2.748&3.046&2.27$\times$$10^{-4}$&-2.27$\times$$10^{-4}$\\
$^{230}$U&0&5.992&6.243&6.706&3.04$\times$$10^{-4}$&-3.04$\times$$10^{-4}$\\
$^{232}$U&0&5.414&9.337&9.786&3.80$\times$$10^{-4}$&-3.80$\times$$10^{-4}$\\
$^{234}$U&0&4.858&12.889&13.245&4.95$\times$$10^{-4}$&-4.95$\times$$10^{-4}$\\
$^{236}$U&0&4.573&14.869&15.328&5.38$\times$$10^{-4}$&-5.38$\times$$10^{-4}$\\
$^{238}$U&0&4.27&17.149&17.742&6.53$\times$$10^{-4}$&-6.53$\times$$10^{-4}$\\
$^{228}$Pu&0&7.94&0.322&-0.285&1.65$\times$$10^{-4}$&-1.65$\times$$10^{-4}$\\
$^{230}$Pu&0&7.178&2.021&2.403&2.08$\times$$10^{-4}$&-2.08$\times$$10^{-4}$\\
$^{234}$Pu&0&6.31&5.723&5.942&2.87$\times$$10^{-4}$&-2.87$\times$$10^{-4}$\\
$^{236}$Pu&0&5.867&7.955&8.16&3.33$\times$$10^{-4}$&-3.32$\times$$10^{-4}$\\
$^{238}$Pu&0&5.593&9.442&9.678&3.80$\times$$10^{-4}$&-3.80$\times$$10^{-4}$\\
$^{240}$Pu&0&5.256&11.316&11.677&4.38$\times$$10^{-4}$&-4.37$\times$$10^{-4}$\\
$^{242}$Pu&0&4.984&13.073&13.445&4.93$\times$$10^{-4}$&-4.93$\times$$10^{-4}$\\
$^{244}$Pu&0&4.666&15.41&15.74&5.70$\times$$10^{-4}$&-5.70$\times$$10^{-4}$\\
$^{234}$Cm&0&7.365&2.285&2.386&2.08$\times$$10^{-4}$&-2.08$\times$$10^{-4}$\\
$^{236}$Cm&0&7.067&3.351&3.537&2.31$\times$$10^{-4}$&-2.31$\times$$10^{-4}$\\
$^{238}$Cm&0&6.67&5.314&5.201&2.67$\times$$10^{-4}$&-2.67$\times$$10^{-4}$\\
$^{240}$Cm&0&6.398&6.419&6.452&2.91$\times$$10^{-4}$&-2.91$\times$$10^{-4}$\\
$^{242}$Cm&0&6.216&7.148&7.334&3.11$\times$$10^{-4}$&-3.11$\times$$10^{-4}$\\
$^{244}$Cm&0&5.902&8.757&8.944&3.47$\times$$10^{-4}$&-3.47$\times$$10^{-4}$\\
$^{246}$Cm&0&5.475&11.172&11.377&4.09$\times$$10^{-4}$&-4.09$\times$$10^{-4}$\\
$^{248}$Cm&0&5.162&13.079&13.352&4.81$\times$$10^{-4}$&-4.81$\times$$10^{-4}$\\
$^{238}$Cf&0&8.133&-0.076&0.47&1.79$\times$$10^{-4}$&-1.79$\times$$10^{-4}$\\
$^{240}$Cf&0&7.711&1.612&1.902&2.00$\times$$10^{-4}$&-2.00$\times$$10^{-4}$\\
$^{242}$Cf&0&7.517&2.534&2.61&2.14$\times$$10^{-4}$&-2.14$\times$$10^{-4}$\\
$^{244}$Cf&0&7.329&3.19&3.311&2.28$\times$$10^{-4}$&-2.27$\times$$10^{-4}$\\
$^{246}$Cf&0&6.862&5.109&5.224&2.63$\times$$10^{-4}$&-2.63$\times$$10^{-4}$\\
$^{248}$Cf&0&6.361&7.46&7.528&3.06$\times$$10^{-4}$&-3.06$\times$$10^{-4}$\\
$^{250}$Cf&0&6.128&8.616&8.692&3.35$\times$$10^{-4}$&-3.35$\times$$10^{-4}$\\
$^{252}$Cf&0&6.217&7.935&8.217&3.29$\times$$10^{-4}$&-3.29$\times$$10^{-4}$\\
$^{254}$Cf&0&5.926&9.224&9.736&3.69$\times$$10^{-4}$&-3.69$\times$$10^{-4}$\\
$^{246}$Fm&0&8.379&0.218&0.379&1.73$\times$$10^{-4}$&-1.73$\times$$10^{-4}$\\
$^{248}$Fm&0&7.995&1.538&1.649&1.95$\times$$10^{-4}$&-1.95$\times$$10^{-4}$\\
$^{250}$Fm&0&7.557&3.27&3.241&2.18$\times$$10^{-4}$&-2.18$\times$$10^{-4}$\\
$^{252}$Fm&0&7.154&4.961&4.834&2.50$\times$$10^{-4}$&-2.50$\times$$10^{-4}$\\
$^{254}$Fm&0&7.307&4.067&4.178&2.41$\times$$10^{-4}$&-2.40$\times$$10^{-4}$\\
$^{256}$Fm&0&7.025&5.064&5.346&2.63$\times$$10^{-4}$&-2.62$\times$$10^{-4}$\\
$^{252}$No&0&8.548&0.562&0.547&1.74$\times$$10^{-4}$&-1.74$\times$$10^{-4}$\\
$^{254}$No&0&8.226&1.755&1.599&1.90$\times$$10^{-4}$&-1.90$\times$$10^{-4}$\\
$^{256}$No&0&8.581&0.466&0.398&1.78$\times$$10^{-4}$&-1.78$\times$$10^{-4}$\\
$^{256}$Rf&0&8.926&0.327&0.094&1.65$\times$$10^{-4}$&-1.65$\times$$10^{-4}$\\
$^{258}$Rf&0&9.196&-0.595&-0.751&1.61$\times$$10^{-4}$&-1.61$\times$$10^{-4}$\\
$^{260}$Sg&0&9.9&-1.772&-2.009&1.37$\times$$10^{-4}$&-1.37$\times$$10^{-4}$\\
$^{266}$Hs&0&10.346&-2.409&-2.543&1.32$\times$$10^{-4}$&-1.32$\times$$10^{-4}$\\
$^{268}$Hs&0&9.765&0.146&-0.988&1.49$\times$$10^{-4}$&-1.49$\times$$10^{-4}$\\
$^{270}$Hs&0&9.065&0.954&1.029&1.82$\times$$10^{-4}$&-1.82$\times$$10^{-4}$\\
$^{270}$Ds&0&11.115&-3.688&-3.775&1.17$\times$$10^{-4}$&-1.17$\times$$10^{-4}$\\
$^{282}$Ds&0&9.145&2.401&1.513&1.82$\times$$10^{-4}$&-1.82$\times$$10^{-4}$\\
$^{286}$Cn&0&9.235&1.477&1.943&1.82$\times$$10^{-4}$&-1.82$\times$$10^{-4}$\\
$^{286}$Fl&0&10.355&-0.658&-0.575&1.46$\times$$10^{-4}$&-1.46$\times$$10^{-4}$\\
$^{288}$Fl&0&10.075&-0.185&0.187&1.51$\times$$10^{-4}$&-1.51$\times$$10^{-4}$\\
$^{290}$Fl&0&9.855&1.903&0.786&1.63$\times$$10^{-4}$&-1.63$\times$$10^{-4}$\\
$^{290}$Lv&0&10.995&-2.046&-1.585&1.31$\times$$10^{-4}$&-1.31$\times$$10^{-4}$\\
$^{292}$Lv&0&10.785&-1.796&-1.064&1.32$\times$$10^{-4}$&-1.32$\times$$10^{-4}$\\
$^{294}$Og&0&11.865&-3.155&-3.053&1.15$\times$$10^{-4}$&-1.15$\times$$10^{-4}$\\
\end{longtable*}
\endgroup

\begin{equation}
\Delta P=\chi^{(1)}+\chi^{(2)}, 
\label{eq45}
\end{equation}
where $\chi^{(1)}$ can be expressed as
\begin{equation}
\begin{split}
\chi^{(1)}=&E(t)\times \frac{(2\mu)^{1/2}Q_{eff}\rm{cos}\theta}{\hbar} \int_{R_{in}}^{R_{out}} \frac{r}{\sqrt{V_0}}\, dr\\
=& B \int_{R_{in}}^{R_{out}} \frac{r}{\sqrt{V_0}}\, dr.
\end{split}
\end{equation}
Here $V_{0}$ represents the difference between the total potential energy $V(r)$ and the decay energy $Q_{\alpha}$, which can be written as
\begin{equation}
V_{0}=\frac{Z_d Z_\alpha e^2}{r}-Q_{\alpha}=Q_{\alpha}(\frac{R_{out}}{r}-1).
\label{eq46}
\end{equation}

We introduce a parameter $\kappa={\rm{arccos}}\sqrt{R_{in}/R_{out}}$, then $\chi^{(1)}$ can be integrated analytically, 
\begin{equation}
\begin{split}
\chi^{(1)}=&B Q_{\alpha}^{-1/2} \int_{R_{in}}^{R_{out}} \frac{r}{\sqrt{\frac{R_{out}}{r}-1}}\, dr\\
=& B Q_{\alpha}^{-5/2} (Z_d Z_\alpha e^2)^2 [\frac{3}{4}\kappa+\frac{1}{4}(3+2\rm{cos}^2\kappa)\rm{sin}\kappa \rm{cos}\kappa].
\label{eq48}
\end{split}
\end{equation}
For $\alpha$ decay, $R_{in} \ll R_{out}$, and we can obtain
\begin{equation}
\kappa \approx \pi/2, \rm{sin}\kappa \approx 1, \rm{cos}\kappa=\sqrt{\emph{R}_{\emph{in}}/\emph{R}_{\emph{out}}}.
\label{eq49}
\end{equation}
Bringing Eq. (\ref{eq49}) into Eq. (\ref{eq48}), we get the analytical solution of $\chi^{(1)}$, 
\begin{equation}
\chi^{(1)} \approx B_1Q_\alpha^{-5/2}+B_2Q_\alpha^{-2}+B_3Q_\alpha^{-1}, 
\label{eq50}
\end{equation}
where $B_1$, $B_2$, and $B_3$ can be expressed as
\begin{equation}
B_1=\frac{3 \pi E(t)\sqrt{2 \mu} Q_{eff} {\rm{cos}} \theta (Z_d Z_\alpha e^2)^2}{8 \hbar}, 
\end{equation}
\begin{equation}
B_2=\frac{3 E(t)\sqrt{2 \mu} Q_{eff} {\rm{cos}}\theta (Z_d Z_\alpha e^2)^{3/2} \sqrt{R_{in}}}{4 \hbar}, 
\end{equation}
\begin{equation}
B_3=\frac{E(t)\sqrt{2 \mu} Q_{eff} {\rm{cos}}\theta (Z_d Z_\alpha e^2)^{1/2} (R_{in})^{3/2}}{2 \hbar}.
\end{equation}
Similarly, $\chi^{(2)}$ can be expressed as
\begin{equation}
\begin{split}
\chi^{(2)}=&E^2(t)\times \frac{(2\mu)^{1/2}(Q_{eff}\rm{cos}\theta)^2}{4\hbar} \int_{R_{in}}^{R_{out}} \frac{r^2}{V_0^{3/2}}\, dr\\
=& C \int_{R_{in}}^{R_{out}} \frac{r^2}{V_0^{3/2}}\, dr.
\label{eq54}
\end{split}
\end{equation}

The integral result of the Eq. (\ref{eq54}) in the integration region $R_{in}$ to $R_{out}$ is divergent. The reason for this violation of the actual law is that the $V_i(r, t, \varphi, \theta) \ll V_{0}$ preconditioner is not satisfied in the most terminal integral region. To obtain the analytic solution of $\chi^{(2)}$, the intersection $R_{c}$ satisfying equation $V_i(r, t, \varphi, \theta) = V_{0}$ is used to replace the upper limit of integration $R_{out}$, which can be expressed as 
\begin{equation}
R_c=\frac{-Q_\alpha+\sqrt{(Q_\alpha^2)+4 (Z_d Z_\alpha e^2) E(t) Q_{eff}}}{2 E(t) Q_{eff}}.
\end{equation}
By calculation, we find that the integral part $R_{out}-R_{c}$ that is rounded off is only one ten-thousandth of the total integral length. Moreover, it is evident from Fig. \ref{fig 1} that the area close to the integral end is much smaller than the rest of the integral, which means  this approximation is reasonable. Then, we introduce a parameter $f=R_{c}/R_{out}$, Eq. (\ref{eq54}) can thus be rewritten as
\begin{equation}
\begin{split}
\chi^{(2)}\approx &C_1 Q_\alpha^{-9/2}, 
\end{split}
\end{equation}
where $C_1$ can be given by
\begin{equation}
C_1=\frac{E^2(t)\sqrt{2 \mu} (Q_{eff} {\rm{cos}\theta)^2} (Z_d Z_\alpha e^2)^3}{4 \hbar}(T_{rc}-\frac{35}{16}\pi).
\end{equation}
The parameter $T_{rc}$ of the above equation can be expressed with $f$
\begin{equation}
T_{rc}=\frac{105-35f-14f^2-8f^3}{24(\frac{1}{f}-1)}\sqrt{\frac{1}{f}-1}+\frac{35}{8}\rm{arccos}\sqrt{f}.
\end{equation}
Now, we obtain the analytical solution of the rate of the relative change of penetration probability $\Delta P$
\begin{equation}
\Delta P \approx B_1Q_\alpha^{-5/2}+B_2Q_\alpha^{-2}+B_3Q_\alpha^{-1}+C_1 Q^{-9/2}.
\label{eq59}
\end{equation}
Similarly, by approximating Eq. (\ref{eq49}), we give here the analytical solution of $\chi^{(0)}$
\begin{equation}
\begin{split}
\chi^{(0)}=&\frac{2(2\mu)^{1/2}}{\hbar} \int_{R_{in}}^{R_{out}} \sqrt{V_0}\, dr\\
\approx& -A_1 Q_\alpha^{-1/2}+A_2, 
\end{split}
\end{equation}
where $A_1$ and $A_2$ can be expressed as
\begin{equation}
A_1=\frac{ \pi \sqrt{2 \mu} Z_d Z_\alpha e^2}{\hbar}, 
\end{equation}
\begin{equation}
A_2=\frac{ 2 \sqrt{2 \mu Z_d Z_\alpha e^2 R_{in} }}{\hbar}, 
\end{equation}
Here $A_1$ is the so-called Gamow constant, and the term $\exp(\chi^{(0)})$ is the $\alpha$ decay penetration probability without laser electric field. The formula for the $\alpha$ decay penetration probability under the influence of laser electric field can be written as

\begin{equation}
\begin{split}
P \approx \exp\!&[-A_1 Q_\alpha^{-1/2}+A_2+B_1Q_\alpha^{-5/2}\\
& +B_2Q_\alpha^{-2}+B_3Q_\alpha^{-1}+C_1 Q_\alpha^{-9/2}].
\end{split}
\end{equation}

\begin{figure}
\includegraphics[width=9.5cm]{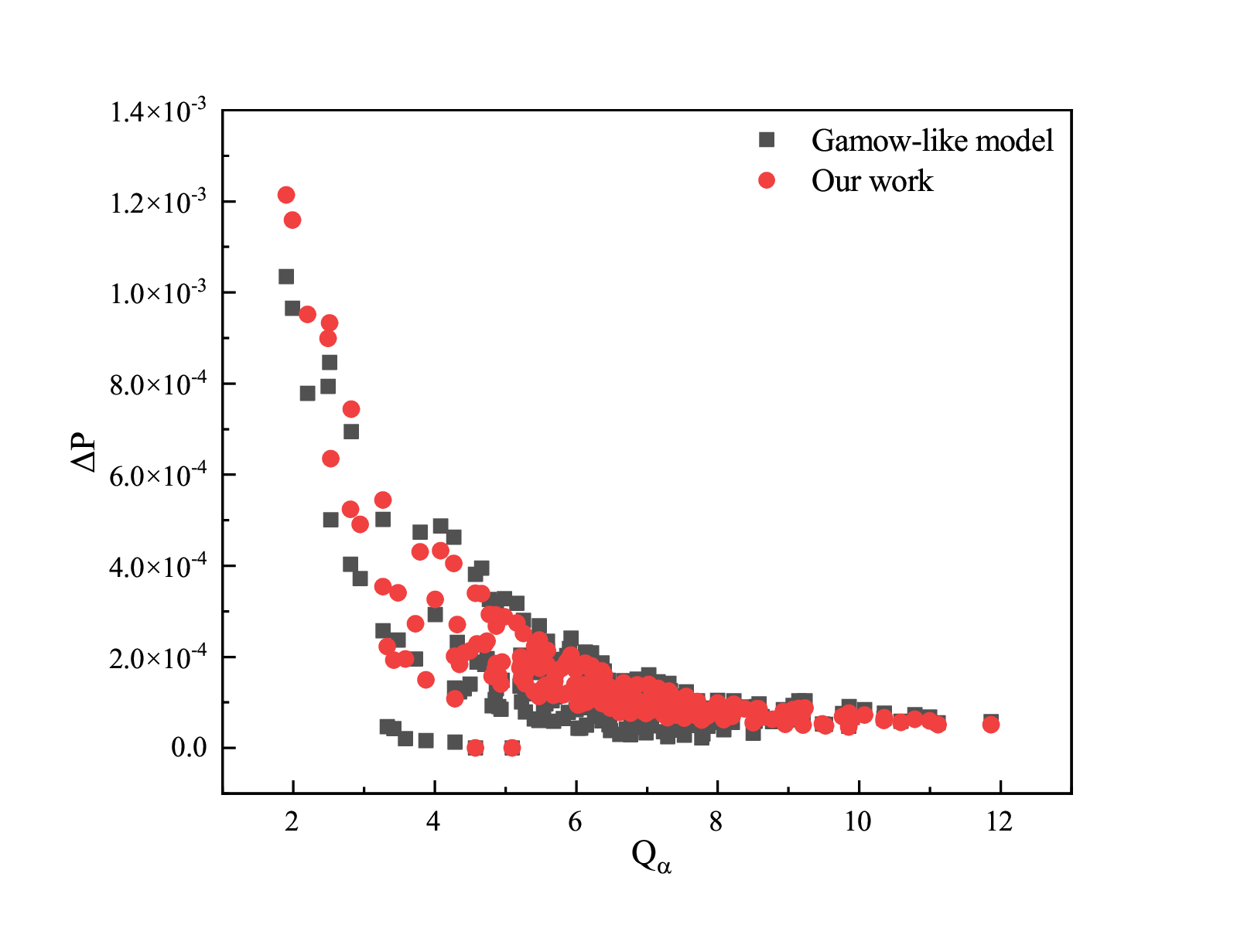}
\caption{(Color online) The rate of the relative change of penetration probability $\Delta P$ for the different $\alpha$ decay energy $Q_\alpha$ calculated with the Gamow-like model and Eq. (\ref{eq59}) in the case of peak laser intensity $I_0=10^{23} {\ } \rm{W}/cm^{2}$ and $\theta=0$, respectively.}
\label{fig 2}
\end{figure}

\begin{figure}
\includegraphics[width=9.5cm]{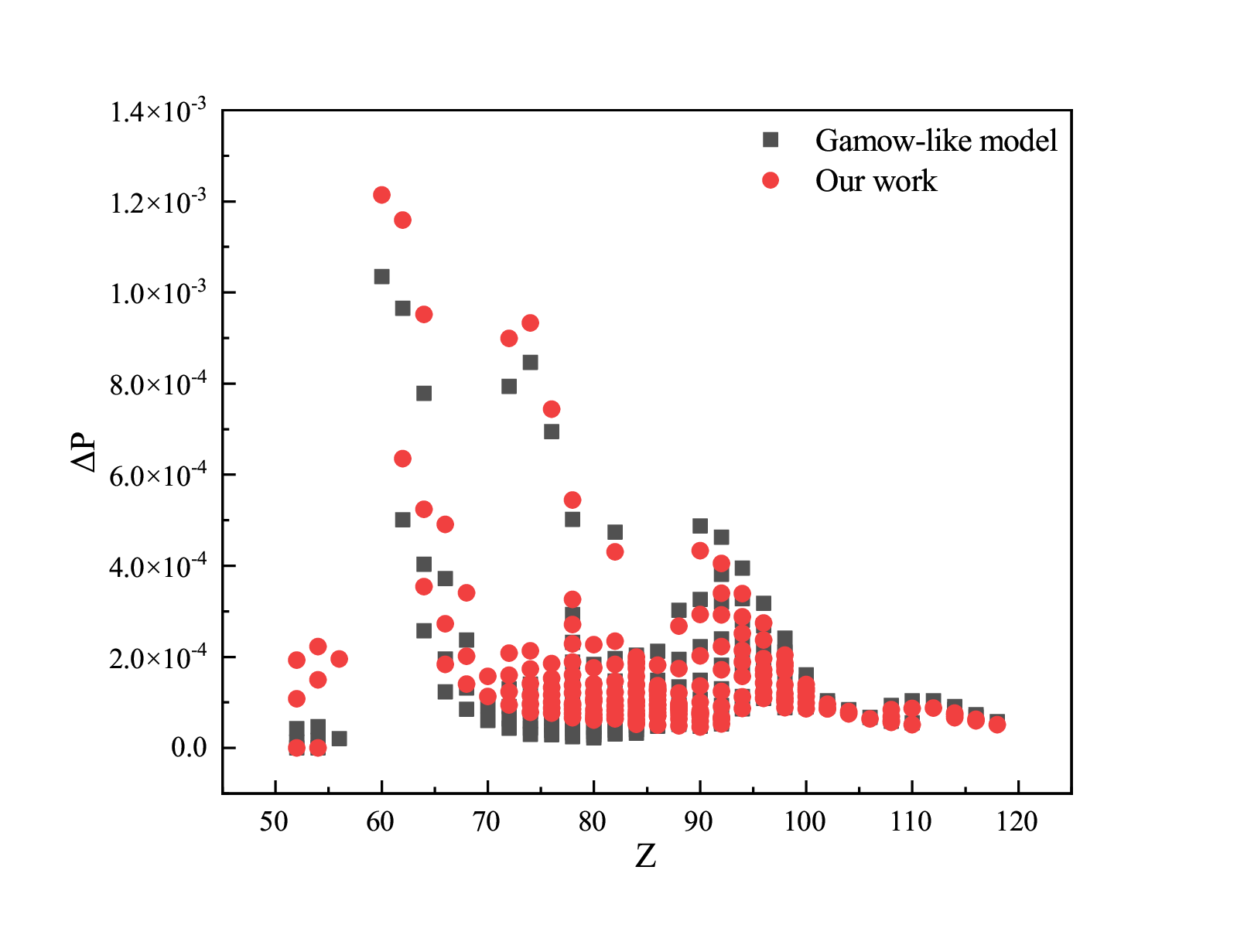}
\caption{(Color online) The rate of the relative change of penetration probability $\Delta P$ for different nuclei calculated with the Gamow-like model and Eq. (\ref{eq59}) in the case of peak laser intensity $I_0=10^{23} {\ } \rm{W}/cm^{2}$ and $\theta=0$, respectively.}
\label{fig 3}
\end{figure}

To verify the correctness of Eq. (\ref{eq59}), we used the Gamow-like model and Eq. (\ref{eq59}) to calculate the rate of the relative change of penetration probability $\Delta P$ for different nuclei in the case of the peak laser intensity $I_0=10^{23} {\ } \rm{W}/cm^{2}$ and $\theta=0$, respectively. The detailed results are shown in Figs. \ref{fig 2} and \ref{fig 3}. The x-axes of Figs. \ref{fig 2} and \ref{fig 3} represent the $\alpha$ decay energy $Q_\alpha$ and the proton number of the parent nucleus, respectively. Their y-axes both represent $\Delta P$. One can see, $\Delta P$ is always positive in the case of the laser intensity of $I_0=10^{23} {\ } \rm{W}/cm^{2}$. The reason for this phenomenon is that the positivity and negativity of $B_1$, $B_2$, and $B_3$ are all related to the positivity and negativity of $E(t)$, and $C_1$ is always positive, which means that $\Delta P$ is positive in the case of positive $E(t)$. Moreover, Eq. (\ref{eq59}) can reproduce the calculation of the Gamow-like model well, and the standard deviation between these two methods is only 0.198. It can also be seen from Fig. \ref{fig 2} that $\Delta P$ is negatively correlated with $Q_\alpha$, which implies that a more pronounced instantaneous penetration probability change rate can be obtained in future experiments using a parent nucleus with smaller decay energies under the same laser conditions. Furthermore, this conclusion also explains why $^{144}$Nd is the most sensitive parent nuclei to the high-intensity laser. Figure \ref{fig 3} shows that the analytical solutions and numerical calculations for nuclei with more significant proton numbers match better, which provides a fast way to estimate $\Delta P$ for superheavy nuclei.

As a comparison,  the rates of change of the $\alpha$ decay penetration probability for some of the parent nuclei given by Ref.\cite{PhysRevLett.124.212505} are listed in Tab. \ref{table 2}, whose calculations are also based on the WKB approximation. In this table, the third column stands for the rate of the relative change of penetration probability $\Delta P$ for the identical parent nuclei obtained by the Eq. (\ref{eq59}) in the case of the peak laser intensity $I_0=10^{26} {\ } \rm{W}/cm^{2}$ and $\theta=0$. It can be seen from Table \ref{table 2} that our computed results are in better agreement with the values given in Ref.\cite{PhysRevLett.124.212505}, which indicates that our proposed analytical formula is plausible. It should be noted that although we can accelerate the $\alpha$ decay within a short laser pulse, the overall change in the $\alpha$ decay of nuclei affected by laser is very insignificant, considering the long dark spans between laser shots. Therefore, obtaining the most significant change of $\alpha$ decay within one pulse becomes a key issue for future experiments.

\begin{table}[hbt!]
\caption{The effect of laser pulse with a peak intensity of $10^{26} {\ } \rm{W}/cm^{2}$ on $\alpha$ decay of the ground state even-even nuclei.}
\label{table 2}
\renewcommand\arraystretch{1.8}
\setlength{\tabcolsep}{4.0mm}
\begin{tabular}{ccc}
\hline
Nucleus&$\Delta P$ (Ref.\cite{PhysRevLett.124.212505}) &$\Delta P$ (Our-work)\\
\hline
$^{106}$Te&5.57$\times$$10^{-4}$&4.01$\times$$10^{-4}$\\
$^{144}$Nd&2.57$\times$$10^{-2}$&3.29$\times$$10^{-2}$\\
$^{162}$W&1.35$\times$$10^{-3}$&1.86$\times$$10^{-3}$\\
$^{212}$Po&1.35$\times$$10^{-3}$&1.96$\times$$10^{-3}$\\
$^{238}$Pu&5.58$\times$$10^{-3}$&7.44$\times$$10^{-3}$\\
$^{238}$U&1.13$\times$$10^{-2}$&1.47$\times$$10^{-2}$\\
\hline
\end{tabular}
\end{table}

\subsection{Parameter influences of laser wavelength and pulse width}

The theoretical calculations in the previous subsection showed that the instantaneous penetration probability change rate is inversely proportional to $Q_\alpha$ for a fixed electric field strength or at some $t$ moment. However, the laser pulse has a duration and specific profile, which may lead to different impacts of the electric field on $\Delta P$ at different moments. The effect of a complete laser pulse on $\alpha$ decay should be of more interest, i.e., the magnitude of the average rate of change of the $\alpha$ decay penetration probability $\Delta P_{avg}$  in the whole laser pulse. Since the laser electric field is a function of time and oscillates back and forth with the change of time, the influence of $\chi_{\varphi}^{(1)}$ in Eq. (\ref{eq 40}) on the $\alpha$ decay penetration probability will be mostly cancelled out. This means that the average rate of change of the $\alpha$ decay penetration probability $\Delta P_{avg}$ within one laser pulse will be much smaller than the maximum instantaneous rate of change of the penetration probability. To obtain a more significant average rate of change of penetration probability, J. Qi \emph{et al. }\cite{PhysRevC.99.044610} and our previous work \cite{PhysRevC.105.024312} proposed experimental schemes based on elliptically polarized laser fields and asymmetric chirped laser pulses, respectively. As seen from Eq. (\ref{eq59}), the rate of change of the penetration probability is related to the properties of the decaying parent nucleus and the laser pulse. The properties of the nucleus itself can hardly be changed, while enhancing the laser intensity strength requires significant experimental effort. In the present work, we studied the effects of easily tunable laser wavelength and pulse width on $\Delta P_{avg}$ in terms of the properties of the laser itself and obtained some feasible methods that can boost $\Delta P_{avg}$ within a single laser pulse.

\begin{figure}
%\begin{tabular}{cc}  
\begin{minipage}{0.48\linewidth}
 \centerline{\includegraphics[width=9cm]{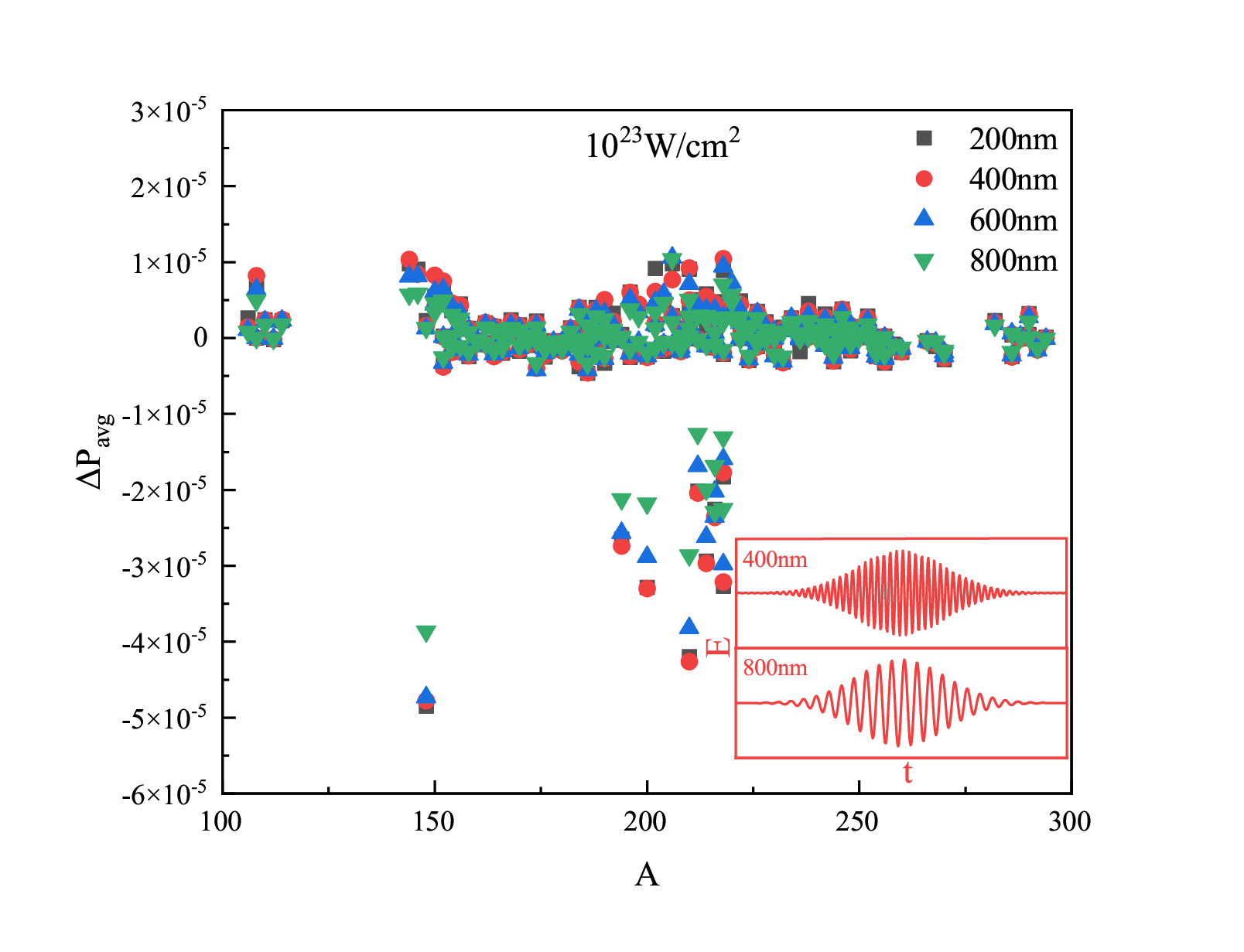}}
 \centerline{(a) The case of $I_0=10^{23} {\ } \rm{W}/cm^{2}$.}
\end{minipage}
\vfill%hiff
\begin{minipage}{0.48\linewidth}
 \centerline{\includegraphics[width=9cm]{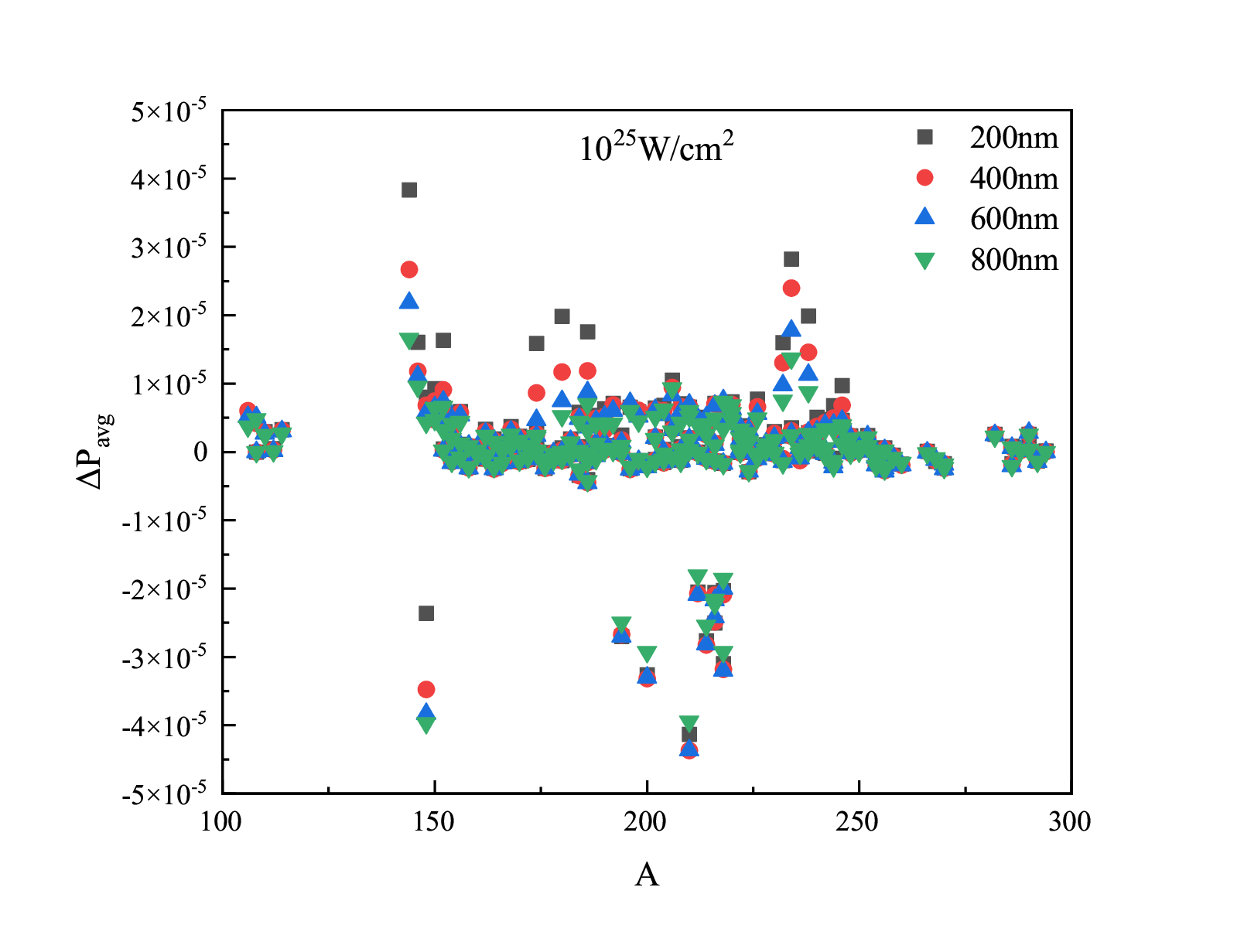}}
 \centerline{(b) The case of $I_0=10^{25} {\ } \rm{W}/cm^{2}$.}
\end{minipage}
\vfill
\begin{minipage}{0.48\linewidth}
 \centerline{\includegraphics[width=9cm]{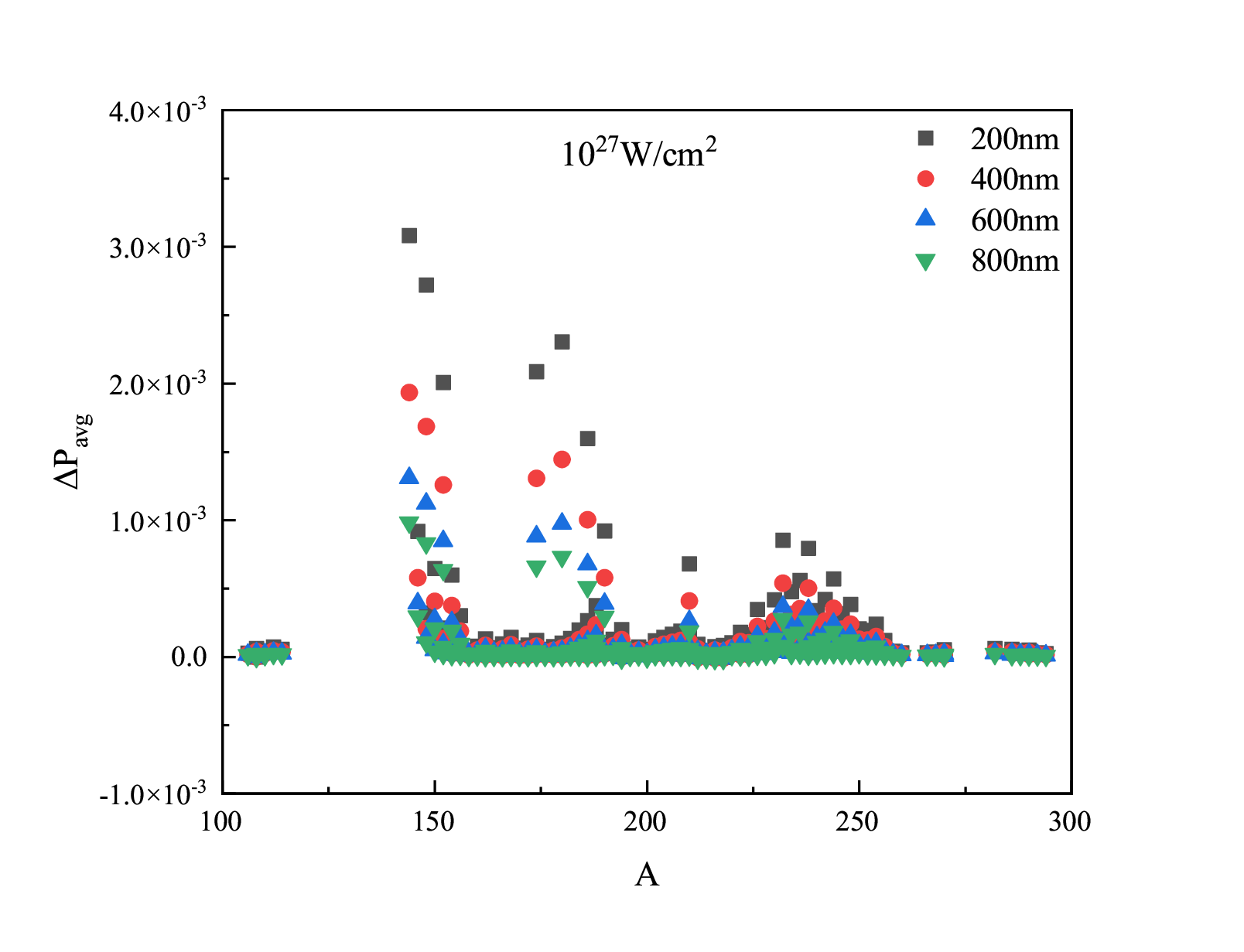}}
 \centerline{(c) The case of $I_0=10^{27} {\ } \rm{W}/cm^{2}$.}
\end{minipage}
%\end{tabular}
\caption{(Color online) The influence of a high-intensity laser on the average rate of change of penetration probability with $\lambda=200 {\ } \rm{nm}, 400 {\ } \rm{nm}, 600 {\ } \rm{nm}$ and $800 {\ } \rm{nm}$, respectively.}
\label{fig 4}
\end{figure}

To investigate the difference between the effects of the individual laser pulse at different wavelengths on $\Delta P_{avg}$, we set the laser pulse width to be a fixed value and calculated $\Delta P_{avg}$ for different nuclei corresponding to individual laser pulse at wavelengths of 200 nm, 400 nm, 600 nm, and 800 nm, respectively. Here the theoretical model in Section \ref{section 2} is used for the calculation, where $\theta=0$ and $\tau=5 T_0$. As a comparison, the peak intensities of the laser pulses were set to be $I_0=10^{23} {\ } \rm{W}/cm^{2}$, $10^{25} {\ } \rm{W}/cm^{2}$, and $10^{27} {\ } \rm{W}/cm^{2}$, respectively. Figure \ref{fig 4} shows the $\Delta P_{avg}$ of different nuclei under the influence of a single laser pulse of different wavelengths. The x-axis represents the mass number of the parent nucleus, and the y-axis represents $\Delta P_{avg}$. Moreover, to better illustrate the wavelength variation under the condition of a constant laser pulse width, a schematic diagram of the laser electric field waveforms corresponding to laser wavelengths of 400 nm and 800 nm is given in the red box on the right side of Fig. \ref{fig 4} (a).

From this figure, one can clearly see that some of the parent nuclei have negative $\Delta P_{avg}$ values in the case of $I_0=10^{23} {\ } \rm{W}/cm^{2}$ and the parent nuclei with negative $\Delta P_{avg}$ become less in the case of $I_0=10^{25} {\ } \rm{W}/cm^{2}$. Once the laser intensity reaches $I_0=10^{27} {\ } \rm{W}/cm^{2}$, all parent nuclei correspond to positive $\Delta P_{avg}$ values. The reason is that the $\chi^{(1)}$ terms oscillate back and forth with time $t$ to cancel each other out, while the $\chi^{(2)}$ terms, which are proportional to the square of the electric field intensity, have less effect in the relatively weak laser. As the laser intensity increases, $\chi^{(2)}$, which is constantly positive, takes up more weight in calculating the rate of change of penetration probability, resulting in a constant positive average rate of change of the $\alpha$ decay penetration probability for all parent nuclei. Figure \ref{fig 4} also shows that smaller wavelengths lead to more significant $\Delta P_{avg}$ for most of the nuclei, which means that shorter wavelength laser pulses are preferable for enhancing the average rate of change of penetration probability in future experiments.

\begin{flushleft}
\begin{figure}[H]
%\begin{tabular}{cc}  
\begin{minipage}{1\linewidth}
 \centerline{\includegraphics[width=9cm]{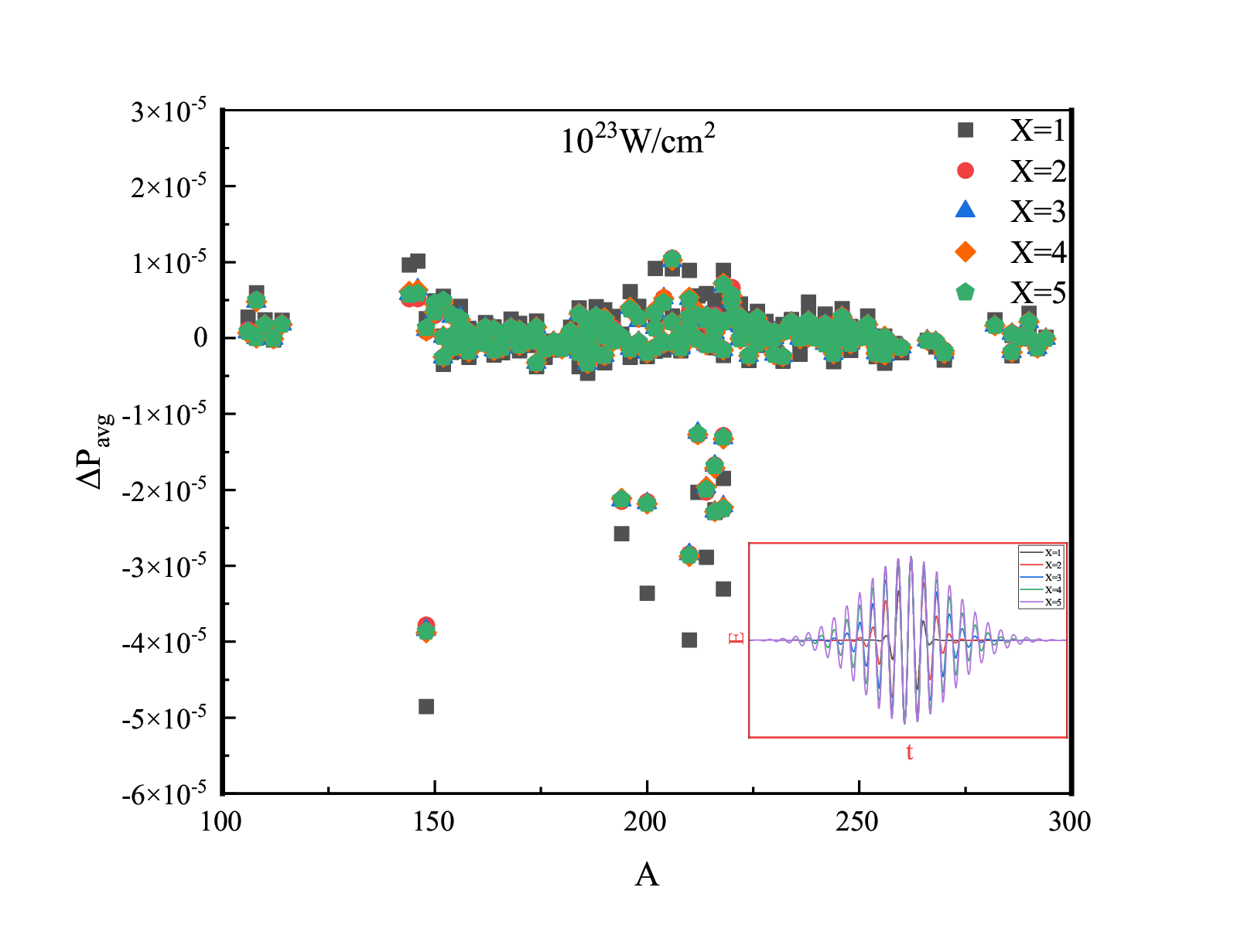}}
 \centerline{(a) The case of $I_0=10^{23} {\ } \rm{W}/cm^{2}$.}
\end{minipage}
\vfill%hiff
\begin{minipage}{1\linewidth}
 \centerline{\includegraphics[width=9cm]{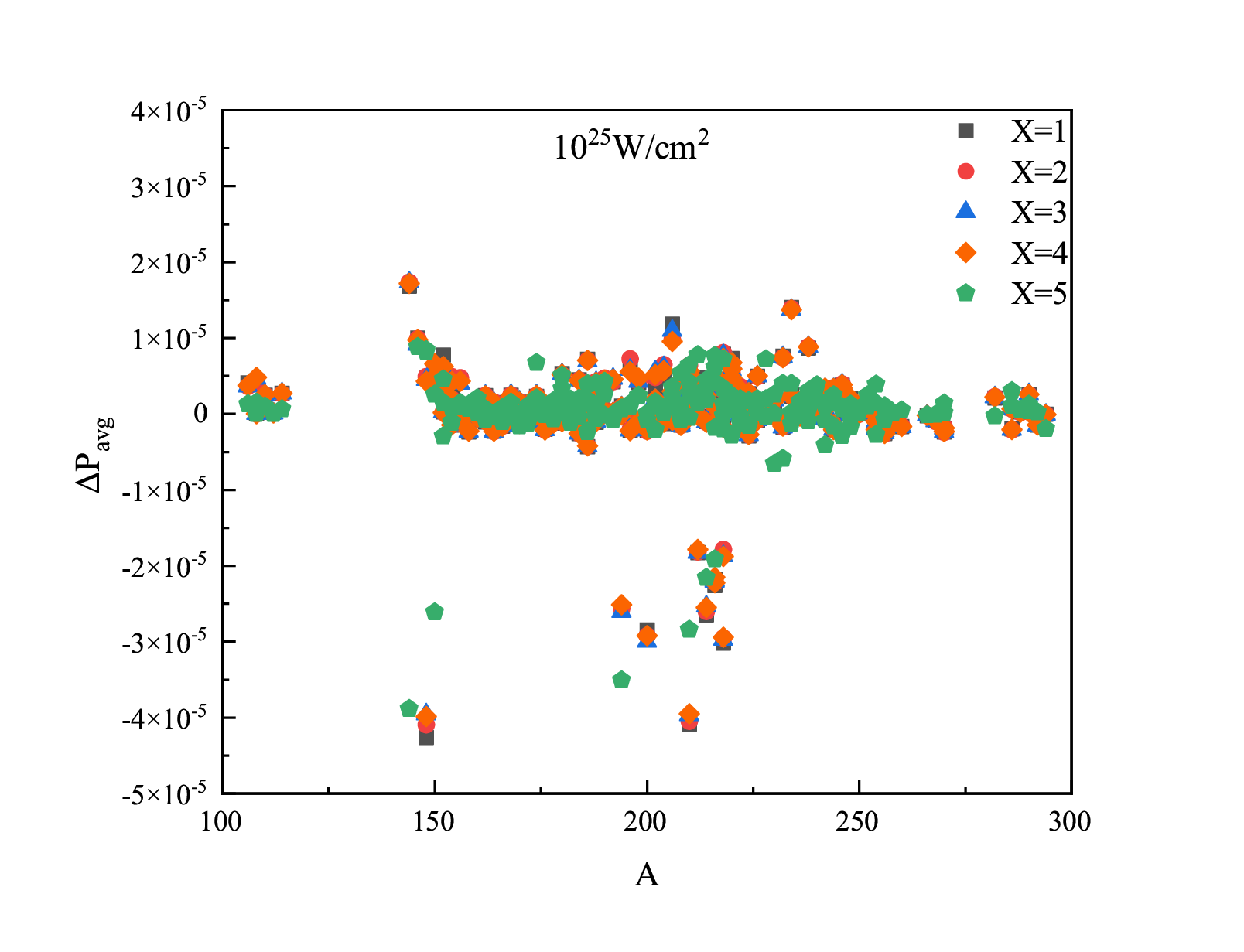}}
 \centerline{(b) The case of $I_0=10^{25} {\ } \rm{W}/cm^{2}$.}
\end{minipage}
\vfill
\begin{minipage}{1\linewidth}
 \centerline{\includegraphics[width=9cm]{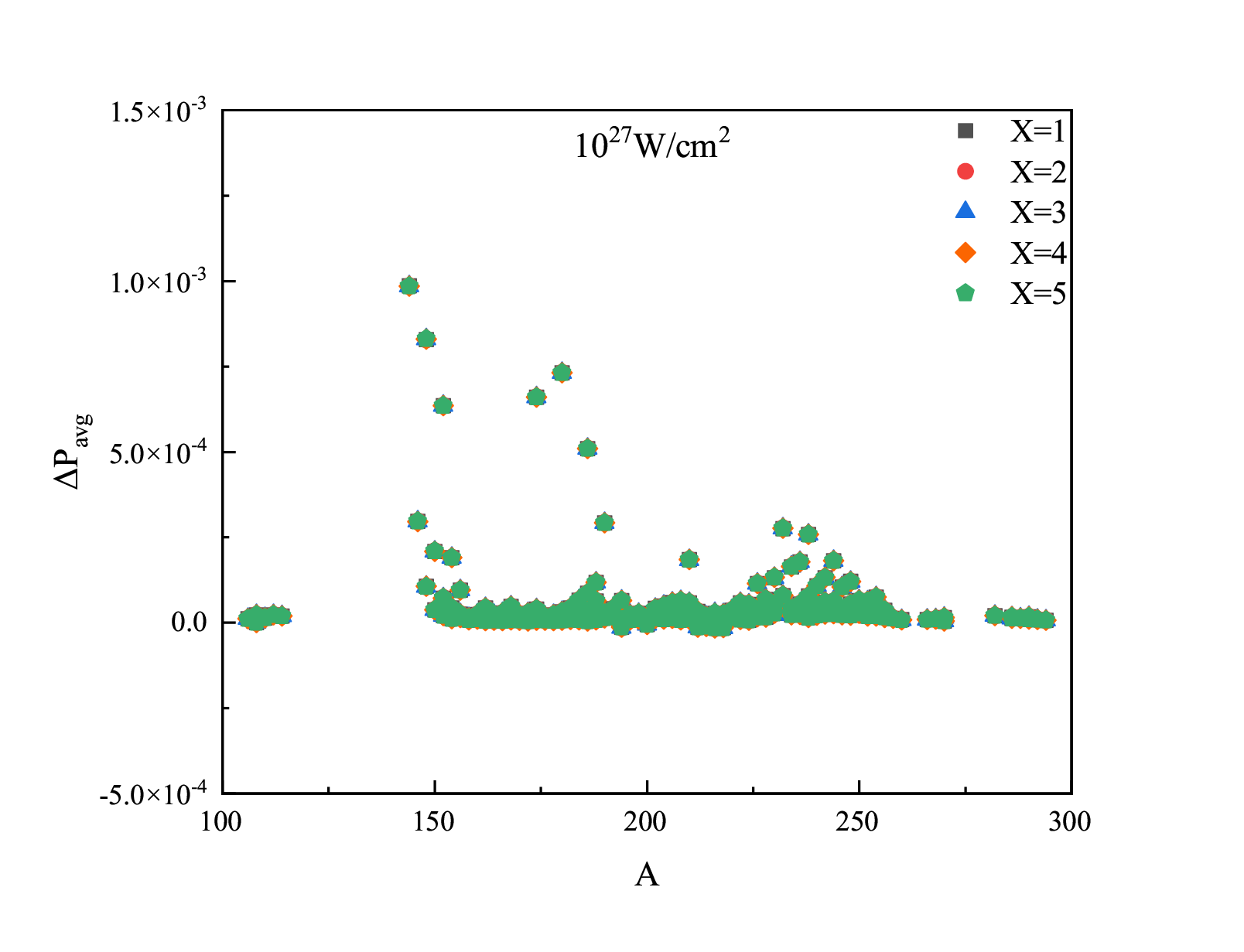}}
 \centerline{(c) The case of $I_0=10^{27} {\ } \rm{W}/cm^{2}$.}
\end{minipage}
%\end{tabular}
\caption{(Color online) The influence of high-intensity laser on the average rate of change of penetration probability with $x=1, 2, 3, 4$ and 5, respectively.}
\label{fig 5}
\end{figure}
\end{flushleft}

We set different laser pulse widths by adjusting the $x$ in Eq. (\ref{eq 29}). Here, $\theta$ is used as in Fig. \ref{fig 2}, and the laser wavelength used for the calculation is fixed at 800 nm. In the present work, the $x$ value describing the pulse width is set from 1 to 5, and a more considerable $x$ value means a more extended pulse width and a longer laser pulse duration. Similarly, the peak intensities of the laser pulses were set to be $I_0=10^{23} {\ } \rm{W}/cm^{2}$, $10^{25} {\ } \rm{W}/cm^{2}$, and $10^{27} {\ } \rm{W}/cm^{2}$, respectively. Figure \ref{fig 5} shows the $\Delta P_{avg}$ of different nuclei under the influence of single laser pulses of different pulse widths. The x-axis and y-axis have the same meaning as in Fig. \ref{fig 4}. The schematic diagram of the laser electric field waveforms with $x$ ranging from 1 to 5 is given in the red box on the right side of Fig. \ref{fig 5} (a).

It is evident from these figures that the effect of the change in pulse width on the average rate of change of the $\alpha$ decay penetration probability is negligible compared to the adjustment of wavelength. A slight $\Delta P_{avg}$ boost can be obtained at low peak laser pulse intensities using long pulse widths for most parent nuclei. There is almost no difference in the average rate of change of the $\alpha$ decay penetration probability corresponding to different pulse widths in the case of high laser pulse intensity. Although a tuned laser pulse width has limited influence on $\Delta P_{avg}$,  we can improve $\Delta P_{avg}$ by tuning the wavelength, e.g., short-wavelength X-ray or free-electron laser.

\section{Summary}
In summary, we systematically study the laser-assisted $\alpha$ decay of the deformed ground state even-even nucleus and aim to obtain achievable quantitative evaluations of the laser influences on $\alpha$ decay. The calculations show that the $\alpha$ decay penetration probability and the $\alpha$ decay half-life of different parent nuclei have different rates of change under the influence of laser intensity of $10^{23} {\ } \rm{W}/cm^{2}$, and the rate of change ranges from $0.0009\%$ to $0.135\%$. Moreover, we obtained analytical formulas for the rate of change of the $\alpha$ decay penetration probability in the ultra-intense laser fields. We also found that the decay energy is negatively related to the rate of change of the penetration probability. Finally, we investigated the effects of laser pulse width and wavelength on the average rate of change of the $\alpha$ decay penetration probability. The results show that using short-wavelength laser pulses in future experiments can obtain a more significant average rate of change of the $\alpha$ decay penetration probability.

\label{section 4}

\begin{acknowledgments}
This work was supported by the National Key R\&D Program of China (Grant No. 2018YFA0404802), National Natural Science Foundation of China (Grant No. 12135009), the Science and Technology Innovation Program of Hunan Province (Grant No. 2020RC4020), and the Hunan Provincial Innovation Foundation for Postgraduate (Grant No. CX20210007). 
\end{acknowledgments}

%\newpage

%\bibliographystyle{apsrev4-1}
%\bibliography{reference}

\begin{thebibliography}{87}%
\makeatletter
\providecommand \@ifxundefined [1]{%
 \@ifx{#1\undefined}
}%
\providecommand \@ifnum [1]{%
 \ifnum #1\expandafter \@firstoftwo
 \else \expandafter \@secondoftwo
 \fi
}%
\providecommand \@ifx [1]{%
 \ifx #1\expandafter \@firstoftwo
 \else \expandafter \@secondoftwo
 \fi
}%
\providecommand \natexlab [1]{#1}%
\providecommand \enquote  [1]{``#1''}%
\providecommand \bibnamefont  [1]{#1}%
\providecommand \bibfnamefont [1]{#1}%
\providecommand \citenamefont [1]{#1}%
\providecommand \href@noop [0]{\@secondoftwo}%
\providecommand \href [0]{\begingroup \@sanitize@url \@href}%
\providecommand \@href[1]{\@@startlink{#1}\@@href}%
\providecommand \@@href[1]{\endgroup#1\@@endlink}%
\providecommand \@sanitize@url [0]{\catcode `\\12\catcode `\$12\catcode
  `\&12\catcode `\#12\catcode `\^12\catcode `\_12\catcode `\%12\relax}%
\providecommand \@@startlink[1]{}%
\providecommand \@@endlink[0]{}%
\providecommand \url  [0]{\begingroup\@sanitize@url \@url }%
\providecommand \@url [1]{\endgroup\@href {#1}{\urlprefix }}%
\providecommand \urlprefix  [0]{URL }%
\providecommand \Eprint [0]{\href }%
\providecommand \doibase [0]{http://dx.doi.org/}%
\providecommand \selectlanguage [0]{\@gobble}%
\providecommand \bibinfo  [0]{\@secondoftwo}%
\providecommand \bibfield  [0]{\@secondoftwo}%
\providecommand \translation [1]{[#1]}%
\providecommand \BibitemOpen [0]{}%
\providecommand \bibitemStop [0]{}%
\providecommand \bibitemNoStop [0]{.\EOS\space}%
\providecommand \EOS [0]{\spacefactor3000\relax}%
\providecommand \BibitemShut  [1]{\csname bibitem#1\endcsname}%
\let\auto@bib@innerbib\@empty
%</preamble>
\bibitem [{\citenamefont {Geesaman}\ \emph {et~al.}(2006)\citenamefont
  {Geesaman}, \citenamefont {Gelbke}, \citenamefont {Janssens},\ and\
  \citenamefont {Sherrill}}]{doi:10.1146/annurev.nucl.55.090704.151604}%
  \BibitemOpen
  \bibfield  {author} {\bibinfo {author} {\bibfnamefont {D.}~\bibnamefont
  {Geesaman}}, \bibinfo {author} {\bibfnamefont {C.}~\bibnamefont {Gelbke}},
  \bibinfo {author} {\bibfnamefont {R.}~\bibnamefont {Janssens}}, \ and\
  \bibinfo {author} {\bibfnamefont {B.}~\bibnamefont {Sherrill}},\ }\href
  {\doibase 10.1146/annurev.nucl.55.090704.151604} {\bibfield  {journal}
  {\bibinfo  {journal} {Annual Review of Nuclear and Particle Science}\
  }\textbf {\bibinfo {volume} {56}},\ \bibinfo {pages} {53} (\bibinfo {year}
  {2006})},\ \Eprint
  {http://arxiv.org/abs/https://doi.org/10.1146/annurev.nucl.55.090704.151604}
  {https://doi.org/10.1146/annurev.nucl.55.090704.151604} \BibitemShut
  {NoStop}%
\bibitem [{\citenamefont {Hofmann}\ and\ \citenamefont
  {M\"unzenberg}(2000)}]{RevModPhys.72.733}%
  \BibitemOpen
  \bibfield  {author} {\bibinfo {author} {\bibfnamefont {S.}~\bibnamefont
  {Hofmann}}\ and\ \bibinfo {author} {\bibfnamefont {G.}~\bibnamefont
  {M\"unzenberg}},\ }\href {\doibase 10.1103/RevModPhys.72.733} {\bibfield
  {journal} {\bibinfo  {journal} {Rev. Mod. Phys.}\ }\textbf {\bibinfo {volume}
  {72}},\ \bibinfo {pages} {733} (\bibinfo {year} {2000})}\BibitemShut
  {NoStop}%
\bibitem [{\citenamefont {Pf\"utzner}\ \emph {et~al.}(2012)\citenamefont
  {Pf\"utzner}, \citenamefont {Karny}, \citenamefont {Grigorenko},\ and\
  \citenamefont {Riisager}}]{RevModPhys.84.567}%
  \BibitemOpen
  \bibfield  {author} {\bibinfo {author} {\bibfnamefont {M.}~\bibnamefont
  {Pf\"utzner}}, \bibinfo {author} {\bibfnamefont {M.}~\bibnamefont {Karny}},
  \bibinfo {author} {\bibfnamefont {L.~V.}\ \bibnamefont {Grigorenko}}, \ and\
  \bibinfo {author} {\bibfnamefont {K.}~\bibnamefont {Riisager}},\ }\href
  {\doibase 10.1103/RevModPhys.84.567} {\bibfield  {journal} {\bibinfo
  {journal} {Rev. Mod. Phys.}\ }\textbf {\bibinfo {volume} {84}},\ \bibinfo
  {pages} {567} (\bibinfo {year} {2012})}\BibitemShut {NoStop}%
\bibitem [{\citenamefont {Andreyev}\ \emph {et~al.}(2013)\citenamefont
  {Andreyev}, \citenamefont {Huyse}, \citenamefont {Van~Duppen}, \citenamefont
  {Qi}, \citenamefont {Liotta}, \citenamefont {Antalic}, \citenamefont
  {Ackermann}, \citenamefont {Franchoo}, \citenamefont {He\ss{}berger},
  \citenamefont {Hofmann}, \citenamefont {Kojouharov}, \citenamefont {Kindler},
  \citenamefont {Kuusiniemi}, \citenamefont {Lesher}, \citenamefont {Lommel},
  \citenamefont {Mann}, \citenamefont {Nishio}, \citenamefont {Page},
  \citenamefont {Streicher}, \citenamefont {\ifmmode~\check{S}\else
  \v{S}\fi{}\'aro}, \citenamefont {Sulignano}, \citenamefont {Wiseman},\ and\
  \citenamefont {Wyss}}]{PhysRevLett.110.242502}%
  \BibitemOpen
  \bibfield  {author} {\bibinfo {author} {\bibfnamefont {A.~N.}\ \bibnamefont
  {Andreyev}}, \bibinfo {author} {\bibfnamefont {M.}~\bibnamefont {Huyse}},
  \bibinfo {author} {\bibfnamefont {P.}~\bibnamefont {Van~Duppen}}, \bibinfo
  {author} {\bibfnamefont {C.}~\bibnamefont {Qi}}, \bibinfo {author}
  {\bibfnamefont {R.~J.}\ \bibnamefont {Liotta}}, \bibinfo {author}
  {\bibfnamefont {S.}~\bibnamefont {Antalic}}, \bibinfo {author} {\bibfnamefont
  {D.}~\bibnamefont {Ackermann}}, \bibinfo {author} {\bibfnamefont
  {S.}~\bibnamefont {Franchoo}}, \bibinfo {author} {\bibfnamefont {F.~P.}\
  \bibnamefont {He\ss{}berger}}, \bibinfo {author} {\bibfnamefont
  {S.}~\bibnamefont {Hofmann}}, \bibinfo {author} {\bibfnamefont
  {I.}~\bibnamefont {Kojouharov}}, \bibinfo {author} {\bibfnamefont
  {B.}~\bibnamefont {Kindler}}, \bibinfo {author} {\bibfnamefont
  {P.}~\bibnamefont {Kuusiniemi}}, \bibinfo {author} {\bibfnamefont {S.~R.}\
  \bibnamefont {Lesher}}, \bibinfo {author} {\bibfnamefont {B.}~\bibnamefont
  {Lommel}}, \bibinfo {author} {\bibfnamefont {R.}~\bibnamefont {Mann}},
  \bibinfo {author} {\bibfnamefont {K.}~\bibnamefont {Nishio}}, \bibinfo
  {author} {\bibfnamefont {R.~D.}\ \bibnamefont {Page}}, \bibinfo {author}
  {\bibfnamefont {B.}~\bibnamefont {Streicher}}, \bibinfo {author}
  {\bibfnamefont {i.~c.~v.}\ \bibnamefont {\ifmmode~\check{S}\else
  \v{S}\fi{}\'aro}}, \bibinfo {author} {\bibfnamefont {B.}~\bibnamefont
  {Sulignano}}, \bibinfo {author} {\bibfnamefont {D.}~\bibnamefont {Wiseman}},
  \ and\ \bibinfo {author} {\bibfnamefont {R.~A.}\ \bibnamefont {Wyss}},\
  }\href {\doibase 10.1103/PhysRevLett.110.242502} {\bibfield  {journal}
  {\bibinfo  {journal} {Phys. Rev. Lett.}\ }\textbf {\bibinfo {volume} {110}},\
  \bibinfo {pages} {242502} (\bibinfo {year} {2013})}\BibitemShut {NoStop}%
\bibitem [{\citenamefont {Kalaninov\'a}\ \emph {et~al.}(2013)\citenamefont
  {Kalaninov\'a}, \citenamefont {Andreyev}, \citenamefont {Antalic},
  \citenamefont {He\ss{}berger}, \citenamefont {Ackermann}, \citenamefont
  {Andel}, \citenamefont {Drummond}, \citenamefont {Hofmann}, \citenamefont
  {Huyse}, \citenamefont {Kindler}, \citenamefont {Lane}, \citenamefont
  {Liberati}, \citenamefont {Lommel}, \citenamefont {Page}, \citenamefont
  {Rapisarda}, \citenamefont {Sandhu}, \citenamefont {\ifmmode~\check{S}\else
  \v{S}\fi{}\'aro}, \citenamefont {Thornthwaite},\ and\ \citenamefont
  {Van~Duppen}}]{PhysRevC.87.044335}%
  \BibitemOpen
  \bibfield  {author} {\bibinfo {author} {\bibfnamefont {Z.}~\bibnamefont
  {Kalaninov\'a}}, \bibinfo {author} {\bibfnamefont {A.~N.}\ \bibnamefont
  {Andreyev}}, \bibinfo {author} {\bibfnamefont {S.}~\bibnamefont {Antalic}},
  \bibinfo {author} {\bibfnamefont {F.~P.}\ \bibnamefont {He\ss{}berger}},
  \bibinfo {author} {\bibfnamefont {D.}~\bibnamefont {Ackermann}}, \bibinfo
  {author} {\bibfnamefont {B.}~\bibnamefont {Andel}}, \bibinfo {author}
  {\bibfnamefont {M.~C.}\ \bibnamefont {Drummond}}, \bibinfo {author}
  {\bibfnamefont {S.}~\bibnamefont {Hofmann}}, \bibinfo {author} {\bibfnamefont
  {M.}~\bibnamefont {Huyse}}, \bibinfo {author} {\bibfnamefont
  {B.}~\bibnamefont {Kindler}}, \bibinfo {author} {\bibfnamefont {J.~F.~W.}\
  \bibnamefont {Lane}}, \bibinfo {author} {\bibfnamefont {V.}~\bibnamefont
  {Liberati}}, \bibinfo {author} {\bibfnamefont {B.}~\bibnamefont {Lommel}},
  \bibinfo {author} {\bibfnamefont {R.~D.}\ \bibnamefont {Page}}, \bibinfo
  {author} {\bibfnamefont {E.}~\bibnamefont {Rapisarda}}, \bibinfo {author}
  {\bibfnamefont {K.}~\bibnamefont {Sandhu}}, \bibinfo {author} {\bibfnamefont
  {i.~c.~v.}\ \bibnamefont {\ifmmode~\check{S}\else \v{S}\fi{}\'aro}}, \bibinfo
  {author} {\bibfnamefont {A.}~\bibnamefont {Thornthwaite}}, \ and\ \bibinfo
  {author} {\bibfnamefont {P.}~\bibnamefont {Van~Duppen}},\ }\href {\doibase
  10.1103/PhysRevC.87.044335} {\bibfield  {journal} {\bibinfo  {journal} {Phys.
  Rev. C}\ }\textbf {\bibinfo {volume} {87}},\ \bibinfo {pages} {044335}
  (\bibinfo {year} {2013})}\BibitemShut {NoStop}%
\bibitem [{\citenamefont {Ma}\ \emph {et~al.}(2015)\citenamefont {Ma},
  \citenamefont {Zhang}, \citenamefont {Gan}, \citenamefont {Yang},
  \citenamefont {Yu}, \citenamefont {Jiang}, \citenamefont {Wang},
  \citenamefont {Tian}, \citenamefont {Wang}, \citenamefont {Guo},
  \citenamefont {Ding}, \citenamefont {Ren}, \citenamefont {Zhou},
  \citenamefont {Zhou}, \citenamefont {Xu},\ and\ \citenamefont
  {Xiao}}]{PhysRevC.91.051302}%
  \BibitemOpen
  \bibfield  {author} {\bibinfo {author} {\bibfnamefont {L.}~\bibnamefont
  {Ma}}, \bibinfo {author} {\bibfnamefont {Z.~Y.}\ \bibnamefont {Zhang}},
  \bibinfo {author} {\bibfnamefont {Z.~G.}\ \bibnamefont {Gan}}, \bibinfo
  {author} {\bibfnamefont {H.~B.}\ \bibnamefont {Yang}}, \bibinfo {author}
  {\bibfnamefont {L.}~\bibnamefont {Yu}}, \bibinfo {author} {\bibfnamefont
  {J.}~\bibnamefont {Jiang}}, \bibinfo {author} {\bibfnamefont {J.~G.}\
  \bibnamefont {Wang}}, \bibinfo {author} {\bibfnamefont {Y.~L.}\ \bibnamefont
  {Tian}}, \bibinfo {author} {\bibfnamefont {Y.~S.}\ \bibnamefont {Wang}},
  \bibinfo {author} {\bibfnamefont {S.}~\bibnamefont {Guo}}, \bibinfo {author}
  {\bibfnamefont {B.}~\bibnamefont {Ding}}, \bibinfo {author} {\bibfnamefont
  {Z.~Z.}\ \bibnamefont {Ren}}, \bibinfo {author} {\bibfnamefont {S.~G.}\
  \bibnamefont {Zhou}}, \bibinfo {author} {\bibfnamefont {X.~H.}\ \bibnamefont
  {Zhou}}, \bibinfo {author} {\bibfnamefont {H.~S.}\ \bibnamefont {Xu}}, \ and\
  \bibinfo {author} {\bibfnamefont {G.~Q.}\ \bibnamefont {Xiao}},\ }\href
  {\doibase 10.1103/PhysRevC.91.051302} {\bibfield  {journal} {\bibinfo
  {journal} {Phys. Rev. C}\ }\textbf {\bibinfo {volume} {91}},\ \bibinfo
  {pages} {051302} (\bibinfo {year} {2015})}\BibitemShut {NoStop}%
\bibitem [{\citenamefont {Yang}\ \emph {et~al.}(2015)\citenamefont {Yang},
  \citenamefont {Zhang}, \citenamefont {Wang}, \citenamefont {Gan},
  \citenamefont {Ma}, \citenamefont {Yu}, \citenamefont {Jiang}, \citenamefont
  {Tian}, \citenamefont {Ding}, \citenamefont {Guo}, \citenamefont {Wang},
  \citenamefont {Huang}, \citenamefont {Sun}, \citenamefont {Wang},
  \citenamefont {Zhou}, \citenamefont {Ren}, \citenamefont {Zhou},
  \citenamefont {Xu},\ and\ \citenamefont {Xiao}}]{10.1140/epja/i2015-15088-9}%
  \BibitemOpen
  \bibfield  {author} {\bibinfo {author} {\bibfnamefont {H.~B.}\ \bibnamefont
  {Yang}}, \bibinfo {author} {\bibfnamefont {Z.~Y.}\ \bibnamefont {Zhang}},
  \bibinfo {author} {\bibfnamefont {J.~G.}\ \bibnamefont {Wang}}, \bibinfo
  {author} {\bibfnamefont {Z.~G.}\ \bibnamefont {Gan}}, \bibinfo {author}
  {\bibfnamefont {L.}~\bibnamefont {Ma}}, \bibinfo {author} {\bibfnamefont
  {L.}~\bibnamefont {Yu}}, \bibinfo {author} {\bibfnamefont {J.}~\bibnamefont
  {Jiang}}, \bibinfo {author} {\bibfnamefont {Y.~L.}\ \bibnamefont {Tian}},
  \bibinfo {author} {\bibfnamefont {B.}~\bibnamefont {Ding}}, \bibinfo {author}
  {\bibfnamefont {S.}~\bibnamefont {Guo}}, \bibinfo {author} {\bibfnamefont
  {Y.~S.}\ \bibnamefont {Wang}}, \bibinfo {author} {\bibfnamefont {T.~H.}\
  \bibnamefont {Huang}}, \bibinfo {author} {\bibfnamefont {M.~D.}\ \bibnamefont
  {Sun}}, \bibinfo {author} {\bibfnamefont {K.~L.}\ \bibnamefont {Wang}},
  \bibinfo {author} {\bibfnamefont {S.~G.}\ \bibnamefont {Zhou}}, \bibinfo
  {author} {\bibfnamefont {Z.~Z.}\ \bibnamefont {Ren}}, \bibinfo {author}
  {\bibfnamefont {X.~H.}\ \bibnamefont {Zhou}}, \bibinfo {author}
  {\bibfnamefont {H.~S.}\ \bibnamefont {Xu}}, \ and\ \bibinfo {author}
  {\bibfnamefont {G.~Q.}\ \bibnamefont {Xiao}},\ }\href {\doibase
  10.1140/epja/i2015-15088-9} {\bibfield  {journal} {\bibinfo  {journal} {Euro.
  Phys. J. A}\ }\textbf {\bibinfo {volume} {51}},\ \bibinfo {pages} {88}
  (\bibinfo {year} {2015})}\BibitemShut {NoStop}%
\bibitem [{\citenamefont {Carroll}\ \emph {et~al.}(2014)\citenamefont
  {Carroll}, \citenamefont {Page}, \citenamefont {Joss}, \citenamefont
  {Uusitalo}, \citenamefont {Darby}, \citenamefont {Andgren}, \citenamefont
  {Cederwall}, \citenamefont {Eeckhaudt}, \citenamefont {Grahn}, \citenamefont
  {Gray-Jones}, \citenamefont {Greenlees}, \citenamefont {Hadinia},
  \citenamefont {Jones}, \citenamefont {Julin}, \citenamefont {Juutinen},
  \citenamefont {Leino}, \citenamefont {Lepp\"anen}, \citenamefont {Nyman},
  \citenamefont {O'Donnell}, \citenamefont {Pakarinen}, \citenamefont
  {Rahkila}, \citenamefont {Sandzelius}, \citenamefont {Sar\'en}, \citenamefont
  {Scholey}, \citenamefont {Seweryniak},\ and\ \citenamefont
  {Simpson}}]{PhysRevLett.112.092501}%
  \BibitemOpen
  \bibfield  {author} {\bibinfo {author} {\bibfnamefont {R.~J.}\ \bibnamefont
  {Carroll}}, \bibinfo {author} {\bibfnamefont {R.~D.}\ \bibnamefont {Page}},
  \bibinfo {author} {\bibfnamefont {D.~T.}\ \bibnamefont {Joss}}, \bibinfo
  {author} {\bibfnamefont {J.}~\bibnamefont {Uusitalo}}, \bibinfo {author}
  {\bibfnamefont {I.~G.}\ \bibnamefont {Darby}}, \bibinfo {author}
  {\bibfnamefont {K.}~\bibnamefont {Andgren}}, \bibinfo {author} {\bibfnamefont
  {B.}~\bibnamefont {Cederwall}}, \bibinfo {author} {\bibfnamefont
  {S.}~\bibnamefont {Eeckhaudt}}, \bibinfo {author} {\bibfnamefont
  {T.}~\bibnamefont {Grahn}}, \bibinfo {author} {\bibfnamefont
  {C.}~\bibnamefont {Gray-Jones}}, \bibinfo {author} {\bibfnamefont {P.~T.}\
  \bibnamefont {Greenlees}}, \bibinfo {author} {\bibfnamefont {B.}~\bibnamefont
  {Hadinia}}, \bibinfo {author} {\bibfnamefont {P.~M.}\ \bibnamefont {Jones}},
  \bibinfo {author} {\bibfnamefont {R.}~\bibnamefont {Julin}}, \bibinfo
  {author} {\bibfnamefont {S.}~\bibnamefont {Juutinen}}, \bibinfo {author}
  {\bibfnamefont {M.}~\bibnamefont {Leino}}, \bibinfo {author} {\bibfnamefont
  {A.-P.}\ \bibnamefont {Lepp\"anen}}, \bibinfo {author} {\bibfnamefont
  {M.}~\bibnamefont {Nyman}}, \bibinfo {author} {\bibfnamefont
  {D.}~\bibnamefont {O'Donnell}}, \bibinfo {author} {\bibfnamefont
  {J.}~\bibnamefont {Pakarinen}}, \bibinfo {author} {\bibfnamefont
  {P.}~\bibnamefont {Rahkila}}, \bibinfo {author} {\bibfnamefont
  {M.}~\bibnamefont {Sandzelius}}, \bibinfo {author} {\bibfnamefont
  {J.}~\bibnamefont {Sar\'en}}, \bibinfo {author} {\bibfnamefont
  {C.}~\bibnamefont {Scholey}}, \bibinfo {author} {\bibfnamefont
  {D.}~\bibnamefont {Seweryniak}}, \ and\ \bibinfo {author} {\bibfnamefont
  {J.}~\bibnamefont {Simpson}},\ }\href {\doibase
  10.1103/PhysRevLett.112.092501} {\bibfield  {journal} {\bibinfo  {journal}
  {Phys. Rev. Lett.}\ }\textbf {\bibinfo {volume} {112}},\ \bibinfo {pages}
  {092501} (\bibinfo {year} {2014})}\BibitemShut {NoStop}%
\bibitem [{\citenamefont {Gamow}(1928)}]{gamow1928quantentheorie}%
  \BibitemOpen
  \bibfield  {author} {\bibinfo {author} {\bibfnamefont {G.}~\bibnamefont
  {Gamow}},\ }\href {\doibase 10.1007/BF01343196} {\bibfield  {journal}
  {\bibinfo  {journal} {Zeitschrift f{\"u}r Physik}\ }\textbf {\bibinfo
  {volume} {51}},\ \bibinfo {pages} {204} (\bibinfo {year} {1928})}\BibitemShut
  {NoStop}%
\bibitem [{\citenamefont {Gurney}\ and\ \citenamefont
  {Condon}(1928)}]{gurney1928wave}%
  \BibitemOpen
  \bibfield  {author} {\bibinfo {author} {\bibfnamefont {R.~W.}\ \bibnamefont
  {Gurney}}\ and\ \bibinfo {author} {\bibfnamefont {E.~U.}\ \bibnamefont
  {Condon}},\ }\href {\doibase 10.1038/122439a0} {\bibfield  {journal}
  {\bibinfo  {journal} {Nature}\ }\textbf {\bibinfo {volume} {122}},\ \bibinfo
  {pages} {439} (\bibinfo {year} {1928})}\BibitemShut {NoStop}%
\bibitem [{\citenamefont {Astier}\ \emph {et~al.}(2010)\citenamefont {Astier},
  \citenamefont {Petkov}, \citenamefont {Porquet}, \citenamefont {Delion},\
  and\ \citenamefont {Schuck}}]{PhysRevLett.104.042701}%
  \BibitemOpen
  \bibfield  {author} {\bibinfo {author} {\bibfnamefont {A.}~\bibnamefont
  {Astier}}, \bibinfo {author} {\bibfnamefont {P.}~\bibnamefont {Petkov}},
  \bibinfo {author} {\bibfnamefont {M.-G.}\ \bibnamefont {Porquet}}, \bibinfo
  {author} {\bibfnamefont {D.~S.}\ \bibnamefont {Delion}}, \ and\ \bibinfo
  {author} {\bibfnamefont {P.}~\bibnamefont {Schuck}},\ }\href {\doibase
  10.1103/PhysRevLett.104.042701} {\bibfield  {journal} {\bibinfo  {journal}
  {Phys. Rev. Lett.}\ }\textbf {\bibinfo {volume} {104}},\ \bibinfo {pages}
  {042701} (\bibinfo {year} {2010})}\BibitemShut {NoStop}%
\bibitem [{\citenamefont {Tohsaki}\ \emph {et~al.}(2001)\citenamefont
  {Tohsaki}, \citenamefont {Horiuchi}, \citenamefont {Schuck},\ and\
  \citenamefont {R\"opke}}]{PhysRevLett.87.192501}%
  \BibitemOpen
  \bibfield  {author} {\bibinfo {author} {\bibfnamefont {A.}~\bibnamefont
  {Tohsaki}}, \bibinfo {author} {\bibfnamefont {H.}~\bibnamefont {Horiuchi}},
  \bibinfo {author} {\bibfnamefont {P.}~\bibnamefont {Schuck}}, \ and\ \bibinfo
  {author} {\bibfnamefont {G.}~\bibnamefont {R\"opke}},\ }\href {\doibase
  10.1103/PhysRevLett.87.192501} {\bibfield  {journal} {\bibinfo  {journal}
  {Phys. Rev. Lett.}\ }\textbf {\bibinfo {volume} {87}},\ \bibinfo {pages}
  {192501} (\bibinfo {year} {2001})}\BibitemShut {NoStop}%
\bibitem [{\citenamefont {Delion}\ \emph {et~al.}(2004)\citenamefont {Delion},
  \citenamefont {Sandulescu},\ and\ \citenamefont
  {Greiner}}]{PhysRevC.69.044318}%
  \BibitemOpen
  \bibfield  {author} {\bibinfo {author} {\bibfnamefont {D.~S.}\ \bibnamefont
  {Delion}}, \bibinfo {author} {\bibfnamefont {A.}~\bibnamefont {Sandulescu}},
  \ and\ \bibinfo {author} {\bibfnamefont {W.}~\bibnamefont {Greiner}},\ }\href
  {\doibase 10.1103/PhysRevC.69.044318} {\bibfield  {journal} {\bibinfo
  {journal} {Phys. Rev. C}\ }\textbf {\bibinfo {volume} {69}},\ \bibinfo
  {pages} {044318} (\bibinfo {year} {2004})}\BibitemShut {NoStop}%
\bibitem [{\citenamefont {Karlgren}\ \emph {et~al.}(2006)\citenamefont
  {Karlgren}, \citenamefont {Liotta}, \citenamefont {Wyss}, \citenamefont
  {Huyse}, \citenamefont {VandeVel},\ and\ \citenamefont
  {VanDuppen}}]{PhysRevC.73.064304}%
  \BibitemOpen
  \bibfield  {author} {\bibinfo {author} {\bibfnamefont {D.}~\bibnamefont
  {Karlgren}}, \bibinfo {author} {\bibfnamefont {R.~J.}\ \bibnamefont
  {Liotta}}, \bibinfo {author} {\bibfnamefont {R.}~\bibnamefont {Wyss}},
  \bibinfo {author} {\bibfnamefont {M.}~\bibnamefont {Huyse}}, \bibinfo
  {author} {\bibfnamefont {K.}~\bibnamefont {VandeVel}}, \ and\ \bibinfo
  {author} {\bibfnamefont {P.}~\bibnamefont {VanDuppen}},\ }\href {\doibase
  10.1103/PhysRevC.73.064304} {\bibfield  {journal} {\bibinfo  {journal} {Phys.
  Rev. C}\ }\textbf {\bibinfo {volume} {73}},\ \bibinfo {pages} {064304}
  (\bibinfo {year} {2006})}\BibitemShut {NoStop}%
\bibitem [{\citenamefont {Kinoshita}\ \emph {et~al.}(2012)\citenamefont
  {Kinoshita}, \citenamefont {Paul}, \citenamefont {Kashiv}, \citenamefont
  {Collon}, \citenamefont {Deibel}, \citenamefont {DiGiovine}, \citenamefont
  {Greene}, \citenamefont {Henderson}, \citenamefont {Jiang}, \citenamefont
  {Marley}, \citenamefont {Nakanishi}, \citenamefont {Pardo}, \citenamefont
  {Rehm}, \citenamefont {Robertson}, \citenamefont {Scott}, \citenamefont
  {Schmitt}, \citenamefont {Tang}, \citenamefont {Vondrasek},\ and\
  \citenamefont {Yokoyama}}]{doi:10.1126/science.1215510}%
  \BibitemOpen
  \bibfield  {author} {\bibinfo {author} {\bibfnamefont {N.}~\bibnamefont
  {Kinoshita}}, \bibinfo {author} {\bibfnamefont {M.}~\bibnamefont {Paul}},
  \bibinfo {author} {\bibfnamefont {Y.}~\bibnamefont {Kashiv}}, \bibinfo
  {author} {\bibfnamefont {P.}~\bibnamefont {Collon}}, \bibinfo {author}
  {\bibfnamefont {C.~M.}\ \bibnamefont {Deibel}}, \bibinfo {author}
  {\bibfnamefont {B.}~\bibnamefont {DiGiovine}}, \bibinfo {author}
  {\bibfnamefont {J.~P.}\ \bibnamefont {Greene}}, \bibinfo {author}
  {\bibfnamefont {D.~J.}\ \bibnamefont {Henderson}}, \bibinfo {author}
  {\bibfnamefont {C.~L.}\ \bibnamefont {Jiang}}, \bibinfo {author}
  {\bibfnamefont {S.~T.}\ \bibnamefont {Marley}}, \bibinfo {author}
  {\bibfnamefont {T.}~\bibnamefont {Nakanishi}}, \bibinfo {author}
  {\bibfnamefont {R.~C.}\ \bibnamefont {Pardo}}, \bibinfo {author}
  {\bibfnamefont {K.~E.}\ \bibnamefont {Rehm}}, \bibinfo {author}
  {\bibfnamefont {D.}~\bibnamefont {Robertson}}, \bibinfo {author}
  {\bibfnamefont {R.}~\bibnamefont {Scott}}, \bibinfo {author} {\bibfnamefont
  {C.}~\bibnamefont {Schmitt}}, \bibinfo {author} {\bibfnamefont {X.~D.}\
  \bibnamefont {Tang}}, \bibinfo {author} {\bibfnamefont {R.}~\bibnamefont
  {Vondrasek}}, \ and\ \bibinfo {author} {\bibfnamefont {A.}~\bibnamefont
  {Yokoyama}},\ }\href {\doibase 10.1126/science.1215510} {\bibfield  {journal}
  {\bibinfo  {journal} {Science}\ }\textbf {\bibinfo {volume} {335}},\ \bibinfo
  {pages} {1614} (\bibinfo {year} {2012})},\ \Eprint
  {http://arxiv.org/abs/https://www.science.org/doi/pdf/10.1126/science.1215510}
  {https://www.science.org/doi/pdf/10.1126/science.1215510} \BibitemShut
  {NoStop}%
\bibitem [{\citenamefont {Wang}\ \emph {et~al.}(2022)\citenamefont {Wang},
  \citenamefont {Liu}, \citenamefont {Lu}, \citenamefont {Chen}, \citenamefont
  {Long}, \citenamefont {Li}, \citenamefont {Chen}, \citenamefont {Chen},
  \citenamefont {Bai}, \citenamefont {Li}, \citenamefont {Peng}, \citenamefont
  {Liu}, \citenamefont {Wu}, \citenamefont {Wang}, \citenamefont {Li},
  \citenamefont {Xu}, \citenamefont {Liang}, \citenamefont {Leng},\ and\
  \citenamefont {Li}}]{9894358}%
  \BibitemOpen
  \bibfield  {author} {\bibinfo {author} {\bibfnamefont {X.}~\bibnamefont
  {Wang}}, \bibinfo {author} {\bibfnamefont {X.}~\bibnamefont {Liu}}, \bibinfo
  {author} {\bibfnamefont {X.}~\bibnamefont {Lu}}, \bibinfo {author}
  {\bibfnamefont {J.}~\bibnamefont {Chen}}, \bibinfo {author} {\bibfnamefont
  {Y.}~\bibnamefont {Long}}, \bibinfo {author} {\bibfnamefont {W.}~\bibnamefont
  {Li}}, \bibinfo {author} {\bibfnamefont {H.}~\bibnamefont {Chen}}, \bibinfo
  {author} {\bibfnamefont {X.}~\bibnamefont {Chen}}, \bibinfo {author}
  {\bibfnamefont {P.}~\bibnamefont {Bai}}, \bibinfo {author} {\bibfnamefont
  {Y.}~\bibnamefont {Li}}, \bibinfo {author} {\bibfnamefont {Y.}~\bibnamefont
  {Peng}}, \bibinfo {author} {\bibfnamefont {Y.}~\bibnamefont {Liu}}, \bibinfo
  {author} {\bibfnamefont {F.}~\bibnamefont {Wu}}, \bibinfo {author}
  {\bibfnamefont {C.}~\bibnamefont {Wang}}, \bibinfo {author} {\bibfnamefont
  {Z.}~\bibnamefont {Li}}, \bibinfo {author} {\bibfnamefont {Y.}~\bibnamefont
  {Xu}}, \bibinfo {author} {\bibfnamefont {X.}~\bibnamefont {Liang}}, \bibinfo
  {author} {\bibfnamefont {Y.}~\bibnamefont {Leng}}, \ and\ \bibinfo {author}
  {\bibfnamefont {R.}~\bibnamefont {Li}},\ }\href {\doibase
  10.34133/2022/9894358} {\bibfield  {journal} {\bibinfo  {journal} {Ultrafast
  Science}\ }\textbf {\bibinfo {volume} {2022}},\ \bibinfo {pages} {9894358}
  (\bibinfo {year} {2022})}\BibitemShut {NoStop}%
\bibitem [{\citenamefont {Strickland}\ and\ \citenamefont
  {Mourou}(1985)}]{STRICKLAND1985219}%
  \BibitemOpen
  \bibfield  {author} {\bibinfo {author} {\bibfnamefont {D.}~\bibnamefont
  {Strickland}}\ and\ \bibinfo {author} {\bibfnamefont {G.}~\bibnamefont
  {Mourou}},\ }\href {\doibase https://doi.org/10.1016/0030-4018(85)90120-8}
  {\bibfield  {journal} {\bibinfo  {journal} {Opt. Comm.}\ }\textbf {\bibinfo
  {volume} {56}},\ \bibinfo {pages} {219} (\bibinfo {year} {1985})}\BibitemShut
  {NoStop}%
\bibitem [{\citenamefont {Yoon}\ \emph {et~al.}(2019)\citenamefont {Yoon},
  \citenamefont {Jeon}, \citenamefont {Shin}, \citenamefont {Lee},
  \citenamefont {Lee}, \citenamefont {Choi}, \citenamefont {Kim}, \citenamefont
  {Sung},\ and\ \citenamefont {Nam}}]{Yoon:19}%
  \BibitemOpen
  \bibfield  {author} {\bibinfo {author} {\bibfnamefont {J.~W.}\ \bibnamefont
  {Yoon}}, \bibinfo {author} {\bibfnamefont {C.}~\bibnamefont {Jeon}}, \bibinfo
  {author} {\bibfnamefont {J.}~\bibnamefont {Shin}}, \bibinfo {author}
  {\bibfnamefont {S.~K.}\ \bibnamefont {Lee}}, \bibinfo {author} {\bibfnamefont
  {H.~W.}\ \bibnamefont {Lee}}, \bibinfo {author} {\bibfnamefont {I.~W.}\
  \bibnamefont {Choi}}, \bibinfo {author} {\bibfnamefont {H.~T.}\ \bibnamefont
  {Kim}}, \bibinfo {author} {\bibfnamefont {J.~H.}\ \bibnamefont {Sung}}, \
  and\ \bibinfo {author} {\bibfnamefont {C.~H.}\ \bibnamefont {Nam}},\ }\href
  {\doibase 10.1364/OE.27.020412} {\bibfield  {journal} {\bibinfo  {journal}
  {Opt. Express}\ }\textbf {\bibinfo {volume} {27}},\ \bibinfo {pages} {20412}
  (\bibinfo {year} {2019})}\BibitemShut {NoStop}%
\bibitem [{\citenamefont {Yoon}\ \emph {et~al.}(2021)\citenamefont {Yoon},
  \citenamefont {Kim}, \citenamefont {Choi}, \citenamefont {Sung},
  \citenamefont {Lee}, \citenamefont {Lee},\ and\ \citenamefont
  {Nam}}]{Yoon:21}%
  \BibitemOpen
  \bibfield  {author} {\bibinfo {author} {\bibfnamefont {J.~W.}\ \bibnamefont
  {Yoon}}, \bibinfo {author} {\bibfnamefont {Y.~G.}\ \bibnamefont {Kim}},
  \bibinfo {author} {\bibfnamefont {I.~W.}\ \bibnamefont {Choi}}, \bibinfo
  {author} {\bibfnamefont {J.~H.}\ \bibnamefont {Sung}}, \bibinfo {author}
  {\bibfnamefont {H.~W.}\ \bibnamefont {Lee}}, \bibinfo {author} {\bibfnamefont
  {S.~K.}\ \bibnamefont {Lee}}, \ and\ \bibinfo {author} {\bibfnamefont
  {C.~H.}\ \bibnamefont {Nam}},\ }\href {\doibase 10.1364/OPTICA.420520}
  {\bibfield  {journal} {\bibinfo  {journal} {Optica}\ }\textbf {\bibinfo
  {volume} {8}},\ \bibinfo {pages} {630} (\bibinfo {year} {2021})}\BibitemShut
  {NoStop}%
\bibitem [{\citenamefont {Li}\ \emph {et~al.}(2018)\citenamefont {Li},
  \citenamefont {Gan}, \citenamefont {Yu}, \citenamefont {Wang}, \citenamefont
  {Liu}, \citenamefont {Guo}, \citenamefont {Xu}, \citenamefont {Xu},
  \citenamefont {Hang}, \citenamefont {Xu}, \citenamefont {Wang}, \citenamefont
  {Huang}, \citenamefont {Cao}, \citenamefont {Yao}, \citenamefont {Zhang},
  \citenamefont {Chen}, \citenamefont {Tang}, \citenamefont {Li}, \citenamefont
  {Liu}, \citenamefont {Li}, \citenamefont {He}, \citenamefont {Yin},
  \citenamefont {Liang}, \citenamefont {Leng}, \citenamefont {Li},\ and\
  \citenamefont {Xu}}]{Li:18}%
  \BibitemOpen
  \bibfield  {author} {\bibinfo {author} {\bibfnamefont {W.}~\bibnamefont
  {Li}}, \bibinfo {author} {\bibfnamefont {Z.}~\bibnamefont {Gan}}, \bibinfo
  {author} {\bibfnamefont {L.}~\bibnamefont {Yu}}, \bibinfo {author}
  {\bibfnamefont {C.}~\bibnamefont {Wang}}, \bibinfo {author} {\bibfnamefont
  {Y.}~\bibnamefont {Liu}}, \bibinfo {author} {\bibfnamefont {Z.}~\bibnamefont
  {Guo}}, \bibinfo {author} {\bibfnamefont {L.}~\bibnamefont {Xu}}, \bibinfo
  {author} {\bibfnamefont {M.}~\bibnamefont {Xu}}, \bibinfo {author}
  {\bibfnamefont {Y.}~\bibnamefont {Hang}}, \bibinfo {author} {\bibfnamefont
  {Y.}~\bibnamefont {Xu}}, \bibinfo {author} {\bibfnamefont {J.}~\bibnamefont
  {Wang}}, \bibinfo {author} {\bibfnamefont {P.}~\bibnamefont {Huang}},
  \bibinfo {author} {\bibfnamefont {H.}~\bibnamefont {Cao}}, \bibinfo {author}
  {\bibfnamefont {B.}~\bibnamefont {Yao}}, \bibinfo {author} {\bibfnamefont
  {X.}~\bibnamefont {Zhang}}, \bibinfo {author} {\bibfnamefont
  {L.}~\bibnamefont {Chen}}, \bibinfo {author} {\bibfnamefont {Y.}~\bibnamefont
  {Tang}}, \bibinfo {author} {\bibfnamefont {S.}~\bibnamefont {Li}}, \bibinfo
  {author} {\bibfnamefont {X.}~\bibnamefont {Liu}}, \bibinfo {author}
  {\bibfnamefont {S.}~\bibnamefont {Li}}, \bibinfo {author} {\bibfnamefont
  {M.}~\bibnamefont {He}}, \bibinfo {author} {\bibfnamefont {D.}~\bibnamefont
  {Yin}}, \bibinfo {author} {\bibfnamefont {X.}~\bibnamefont {Liang}}, \bibinfo
  {author} {\bibfnamefont {Y.}~\bibnamefont {Leng}}, \bibinfo {author}
  {\bibfnamefont {R.}~\bibnamefont {Li}}, \ and\ \bibinfo {author}
  {\bibfnamefont {Z.}~\bibnamefont {Xu}},\ }\href {\doibase
  10.1364/OL.43.005681} {\bibfield  {journal} {\bibinfo  {journal} {Opt.
  Lett.}\ }\textbf {\bibinfo {volume} {43}},\ \bibinfo {pages} {5681} (\bibinfo
  {year} {2018})}\BibitemShut {NoStop}%
\bibitem [{\citenamefont {Yu}\ \emph {et~al.}(2018)\citenamefont {Yu},
  \citenamefont {Xu}, \citenamefont {Liu}, \citenamefont {Li}, \citenamefont
  {Li}, \citenamefont {Liu}, \citenamefont {Li}, \citenamefont {Wu},
  \citenamefont {Yang}, \citenamefont {Yang}, \citenamefont {Wang},
  \citenamefont {Lu}, \citenamefont {Leng}, \citenamefont {Li},\ and\
  \citenamefont {Xu}}]{Yu:18}%
  \BibitemOpen
  \bibfield  {author} {\bibinfo {author} {\bibfnamefont {L.}~\bibnamefont
  {Yu}}, \bibinfo {author} {\bibfnamefont {Y.}~\bibnamefont {Xu}}, \bibinfo
  {author} {\bibfnamefont {Y.}~\bibnamefont {Liu}}, \bibinfo {author}
  {\bibfnamefont {Y.}~\bibnamefont {Li}}, \bibinfo {author} {\bibfnamefont
  {S.}~\bibnamefont {Li}}, \bibinfo {author} {\bibfnamefont {Z.}~\bibnamefont
  {Liu}}, \bibinfo {author} {\bibfnamefont {W.}~\bibnamefont {Li}}, \bibinfo
  {author} {\bibfnamefont {F.}~\bibnamefont {Wu}}, \bibinfo {author}
  {\bibfnamefont {X.}~\bibnamefont {Yang}}, \bibinfo {author} {\bibfnamefont
  {Y.}~\bibnamefont {Yang}}, \bibinfo {author} {\bibfnamefont {C.}~\bibnamefont
  {Wang}}, \bibinfo {author} {\bibfnamefont {X.}~\bibnamefont {Lu}}, \bibinfo
  {author} {\bibfnamefont {Y.}~\bibnamefont {Leng}}, \bibinfo {author}
  {\bibfnamefont {R.}~\bibnamefont {Li}}, \ and\ \bibinfo {author}
  {\bibfnamefont {Z.}~\bibnamefont {Xu}},\ }\href {\doibase
  10.1364/OE.26.002625} {\bibfield  {journal} {\bibinfo  {journal} {Opt.
  Express}\ }\textbf {\bibinfo {volume} {26}},\ \bibinfo {pages} {2625}
  (\bibinfo {year} {2018})}\BibitemShut {NoStop}%
\bibitem [{\citenamefont {Mi{\c{s}}icu}\ and\ \citenamefont
  {Rizea}(2019)}]{Mi_icu_2019}%
  \BibitemOpen
  \bibfield  {author} {\bibinfo {author} {\bibfnamefont {{\c{S}}.}~\bibnamefont
  {Mi{\c{s}}icu}}\ and\ \bibinfo {author} {\bibfnamefont {M.}~\bibnamefont
  {Rizea}},\ }\href {\doibase 10.1088/1361-6471/ab1d7c} {\bibfield  {journal}
  {\bibinfo  {journal} {J. Phys. G}\ }\textbf {\bibinfo {volume} {46}},\
  \bibinfo {pages} {115106} (\bibinfo {year} {2019})}\BibitemShut {NoStop}%
\bibitem [{\citenamefont {Tanaka}\ \emph {et~al.}(2020)\citenamefont {Tanaka},
  \citenamefont {Spohr}, \citenamefont {Balabanski}, \citenamefont {Balascuta},
  \citenamefont {Capponi}, \citenamefont {Cernaianu}, \citenamefont {Cuciuc},
  \citenamefont {Cucoanes}, \citenamefont {Dancus}, \citenamefont {Dhal},
  \citenamefont {Diaconescu}, \citenamefont {Doria}, \citenamefont {Ghenuche},
  \citenamefont {Ghita}, \citenamefont {Kisyov}, \citenamefont {Nastasa},
  \citenamefont {Ong}, \citenamefont {Rotaru}, \citenamefont {Sangwan},
  \citenamefont {Söderström}, \citenamefont {Stutman}, \citenamefont
  {Suliman}, \citenamefont {Tesileanu}, \citenamefont {Tudor}, \citenamefont
  {Tsoneva}, \citenamefont {Ur}, \citenamefont {Ursescu},\ and\ \citenamefont
  {Zamfir}}]{doi:10.1063/1.5093535}%
  \BibitemOpen
  \bibfield  {author} {\bibinfo {author} {\bibfnamefont {K.~A.}\ \bibnamefont
  {Tanaka}}, \bibinfo {author} {\bibfnamefont {K.~M.}\ \bibnamefont {Spohr}},
  \bibinfo {author} {\bibfnamefont {D.~L.}\ \bibnamefont {Balabanski}},
  \bibinfo {author} {\bibfnamefont {S.}~\bibnamefont {Balascuta}}, \bibinfo
  {author} {\bibfnamefont {L.}~\bibnamefont {Capponi}}, \bibinfo {author}
  {\bibfnamefont {M.~O.}\ \bibnamefont {Cernaianu}}, \bibinfo {author}
  {\bibfnamefont {M.}~\bibnamefont {Cuciuc}}, \bibinfo {author} {\bibfnamefont
  {A.}~\bibnamefont {Cucoanes}}, \bibinfo {author} {\bibfnamefont
  {I.}~\bibnamefont {Dancus}}, \bibinfo {author} {\bibfnamefont
  {A.}~\bibnamefont {Dhal}}, \bibinfo {author} {\bibfnamefont {B.}~\bibnamefont
  {Diaconescu}}, \bibinfo {author} {\bibfnamefont {D.}~\bibnamefont {Doria}},
  \bibinfo {author} {\bibfnamefont {P.}~\bibnamefont {Ghenuche}}, \bibinfo
  {author} {\bibfnamefont {D.~G.}\ \bibnamefont {Ghita}}, \bibinfo {author}
  {\bibfnamefont {S.}~\bibnamefont {Kisyov}}, \bibinfo {author} {\bibfnamefont
  {V.}~\bibnamefont {Nastasa}}, \bibinfo {author} {\bibfnamefont {J.~F.}\
  \bibnamefont {Ong}}, \bibinfo {author} {\bibfnamefont {F.}~\bibnamefont
  {Rotaru}}, \bibinfo {author} {\bibfnamefont {D.}~\bibnamefont {Sangwan}},
  \bibinfo {author} {\bibfnamefont {P.-A.}\ \bibnamefont {Söderström}},
  \bibinfo {author} {\bibfnamefont {D.}~\bibnamefont {Stutman}}, \bibinfo
  {author} {\bibfnamefont {G.}~\bibnamefont {Suliman}}, \bibinfo {author}
  {\bibfnamefont {O.}~\bibnamefont {Tesileanu}}, \bibinfo {author}
  {\bibfnamefont {L.}~\bibnamefont {Tudor}}, \bibinfo {author} {\bibfnamefont
  {N.}~\bibnamefont {Tsoneva}}, \bibinfo {author} {\bibfnamefont {C.~A.}\
  \bibnamefont {Ur}}, \bibinfo {author} {\bibfnamefont {D.}~\bibnamefont
  {Ursescu}}, \ and\ \bibinfo {author} {\bibfnamefont {N.~V.}\ \bibnamefont
  {Zamfir}},\ }\href {\doibase 10.1063/1.5093535} {\bibfield  {journal}
  {\bibinfo  {journal} {Matter and Radiation at Extremes}\ }\textbf {\bibinfo
  {volume} {5}},\ \bibinfo {pages} {024402} (\bibinfo {year} {2020})},\ \Eprint
  {http://arxiv.org/abs/https://doi.org/10.1063/1.5093535}
  {https://doi.org/10.1063/1.5093535} \BibitemShut {NoStop}%
\bibitem [{\citenamefont {Wang}\ \emph
  {et~al.}(2021{\natexlab{a}})\citenamefont {Wang}, \citenamefont {Zhou},
  \citenamefont {Liu},\ and\ \citenamefont {Wang}}]{PhysRevLett.127.052501}%
  \BibitemOpen
  \bibfield  {author} {\bibinfo {author} {\bibfnamefont {W.}~\bibnamefont
  {Wang}}, \bibinfo {author} {\bibfnamefont {J.}~\bibnamefont {Zhou}}, \bibinfo
  {author} {\bibfnamefont {B.}~\bibnamefont {Liu}}, \ and\ \bibinfo {author}
  {\bibfnamefont {X.}~\bibnamefont {Wang}},\ }\href {\doibase
  10.1103/PhysRevLett.127.052501} {\bibfield  {journal} {\bibinfo  {journal}
  {Phys. Rev. Lett.}\ }\textbf {\bibinfo {volume} {127}},\ \bibinfo {pages}
  {052501} (\bibinfo {year} {2021}{\natexlab{a}})}\BibitemShut {NoStop}%
\bibitem [{\citenamefont {Lv}\ \emph {et~al.}(2019)\citenamefont {Lv},
  \citenamefont {Duan},\ and\ \citenamefont {Liu}}]{PhysRevC.100.064610}%
  \BibitemOpen
  \bibfield  {author} {\bibinfo {author} {\bibfnamefont {W.}~\bibnamefont
  {Lv}}, \bibinfo {author} {\bibfnamefont {H.}~\bibnamefont {Duan}}, \ and\
  \bibinfo {author} {\bibfnamefont {J.}~\bibnamefont {Liu}},\ }\href {\doibase
  10.1103/PhysRevC.100.064610} {\bibfield  {journal} {\bibinfo  {journal}
  {Phys. Rev. C}\ }\textbf {\bibinfo {volume} {100}},\ \bibinfo {pages}
  {064610} (\bibinfo {year} {2019})}\BibitemShut {NoStop}%
\bibitem [{\citenamefont {Ghinescu}\ and\ \citenamefont
  {Delion}(2020)}]{PhysRevC.101.044304}%
  \BibitemOpen
  \bibfield  {author} {\bibinfo {author} {\bibfnamefont {S.~A.}\ \bibnamefont
  {Ghinescu}}\ and\ \bibinfo {author} {\bibfnamefont {D.~S.}\ \bibnamefont
  {Delion}},\ }\href {\doibase 10.1103/PhysRevC.101.044304} {\bibfield
  {journal} {\bibinfo  {journal} {Phys. Rev. C}\ }\textbf {\bibinfo {volume}
  {101}},\ \bibinfo {pages} {044304} (\bibinfo {year} {2020})}\BibitemShut
  {NoStop}%
\bibitem [{\citenamefont {von~der Wense}\ \emph {et~al.}(2020)\citenamefont
  {von~der Wense}, \citenamefont {Bilous}, \citenamefont {Seiferle},
  \citenamefont {Stellmer}, \citenamefont {Weitenberg}, \citenamefont
  {Thirolf},\ and\ \citenamefont {A.~Palffy}}]{V2020}%
  \BibitemOpen
  \bibfield  {author} {\bibinfo {author} {\bibfnamefont {L.}~\bibnamefont
  {von~der Wense}}, \bibinfo {author} {\bibfnamefont {P.~V.}\ \bibnamefont
  {Bilous}}, \bibinfo {author} {\bibfnamefont {B.}~\bibnamefont {Seiferle}},
  \bibinfo {author} {\bibfnamefont {S.}~\bibnamefont {Stellmer}}, \bibinfo
  {author} {\bibfnamefont {J.}~\bibnamefont {Weitenberg}}, \bibinfo {author}
  {\bibfnamefont {P.~G.}\ \bibnamefont {Thirolf}}, \ and\ \bibinfo {author}
  {\bibfnamefont {G.~K.}\ \bibnamefont {A.~Palffy}},\ }\href {\doibase
  10.1140/epja/s10050-020-00177-x} {\bibfield  {journal} {\bibinfo  {journal}
  {Euro. Phys. J. A}\ }\textbf {\bibinfo {volume} {56}},\ \bibinfo {pages}
  {176} (\bibinfo {year} {2020})}\BibitemShut {NoStop}%
\bibitem [{\citenamefont {Mi\ifmmode~\mbox{\c{s}}\else
  \c{s}\fi{}icu}(2022)}]{PhysRevC.106.034612}%
  \BibitemOpen
  \bibfield  {author} {\bibinfo {author} {\bibfnamefont {i.~m.~c.}\
  \bibnamefont {Mi\ifmmode~\mbox{\c{s}}\else \c{s}\fi{}icu}},\ }\href {\doibase
  10.1103/PhysRevC.106.034612} {\bibfield  {journal} {\bibinfo  {journal}
  {Phys. Rev. C}\ }\textbf {\bibinfo {volume} {106}},\ \bibinfo {pages}
  {034612} (\bibinfo {year} {2022})}\BibitemShut {NoStop}%
\bibitem [{\citenamefont {Wang}(2022)}]{PhysRevC.106.024606}%
  \BibitemOpen
  \bibfield  {author} {\bibinfo {author} {\bibfnamefont {X.}~\bibnamefont
  {Wang}},\ }\href {\doibase 10.1103/PhysRevC.106.024606} {\bibfield  {journal}
  {\bibinfo  {journal} {Phys. Rev. C}\ }\textbf {\bibinfo {volume} {106}},\
  \bibinfo {pages} {024606} (\bibinfo {year} {2022})}\BibitemShut {NoStop}%
\bibitem [{\citenamefont {Bekx}\ \emph {et~al.}(2022)\citenamefont {Bekx},
  \citenamefont {Lindsey}, \citenamefont {Glenzer},\ and\ \citenamefont
  {Schlesinger}}]{PhysRevC.105.054001}%
  \BibitemOpen
  \bibfield  {author} {\bibinfo {author} {\bibfnamefont {J.~J.}\ \bibnamefont
  {Bekx}}, \bibinfo {author} {\bibfnamefont {M.~L.}\ \bibnamefont {Lindsey}},
  \bibinfo {author} {\bibfnamefont {S.~H.}\ \bibnamefont {Glenzer}}, \ and\
  \bibinfo {author} {\bibfnamefont {K.-G.}\ \bibnamefont {Schlesinger}},\
  }\href {\doibase 10.1103/PhysRevC.105.054001} {\bibfield  {journal} {\bibinfo
   {journal} {Phys. Rev. C}\ }\textbf {\bibinfo {volume} {105}},\ \bibinfo
  {pages} {054001} (\bibinfo {year} {2022})}\BibitemShut {NoStop}%
\bibitem [{\citenamefont {Qi}\ \emph {et~al.}(2019)\citenamefont {Qi},
  \citenamefont {Li}, \citenamefont {Xu}, \citenamefont {Fu},\ and\
  \citenamefont {Wang}}]{PhysRevC.99.044610}%
  \BibitemOpen
  \bibfield  {author} {\bibinfo {author} {\bibfnamefont {J.}~\bibnamefont
  {Qi}}, \bibinfo {author} {\bibfnamefont {T.}~\bibnamefont {Li}}, \bibinfo
  {author} {\bibfnamefont {R.}~\bibnamefont {Xu}}, \bibinfo {author}
  {\bibfnamefont {L.}~\bibnamefont {Fu}}, \ and\ \bibinfo {author}
  {\bibfnamefont {X.}~\bibnamefont {Wang}},\ }\href {\doibase
  10.1103/PhysRevC.99.044610} {\bibfield  {journal} {\bibinfo  {journal} {Phys.
  Rev. C}\ }\textbf {\bibinfo {volume} {99}},\ \bibinfo {pages} {044610}
  (\bibinfo {year} {2019})}\BibitemShut {NoStop}%
\bibitem [{\citenamefont {Queisser}\ and\ \citenamefont
  {Sch\"utzhold}(2019)}]{PhysRevC.100.041601}%
  \BibitemOpen
  \bibfield  {author} {\bibinfo {author} {\bibfnamefont {F.}~\bibnamefont
  {Queisser}}\ and\ \bibinfo {author} {\bibfnamefont {R.}~\bibnamefont
  {Sch\"utzhold}},\ }\href {\doibase 10.1103/PhysRevC.100.041601} {\bibfield
  {journal} {\bibinfo  {journal} {Phys. Rev. C}\ }\textbf {\bibinfo {volume}
  {100}},\ \bibinfo {pages} {041601} (\bibinfo {year} {2019})}\BibitemShut
  {NoStop}%
\bibitem [{\citenamefont {Li}\ and\ \citenamefont {Wang}(2021)}]{Li_2021}%
  \BibitemOpen
  \bibfield  {author} {\bibinfo {author} {\bibfnamefont {T.}~\bibnamefont
  {Li}}\ and\ \bibinfo {author} {\bibfnamefont {X.}~\bibnamefont {Wang}},\
  }\href {\doibase 10.1088/1361-6471/ac1712} {\bibfield  {journal} {\bibinfo
  {journal} {Journal of Physics G: Nuclear and Particle Physics}\ }\textbf
  {\bibinfo {volume} {48}},\ \bibinfo {pages} {095105} (\bibinfo {year}
  {2021})}\BibitemShut {NoStop}%
\bibitem [{\citenamefont {Liu}\ \emph {et~al.}(2021)\citenamefont {Liu},
  \citenamefont {Duan}, \citenamefont {Ye},\ and\ \citenamefont
  {Liu}}]{PhysRevC.104.044614}%
  \BibitemOpen
  \bibfield  {author} {\bibinfo {author} {\bibfnamefont {S.}~\bibnamefont
  {Liu}}, \bibinfo {author} {\bibfnamefont {H.}~\bibnamefont {Duan}}, \bibinfo
  {author} {\bibfnamefont {D.}~\bibnamefont {Ye}}, \ and\ \bibinfo {author}
  {\bibfnamefont {J.}~\bibnamefont {Liu}},\ }\href {\doibase
  10.1103/PhysRevC.104.044614} {\bibfield  {journal} {\bibinfo  {journal}
  {Phys. Rev. C}\ }\textbf {\bibinfo {volume} {104}},\ \bibinfo {pages}
  {044614} (\bibinfo {year} {2021})}\BibitemShut {NoStop}%
\bibitem [{\citenamefont {Lv}\ \emph {et~al.}(2022)\citenamefont {Lv},
  \citenamefont {Wu}, \citenamefont {Duan}, \citenamefont {Liu},\ and\
  \citenamefont {Liu}}]{Lv2020}%
  \BibitemOpen
  \bibfield  {author} {\bibinfo {author} {\bibfnamefont {W.~J.}\ \bibnamefont
  {Lv}}, \bibinfo {author} {\bibfnamefont {B.~B.}\ \bibnamefont {Wu}}, \bibinfo
  {author} {\bibfnamefont {H.}~\bibnamefont {Duan}}, \bibinfo {author}
  {\bibfnamefont {S.~W.}\ \bibnamefont {Liu}}, \ and\ \bibinfo {author}
  {\bibfnamefont {J.}~\bibnamefont {Liu}},\ }\href {\doibase
  10.1140/epja/s10050-022-00697-8} {\bibfield  {journal} {\bibinfo  {journal}
  {Euro. Phys. J. A}\ }\textbf {\bibinfo {volume} {58}},\ \bibinfo {pages} {54}
  (\bibinfo {year} {2022})}\BibitemShut {NoStop}%
\bibitem [{\citenamefont {Wang}(2020)}]{PhysRevC.102.011601}%
  \BibitemOpen
  \bibfield  {author} {\bibinfo {author} {\bibfnamefont {X.}~\bibnamefont
  {Wang}},\ }\href {\doibase 10.1103/PhysRevC.102.011601} {\bibfield  {journal}
  {\bibinfo  {journal} {Phys. Rev. C}\ }\textbf {\bibinfo {volume} {102}},\
  \bibinfo {pages} {011601} (\bibinfo {year} {2020})}\BibitemShut {NoStop}%
\bibitem [{\citenamefont {Qi}\ \emph {et~al.}(2020)\citenamefont {Qi},
  \citenamefont {Fu},\ and\ \citenamefont {Wang}}]{PhysRevC.102.064629}%
  \BibitemOpen
  \bibfield  {author} {\bibinfo {author} {\bibfnamefont {J.}~\bibnamefont
  {Qi}}, \bibinfo {author} {\bibfnamefont {L.}~\bibnamefont {Fu}}, \ and\
  \bibinfo {author} {\bibfnamefont {X.}~\bibnamefont {Wang}},\ }\href {\doibase
  10.1103/PhysRevC.102.064629} {\bibfield  {journal} {\bibinfo  {journal}
  {Phys. Rev. C}\ }\textbf {\bibinfo {volume} {102}},\ \bibinfo {pages}
  {064629} (\bibinfo {year} {2020})}\BibitemShut {NoStop}%
\bibitem [{\citenamefont {Feng}\ \emph {et~al.}(2022)\citenamefont {Feng},
  \citenamefont {Wang}, \citenamefont {Fu}, \citenamefont {Chen}, \citenamefont
  {Tan}, \citenamefont {Li}, \citenamefont {Wang}, \citenamefont {Li},
  \citenamefont {Zhang}, \citenamefont {Ma},\ and\ \citenamefont
  {Zhang}}]{PhysRevLett.128.052501}%
  \BibitemOpen
  \bibfield  {author} {\bibinfo {author} {\bibfnamefont {J.}~\bibnamefont
  {Feng}}, \bibinfo {author} {\bibfnamefont {W.}~\bibnamefont {Wang}}, \bibinfo
  {author} {\bibfnamefont {C.}~\bibnamefont {Fu}}, \bibinfo {author}
  {\bibfnamefont {L.}~\bibnamefont {Chen}}, \bibinfo {author} {\bibfnamefont
  {J.}~\bibnamefont {Tan}}, \bibinfo {author} {\bibfnamefont {Y.}~\bibnamefont
  {Li}}, \bibinfo {author} {\bibfnamefont {J.}~\bibnamefont {Wang}}, \bibinfo
  {author} {\bibfnamefont {Y.}~\bibnamefont {Li}}, \bibinfo {author}
  {\bibfnamefont {G.}~\bibnamefont {Zhang}}, \bibinfo {author} {\bibfnamefont
  {Y.}~\bibnamefont {Ma}}, \ and\ \bibinfo {author} {\bibfnamefont
  {J.}~\bibnamefont {Zhang}},\ }\href {\doibase 10.1103/PhysRevLett.128.052501}
  {\bibfield  {journal} {\bibinfo  {journal} {Phys. Rev. Lett.}\ }\textbf
  {\bibinfo {volume} {128}},\ \bibinfo {pages} {052501} (\bibinfo {year}
  {2022})}\BibitemShut {NoStop}%
\bibitem [{\citenamefont {Mi{\c{s}}icu}\ and\ \citenamefont
  {Rizea}(2013)}]{Mi_icu_2013}%
  \BibitemOpen
  \bibfield  {author} {\bibinfo {author} {\bibfnamefont {{\c{S}}.}~\bibnamefont
  {Mi{\c{s}}icu}}\ and\ \bibinfo {author} {\bibfnamefont {M.}~\bibnamefont
  {Rizea}},\ }\href {\doibase 10.1088/0954-3899/40/9/095101} {\bibfield
  {journal} {\bibinfo  {journal} {J. Phys. G}\ }\textbf {\bibinfo {volume}
  {40}},\ \bibinfo {pages} {095101} (\bibinfo {year} {2013})}\BibitemShut
  {NoStop}%
\bibitem [{\citenamefont {Delion}\ and\ \citenamefont
  {Ghinescu}(2017)}]{PhysRevLett.119.202501}%
  \BibitemOpen
  \bibfield  {author} {\bibinfo {author} {\bibfnamefont {D.~S.}\ \bibnamefont
  {Delion}}\ and\ \bibinfo {author} {\bibfnamefont {S.~A.}\ \bibnamefont
  {Ghinescu}},\ }\href {\doibase 10.1103/PhysRevLett.119.202501} {\bibfield
  {journal} {\bibinfo  {journal} {Phys. Rev. Lett.}\ }\textbf {\bibinfo
  {volume} {119}},\ \bibinfo {pages} {202501} (\bibinfo {year}
  {2017})}\BibitemShut {NoStop}%
\bibitem [{\citenamefont {Kis}\ and\ \citenamefont
  {Szilvasi}(2018)}]{Kis_2018}%
  \BibitemOpen
  \bibfield  {author} {\bibinfo {author} {\bibfnamefont {D.~P.}\ \bibnamefont
  {Kis}}\ and\ \bibinfo {author} {\bibfnamefont {R.}~\bibnamefont {Szilvasi}},\
  }\href {\doibase 10.1088/1361-6471/aab0d5} {\bibfield  {journal} {\bibinfo
  {journal} {J. Phys. G}\ }\textbf {\bibinfo {volume} {45}},\ \bibinfo {pages}
  {045103} (\bibinfo {year} {2018})}\BibitemShut {NoStop}%
\bibitem [{\citenamefont {Bai}\ \emph {et~al.}(2018)\citenamefont {Bai},
  \citenamefont {Deng},\ and\ \citenamefont {Ren}}]{BAI201823}%
  \BibitemOpen
  \bibfield  {author} {\bibinfo {author} {\bibfnamefont {D.}~\bibnamefont
  {Bai}}, \bibinfo {author} {\bibfnamefont {D.}~\bibnamefont {Deng}}, \ and\
  \bibinfo {author} {\bibfnamefont {Z.}~\bibnamefont {Ren}},\ }\href {\doibase
  https://doi.org/10.1016/j.nuclphysa.2018.05.004} {\bibfield  {journal}
  {\bibinfo  {journal} {Nucl. Phys. A}\ }\textbf {\bibinfo {volume} {976}},\
  \bibinfo {pages} {23} (\bibinfo {year} {2018})}\BibitemShut {NoStop}%
\bibitem [{\citenamefont {P\'alffy}\ and\ \citenamefont
  {Popruzhenko}(2020)}]{PhysRevLett.124.212505}%
  \BibitemOpen
  \bibfield  {author} {\bibinfo {author} {\bibfnamefont {A.}~\bibnamefont
  {P\'alffy}}\ and\ \bibinfo {author} {\bibfnamefont {S.~V.}\ \bibnamefont
  {Popruzhenko}},\ }\href {\doibase 10.1103/PhysRevLett.124.212505} {\bibfield
  {journal} {\bibinfo  {journal} {Phys. Rev. Lett.}\ }\textbf {\bibinfo
  {volume} {124}},\ \bibinfo {pages} {212505} (\bibinfo {year}
  {2020})}\BibitemShut {NoStop}%
\bibitem [{\citenamefont {Soylu}\ and\ \citenamefont
  {Evlice}(2015)}]{SOYLU201559}%
  \BibitemOpen
  \bibfield  {author} {\bibinfo {author} {\bibfnamefont {A.}~\bibnamefont
  {Soylu}}\ and\ \bibinfo {author} {\bibfnamefont {S.}~\bibnamefont {Evlice}},\
  }\href {\doibase https://doi.org/10.1016/j.nuclphysa.2015.01.008} {\bibfield
  {journal} {\bibinfo  {journal} {Nuclear Physics A}\ }\textbf {\bibinfo
  {volume} {936}},\ \bibinfo {pages} {59} (\bibinfo {year} {2015})}\BibitemShut
  {NoStop}%
\bibitem [{\citenamefont {Ni}\ and\ \citenamefont
  {Ren}(2011)}]{PhysRevC.83.067302}%
  \BibitemOpen
  \bibfield  {author} {\bibinfo {author} {\bibfnamefont {D.}~\bibnamefont
  {Ni}}\ and\ \bibinfo {author} {\bibfnamefont {Z.}~\bibnamefont {Ren}},\
  }\href {\doibase 10.1103/PhysRevC.83.067302} {\bibfield  {journal} {\bibinfo
  {journal} {Phys. Rev. C}\ }\textbf {\bibinfo {volume} {83}},\ \bibinfo
  {pages} {067302} (\bibinfo {year} {2011})}\BibitemShut {NoStop}%
\bibitem [{\citenamefont {Coban}\ \emph {et~al.}(2012)\citenamefont {Coban},
  \citenamefont {Bayrak}, \citenamefont {Soylu},\ and\ \citenamefont
  {Boztosun}}]{PhysRevC.85.044324}%
  \BibitemOpen
  \bibfield  {author} {\bibinfo {author} {\bibfnamefont {A.}~\bibnamefont
  {Coban}}, \bibinfo {author} {\bibfnamefont {O.}~\bibnamefont {Bayrak}},
  \bibinfo {author} {\bibfnamefont {A.}~\bibnamefont {Soylu}}, \ and\ \bibinfo
  {author} {\bibfnamefont {I.}~\bibnamefont {Boztosun}},\ }\href {\doibase
  10.1103/PhysRevC.85.044324} {\bibfield  {journal} {\bibinfo  {journal} {Phys.
  Rev. C}\ }\textbf {\bibinfo {volume} {85}},\ \bibinfo {pages} {044324}
  (\bibinfo {year} {2012})}\BibitemShut {NoStop}%
\bibitem [{\citenamefont {Zdeb}\ \emph {et~al.}(2013)\citenamefont {Zdeb},
  \citenamefont {Warda},\ and\ \citenamefont {Pomorski}}]{PhysRevC.87.024308}%
  \BibitemOpen
  \bibfield  {author} {\bibinfo {author} {\bibfnamefont {A.}~\bibnamefont
  {Zdeb}}, \bibinfo {author} {\bibfnamefont {M.}~\bibnamefont {Warda}}, \ and\
  \bibinfo {author} {\bibfnamefont {K.}~\bibnamefont {Pomorski}},\ }\href
  {\doibase 10.1103/PhysRevC.87.024308} {\bibfield  {journal} {\bibinfo
  {journal} {Phys. Rev. C}\ }\textbf {\bibinfo {volume} {87}},\ \bibinfo
  {pages} {024308} (\bibinfo {year} {2013})}\BibitemShut {NoStop}%
\bibitem [{\citenamefont {Cheng}\ \emph {et~al.}(2019)\citenamefont {Cheng},
  \citenamefont {Chen}, \citenamefont {Deng}, \citenamefont {Wu}, \citenamefont
  {Li},\ and\ \citenamefont {Chu}}]{CHENG2019350}%
  \BibitemOpen
  \bibfield  {author} {\bibinfo {author} {\bibfnamefont {J.-H.}\ \bibnamefont
  {Cheng}}, \bibinfo {author} {\bibfnamefont {J.-L.}\ \bibnamefont {Chen}},
  \bibinfo {author} {\bibfnamefont {J.-G.}\ \bibnamefont {Deng}}, \bibinfo
  {author} {\bibfnamefont {X.-J.}\ \bibnamefont {Wu}}, \bibinfo {author}
  {\bibfnamefont {X.-H.}\ \bibnamefont {Li}}, \ and\ \bibinfo {author}
  {\bibfnamefont {P.-C.}\ \bibnamefont {Chu}},\ }\href {\doibase
  https://doi.org/10.1016/j.nuclphysa.2019.05.002} {\bibfield  {journal}
  {\bibinfo  {journal} {Nuclear Physics A}\ }\textbf {\bibinfo {volume}
  {987}},\ \bibinfo {pages} {350} (\bibinfo {year} {2019})}\BibitemShut
  {NoStop}%
\bibitem [{\citenamefont {Guo}\ \emph {et~al.}(2015)\citenamefont {Guo},
  \citenamefont {Bao}, \citenamefont {Gao}, \citenamefont {Li},\ and\
  \citenamefont {Zhang}}]{GUO2015110}%
  \BibitemOpen
  \bibfield  {author} {\bibinfo {author} {\bibfnamefont {S.}~\bibnamefont
  {Guo}}, \bibinfo {author} {\bibfnamefont {X.}~\bibnamefont {Bao}}, \bibinfo
  {author} {\bibfnamefont {Y.}~\bibnamefont {Gao}}, \bibinfo {author}
  {\bibfnamefont {J.}~\bibnamefont {Li}}, \ and\ \bibinfo {author}
  {\bibfnamefont {H.}~\bibnamefont {Zhang}},\ }\href {\doibase
  https://doi.org/10.1016/j.nuclphysa.2014.12.001} {\bibfield  {journal}
  {\bibinfo  {journal} {Nuclear Physics A}\ }\textbf {\bibinfo {volume}
  {934}},\ \bibinfo {pages} {110} (\bibinfo {year} {2015})}\BibitemShut
  {NoStop}%
\bibitem [{\citenamefont {Zhang}\ \emph {et~al.}(2006)\citenamefont {Zhang},
  \citenamefont {Zuo}, \citenamefont {Li},\ and\ \citenamefont
  {Royer}}]{PhysRevC.74.017304}%
  \BibitemOpen
  \bibfield  {author} {\bibinfo {author} {\bibfnamefont {H.}~\bibnamefont
  {Zhang}}, \bibinfo {author} {\bibfnamefont {W.}~\bibnamefont {Zuo}}, \bibinfo
  {author} {\bibfnamefont {J.}~\bibnamefont {Li}}, \ and\ \bibinfo {author}
  {\bibfnamefont {G.}~\bibnamefont {Royer}},\ }\href {\doibase
  10.1103/PhysRevC.74.017304} {\bibfield  {journal} {\bibinfo  {journal} {Phys.
  Rev. C}\ }\textbf {\bibinfo {volume} {74}},\ \bibinfo {pages} {017304}
  (\bibinfo {year} {2006})}\BibitemShut {NoStop}%
\bibitem [{\citenamefont {Gon\ifmmode~\mbox{\c{c}}\else \c{c}\fi{}alves}\ and\
  \citenamefont {Duarte}(1993)}]{PhysRevC.48.2409}%
  \BibitemOpen
  \bibfield  {author} {\bibinfo {author} {\bibfnamefont {M.}~\bibnamefont
  {Gon\ifmmode~\mbox{\c{c}}\else \c{c}\fi{}alves}}\ and\ \bibinfo {author}
  {\bibfnamefont {S.~B.}\ \bibnamefont {Duarte}},\ }\href {\doibase
  10.1103/PhysRevC.48.2409} {\bibfield  {journal} {\bibinfo  {journal} {Phys.
  Rev. C}\ }\textbf {\bibinfo {volume} {48}},\ \bibinfo {pages} {2409}
  (\bibinfo {year} {1993})}\BibitemShut {NoStop}%
\bibitem [{\citenamefont {Buck}\ \emph {et~al.}(1990)\citenamefont {Buck},
  \citenamefont {Merchant},\ and\ \citenamefont {Perez}}]{PhysRevLett.65.2975}%
  \BibitemOpen
  \bibfield  {author} {\bibinfo {author} {\bibfnamefont {B.}~\bibnamefont
  {Buck}}, \bibinfo {author} {\bibfnamefont {A.~C.}\ \bibnamefont {Merchant}},
  \ and\ \bibinfo {author} {\bibfnamefont {S.~M.}\ \bibnamefont {Perez}},\
  }\href {\doibase 10.1103/PhysRevLett.65.2975} {\bibfield  {journal} {\bibinfo
   {journal} {Phys. Rev. Lett.}\ }\textbf {\bibinfo {volume} {65}},\ \bibinfo
  {pages} {2975} (\bibinfo {year} {1990})}\BibitemShut {NoStop}%
\bibitem [{\citenamefont {Xu}\ and\ \citenamefont
  {Ren}(2006{\natexlab{a}})}]{PhysRevC.74.014304}%
  \BibitemOpen
  \bibfield  {author} {\bibinfo {author} {\bibfnamefont {C.}~\bibnamefont
  {Xu}}\ and\ \bibinfo {author} {\bibfnamefont {Z.}~\bibnamefont {Ren}},\
  }\href {\doibase 10.1103/PhysRevC.74.014304} {\bibfield  {journal} {\bibinfo
  {journal} {Phys. Rev. C}\ }\textbf {\bibinfo {volume} {74}},\ \bibinfo
  {pages} {014304} (\bibinfo {year} {2006}{\natexlab{a}})}\BibitemShut
  {NoStop}%
\bibitem [{\citenamefont {Xu}\ and\ \citenamefont {Ren}(2005)}]{XU2005303}%
  \BibitemOpen
  \bibfield  {author} {\bibinfo {author} {\bibfnamefont {C.}~\bibnamefont
  {Xu}}\ and\ \bibinfo {author} {\bibfnamefont {Z.}~\bibnamefont {Ren}},\
  }\href {\doibase https://doi.org/10.1016/j.nuclphysa.2005.06.011} {\bibfield
  {journal} {\bibinfo  {journal} {Nuclear Physics A}\ }\textbf {\bibinfo
  {volume} {760}},\ \bibinfo {pages} {303} (\bibinfo {year}
  {2005})}\BibitemShut {NoStop}%
\bibitem [{\citenamefont {Dzyublik}(2017)}]{Dzyublik2017}%
  \BibitemOpen
  \bibfield  {author} {\bibinfo {author} {\bibfnamefont {A.~Y.}\ \bibnamefont
  {Dzyublik}},\ }\href {\doibase https://doi.org/10.5506/APhysPolBSupp.10.69}
  {\bibfield  {journal} {\bibinfo  {journal} {Acta Physica Polonica B
  Proceedings Supplement}\ }\textbf {\bibinfo {volume} {10}},\ \bibinfo {pages}
  {69} (\bibinfo {year} {2017})}\BibitemShut {NoStop}%
\bibitem [{\citenamefont {Delion}\ \emph {et~al.}(2015)\citenamefont {Delion},
  \citenamefont {Liotta},\ and\ \citenamefont {Wyss}}]{PhysRevC.92.051301}%
  \BibitemOpen
  \bibfield  {author} {\bibinfo {author} {\bibfnamefont {D.~S.}\ \bibnamefont
  {Delion}}, \bibinfo {author} {\bibfnamefont {R.~J.}\ \bibnamefont {Liotta}},
  \ and\ \bibinfo {author} {\bibfnamefont {R.}~\bibnamefont {Wyss}},\ }\href
  {\doibase 10.1103/PhysRevC.92.051301} {\bibfield  {journal} {\bibinfo
  {journal} {Phys. Rev. C}\ }\textbf {\bibinfo {volume} {92}},\ \bibinfo
  {pages} {051301} (\bibinfo {year} {2015})}\BibitemShut {NoStop}%
\bibitem [{\citenamefont {Delion}\ \emph {et~al.}(2006)\citenamefont {Delion},
  \citenamefont {Peltonen},\ and\ \citenamefont
  {Suhonen}}]{PhysRevC.73.014315}%
  \BibitemOpen
  \bibfield  {author} {\bibinfo {author} {\bibfnamefont {D.~S.}\ \bibnamefont
  {Delion}}, \bibinfo {author} {\bibfnamefont {S.}~\bibnamefont {Peltonen}}, \
  and\ \bibinfo {author} {\bibfnamefont {J.}~\bibnamefont {Suhonen}},\ }\href
  {\doibase 10.1103/PhysRevC.73.014315} {\bibfield  {journal} {\bibinfo
  {journal} {Phys. Rev. C}\ }\textbf {\bibinfo {volume} {73}},\ \bibinfo
  {pages} {014315} (\bibinfo {year} {2006})}\BibitemShut {NoStop}%
\bibitem [{\citenamefont {Peltonen}\ \emph {et~al.}(2008)\citenamefont
  {Peltonen}, \citenamefont {Delion},\ and\ \citenamefont
  {Suhonen}}]{PhysRevC.78.034608}%
  \BibitemOpen
  \bibfield  {author} {\bibinfo {author} {\bibfnamefont {S.}~\bibnamefont
  {Peltonen}}, \bibinfo {author} {\bibfnamefont {D.~S.}\ \bibnamefont
  {Delion}}, \ and\ \bibinfo {author} {\bibfnamefont {J.}~\bibnamefont
  {Suhonen}},\ }\href {\doibase 10.1103/PhysRevC.78.034608} {\bibfield
  {journal} {\bibinfo  {journal} {Phys. Rev. C}\ }\textbf {\bibinfo {volume}
  {78}},\ \bibinfo {pages} {034608} (\bibinfo {year} {2008})}\BibitemShut
  {NoStop}%
\bibitem [{\citenamefont {Xu}\ and\ \citenamefont
  {Ren}(2006{\natexlab{b}})}]{PhysRevC.73.041301}%
  \BibitemOpen
  \bibfield  {author} {\bibinfo {author} {\bibfnamefont {C.}~\bibnamefont
  {Xu}}\ and\ \bibinfo {author} {\bibfnamefont {Z.}~\bibnamefont {Ren}},\
  }\href {\doibase 10.1103/PhysRevC.73.041301} {\bibfield  {journal} {\bibinfo
  {journal} {Phys. Rev. C}\ }\textbf {\bibinfo {volume} {73}},\ \bibinfo
  {pages} {041301} (\bibinfo {year} {2006}{\natexlab{b}})}\BibitemShut
  {NoStop}%
\bibitem [{\citenamefont {Ismail}\ \emph {et~al.}(2017)\citenamefont {Ismail},
  \citenamefont {Seif}, \citenamefont {Adel},\ and\ \citenamefont
  {Abdurrahman}}]{ISMAIL2017202}%
  \BibitemOpen
  \bibfield  {author} {\bibinfo {author} {\bibfnamefont {M.}~\bibnamefont
  {Ismail}}, \bibinfo {author} {\bibfnamefont {W.}~\bibnamefont {Seif}},
  \bibinfo {author} {\bibfnamefont {A.}~\bibnamefont {Adel}}, \ and\ \bibinfo
  {author} {\bibfnamefont {A.}~\bibnamefont {Abdurrahman}},\ }\href {\doibase
  https://doi.org/10.1016/j.nuclphysa.2016.11.010} {\bibfield  {journal}
  {\bibinfo  {journal} {Nuclear Physics A}\ }\textbf {\bibinfo {volume}
  {958}},\ \bibinfo {pages} {202} (\bibinfo {year} {2017})}\BibitemShut
  {NoStop}%
\bibitem [{\citenamefont {REN}\ and\ \citenamefont
  {XU}(2008)}]{doi:10.1142/S0217732308029885}%
  \BibitemOpen
  \bibfield  {author} {\bibinfo {author} {\bibfnamefont {Z.}~\bibnamefont
  {REN}}\ and\ \bibinfo {author} {\bibfnamefont {C.}~\bibnamefont {XU}},\
  }\href {\doibase 10.1142/S0217732308029885} {\bibfield  {journal} {\bibinfo
  {journal} {Modern Physics Letters A}\ }\textbf {\bibinfo {volume} {23}},\
  \bibinfo {pages} {2597} (\bibinfo {year} {2008})},\ \Eprint
  {http://arxiv.org/abs/https://doi.org/10.1142/S0217732308029885}
  {https://doi.org/10.1142/S0217732308029885} \BibitemShut {NoStop}%
\bibitem [{\citenamefont {Gurvitz}\ and\ \citenamefont
  {Kalbermann}(1987)}]{PhysRevLett.59.262}%
  \BibitemOpen
  \bibfield  {author} {\bibinfo {author} {\bibfnamefont {S.~A.}\ \bibnamefont
  {Gurvitz}}\ and\ \bibinfo {author} {\bibfnamefont {G.}~\bibnamefont
  {Kalbermann}},\ }\href {\doibase 10.1103/PhysRevLett.59.262} {\bibfield
  {journal} {\bibinfo  {journal} {Phys. Rev. Lett.}\ }\textbf {\bibinfo
  {volume} {59}},\ \bibinfo {pages} {262} (\bibinfo {year} {1987})}\BibitemShut
  {NoStop}%
\bibitem [{\citenamefont {Ni}\ and\ \citenamefont
  {Ren}(2010)}]{PhysRevC.81.064318}%
  \BibitemOpen
  \bibfield  {author} {\bibinfo {author} {\bibfnamefont {D.}~\bibnamefont
  {Ni}}\ and\ \bibinfo {author} {\bibfnamefont {Z.}~\bibnamefont {Ren}},\
  }\href {\doibase 10.1103/PhysRevC.81.064318} {\bibfield  {journal} {\bibinfo
  {journal} {Phys. Rev. C}\ }\textbf {\bibinfo {volume} {81}},\ \bibinfo
  {pages} {064318} (\bibinfo {year} {2010})}\BibitemShut {NoStop}%
\bibitem [{\citenamefont {Xu}\ and\ \citenamefont
  {Ren}(2008)}]{PhysRevC.78.057302}%
  \BibitemOpen
  \bibfield  {author} {\bibinfo {author} {\bibfnamefont {C.}~\bibnamefont
  {Xu}}\ and\ \bibinfo {author} {\bibfnamefont {Z.}~\bibnamefont {Ren}},\
  }\href {\doibase 10.1103/PhysRevC.78.057302} {\bibfield  {journal} {\bibinfo
  {journal} {Phys. Rev. C}\ }\textbf {\bibinfo {volume} {78}},\ \bibinfo
  {pages} {057302} (\bibinfo {year} {2008})}\BibitemShut {NoStop}%
\bibitem [{\citenamefont {Qi}\ \emph {et~al.}(2014)\citenamefont {Qi},
  \citenamefont {Andreyev}, \citenamefont {Huyse}, \citenamefont {Liotta},
  \citenamefont {{Van Duppen}},\ and\ \citenamefont {Wyss}}]{QI2014203}%
  \BibitemOpen
  \bibfield  {author} {\bibinfo {author} {\bibfnamefont {C.}~\bibnamefont
  {Qi}}, \bibinfo {author} {\bibfnamefont {A.}~\bibnamefont {Andreyev}},
  \bibinfo {author} {\bibfnamefont {M.}~\bibnamefont {Huyse}}, \bibinfo
  {author} {\bibfnamefont {R.}~\bibnamefont {Liotta}}, \bibinfo {author}
  {\bibfnamefont {P.}~\bibnamefont {{Van Duppen}}}, \ and\ \bibinfo {author}
  {\bibfnamefont {R.}~\bibnamefont {Wyss}},\ }\href {\doibase
  https://doi.org/10.1016/j.physletb.2014.05.066} {\bibfield  {journal}
  {\bibinfo  {journal} {Physics Letters B}\ }\textbf {\bibinfo {volume}
  {734}},\ \bibinfo {pages} {203} (\bibinfo {year} {2014})}\BibitemShut
  {NoStop}%
\bibitem [{\citenamefont {Deng}\ \emph {et~al.}(2020)\citenamefont {Deng},
  \citenamefont {Zhang},\ and\ \citenamefont {Royer}}]{PhysRevC.101.034307}%
  \BibitemOpen
  \bibfield  {author} {\bibinfo {author} {\bibfnamefont {J.-G.}\ \bibnamefont
  {Deng}}, \bibinfo {author} {\bibfnamefont {H.-F.}\ \bibnamefont {Zhang}}, \
  and\ \bibinfo {author} {\bibfnamefont {G.}~\bibnamefont {Royer}},\ }\href
  {\doibase 10.1103/PhysRevC.101.034307} {\bibfield  {journal} {\bibinfo
  {journal} {Phys. Rev. C}\ }\textbf {\bibinfo {volume} {101}},\ \bibinfo
  {pages} {034307} (\bibinfo {year} {2020})}\BibitemShut {NoStop}%
\bibitem [{\citenamefont {Deng}\ and\ \citenamefont
  {Zhang}(2020)}]{PhysRevC.102.044314}%
  \BibitemOpen
  \bibfield  {author} {\bibinfo {author} {\bibfnamefont {J.-G.}\ \bibnamefont
  {Deng}}\ and\ \bibinfo {author} {\bibfnamefont {H.-F.}\ \bibnamefont
  {Zhang}},\ }\href {\doibase 10.1103/PhysRevC.102.044314} {\bibfield
  {journal} {\bibinfo  {journal} {Phys. Rev. C}\ }\textbf {\bibinfo {volume}
  {102}},\ \bibinfo {pages} {044314} (\bibinfo {year} {2020})}\BibitemShut
  {NoStop}%
\bibitem [{\citenamefont {Deng}\ and\ \citenamefont
  {Zhang}(2021{\natexlab{a}})}]{Deng_2021}%
  \BibitemOpen
  \bibfield  {author} {\bibinfo {author} {\bibfnamefont {J.-G.}\ \bibnamefont
  {Deng}}\ and\ \bibinfo {author} {\bibfnamefont {H.-F.}\ \bibnamefont
  {Zhang}},\ }\href {\doibase 10.1088/1674-1137/abcc5a} {\bibfield  {journal}
  {\bibinfo  {journal} {Chin. Phys. C}\ }\textbf {\bibinfo {volume} {45}},\
  \bibinfo {pages} {024104} (\bibinfo {year} {2021}{\natexlab{a}})}\BibitemShut
  {NoStop}%
\bibitem [{\citenamefont {Deng}\ and\ \citenamefont
  {Zhang}(2021{\natexlab{b}})}]{DENG2021136247}%
  \BibitemOpen
  \bibfield  {author} {\bibinfo {author} {\bibfnamefont {J.-G.}\ \bibnamefont
  {Deng}}\ and\ \bibinfo {author} {\bibfnamefont {H.-F.}\ \bibnamefont
  {Zhang}},\ }\href {\doibase https://doi.org/10.1016/j.physletb.2021.136247}
  {\bibfield  {journal} {\bibinfo  {journal} {Physics Letters B}\ }\textbf
  {\bibinfo {volume} {816}},\ \bibinfo {pages} {136247} (\bibinfo {year}
  {2021}{\natexlab{b}})}\BibitemShut {NoStop}%
\bibitem [{\citenamefont {Cheng}\ \emph {et~al.}(2022)\citenamefont {Cheng},
  \citenamefont {Li},\ and\ \citenamefont {Yu}}]{PhysRevC.105.024312}%
  \BibitemOpen
  \bibfield  {author} {\bibinfo {author} {\bibfnamefont {J.-H.}\ \bibnamefont
  {Cheng}}, \bibinfo {author} {\bibfnamefont {Y.}~\bibnamefont {Li}}, \ and\
  \bibinfo {author} {\bibfnamefont {T.-P.}\ \bibnamefont {Yu}},\ }\href
  {\doibase 10.1103/PhysRevC.105.024312} {\bibfield  {journal} {\bibinfo
  {journal} {Phys. Rev. C}\ }\textbf {\bibinfo {volume} {105}},\ \bibinfo
  {pages} {024312} (\bibinfo {year} {2022})}\BibitemShut {NoStop}%
\bibitem [{\citenamefont {Ismail}\ \emph {et~al.}(2012)\citenamefont {Ismail},
  \citenamefont {Ellithi}, \citenamefont {Botros},\ and\ \citenamefont
  {Abdurrahman}}]{PhysRevC.86.044317}%
  \BibitemOpen
  \bibfield  {author} {\bibinfo {author} {\bibfnamefont {M.}~\bibnamefont
  {Ismail}}, \bibinfo {author} {\bibfnamefont {A.~Y.}\ \bibnamefont {Ellithi}},
  \bibinfo {author} {\bibfnamefont {M.~M.}\ \bibnamefont {Botros}}, \ and\
  \bibinfo {author} {\bibfnamefont {A.}~\bibnamefont {Abdurrahman}},\ }\href
  {\doibase 10.1103/PhysRevC.86.044317} {\bibfield  {journal} {\bibinfo
  {journal} {Phys. Rev. C}\ }\textbf {\bibinfo {volume} {86}},\ \bibinfo
  {pages} {044317} (\bibinfo {year} {2012})}\BibitemShut {NoStop}%
\bibitem [{\citenamefont {Ni}\ and\ \citenamefont {Ren}(2015)}]{NI2015108}%
  \BibitemOpen
  \bibfield  {author} {\bibinfo {author} {\bibfnamefont {D.}~\bibnamefont
  {Ni}}\ and\ \bibinfo {author} {\bibfnamefont {Z.}~\bibnamefont {Ren}},\
  }\href {\doibase https://doi.org/10.1016/j.aop.2015.03.001} {\bibfield
  {journal} {\bibinfo  {journal} {Annals of Physics}\ }\textbf {\bibinfo
  {volume} {358}},\ \bibinfo {pages} {108} (\bibinfo {year} {2015})},\ \bibinfo
  {note} {school of Physics at Nanjing University}\BibitemShut {NoStop}%
\bibitem [{\citenamefont {Qian}\ \emph {et~al.}(2011)\citenamefont {Qian},
  \citenamefont {Ren},\ and\ \citenamefont {Ni}}]{PhysRevC.83.044317}%
  \BibitemOpen
  \bibfield  {author} {\bibinfo {author} {\bibfnamefont {Y.}~\bibnamefont
  {Qian}}, \bibinfo {author} {\bibfnamefont {Z.}~\bibnamefont {Ren}}, \ and\
  \bibinfo {author} {\bibfnamefont {D.}~\bibnamefont {Ni}},\ }\href {\doibase
  10.1103/PhysRevC.83.044317} {\bibfield  {journal} {\bibinfo  {journal} {Phys.
  Rev. C}\ }\textbf {\bibinfo {volume} {83}},\ \bibinfo {pages} {044317}
  (\bibinfo {year} {2011})}\BibitemShut {NoStop}%
\bibitem [{\citenamefont {Stewart}\ \emph {et~al.}(1996)\citenamefont
  {Stewart}, \citenamefont {Kermode}, \citenamefont {Beachey}, \citenamefont
  {Rowley}, \citenamefont {Grant},\ and\ \citenamefont
  {Kruppa}}]{STEWART1996332}%
  \BibitemOpen
  \bibfield  {author} {\bibinfo {author} {\bibfnamefont {T.}~\bibnamefont
  {Stewart}}, \bibinfo {author} {\bibfnamefont {M.}~\bibnamefont {Kermode}},
  \bibinfo {author} {\bibfnamefont {D.}~\bibnamefont {Beachey}}, \bibinfo
  {author} {\bibfnamefont {N.}~\bibnamefont {Rowley}}, \bibinfo {author}
  {\bibfnamefont {I.}~\bibnamefont {Grant}}, \ and\ \bibinfo {author}
  {\bibfnamefont {A.}~\bibnamefont {Kruppa}},\ }\href {\doibase
  https://doi.org/10.1016/S0375-9474(96)00404-6} {\bibfield  {journal}
  {\bibinfo  {journal} {Nuclear Physics A}\ }\textbf {\bibinfo {volume}
  {611}},\ \bibinfo {pages} {332} (\bibinfo {year} {1996})}\BibitemShut
  {NoStop}%
\bibitem [{\citenamefont {Buck}\ \emph {et~al.}(1996)\citenamefont {Buck},
  \citenamefont {Johnston}, \citenamefont {Merchant},\ and\ \citenamefont
  {Perez}}]{PhysRevC.53.2841}%
  \BibitemOpen
  \bibfield  {author} {\bibinfo {author} {\bibfnamefont {B.}~\bibnamefont
  {Buck}}, \bibinfo {author} {\bibfnamefont {J.~C.}\ \bibnamefont {Johnston}},
  \bibinfo {author} {\bibfnamefont {A.~C.}\ \bibnamefont {Merchant}}, \ and\
  \bibinfo {author} {\bibfnamefont {S.~M.}\ \bibnamefont {Perez}},\ }\href
  {\doibase 10.1103/PhysRevC.53.2841} {\bibfield  {journal} {\bibinfo
  {journal} {Phys. Rev. C}\ }\textbf {\bibinfo {volume} {53}},\ \bibinfo
  {pages} {2841} (\bibinfo {year} {1996})}\BibitemShut {NoStop}%
\bibitem [{\citenamefont {Möller}\ \emph {et~al.}(2016)\citenamefont
  {Möller}, \citenamefont {Sierk}, \citenamefont {Ichikawa},\ and\
  \citenamefont {Sagawa}}]{MOLLER20161}%
  \BibitemOpen
  \bibfield  {author} {\bibinfo {author} {\bibfnamefont {P.}~\bibnamefont
  {Möller}}, \bibinfo {author} {\bibfnamefont {A.}~\bibnamefont {Sierk}},
  \bibinfo {author} {\bibfnamefont {T.}~\bibnamefont {Ichikawa}}, \ and\
  \bibinfo {author} {\bibfnamefont {H.}~\bibnamefont {Sagawa}},\ }\href
  {\doibase https://doi.org/10.1016/j.adt.2015.10.002} {\bibfield  {journal}
  {\bibinfo  {journal} {Atomic Data and Nuclear Data Tables}\ }\textbf
  {\bibinfo {volume} {109-110}},\ \bibinfo {pages} {1} (\bibinfo {year}
  {2016})}\BibitemShut {NoStop}%
\bibitem [{\citenamefont {Takigawa}\ \emph {et~al.}(2000)\citenamefont
  {Takigawa}, \citenamefont {Rumin},\ and\ \citenamefont
  {Ihara}}]{PhysRevC.61.044607}%
  \BibitemOpen
  \bibfield  {author} {\bibinfo {author} {\bibfnamefont {N.}~\bibnamefont
  {Takigawa}}, \bibinfo {author} {\bibfnamefont {T.}~\bibnamefont {Rumin}}, \
  and\ \bibinfo {author} {\bibfnamefont {N.}~\bibnamefont {Ihara}},\ }\href
  {\doibase 10.1103/PhysRevC.61.044607} {\bibfield  {journal} {\bibinfo
  {journal} {Phys. Rev. C}\ }\textbf {\bibinfo {volume} {61}},\ \bibinfo
  {pages} {044607} (\bibinfo {year} {2000})}\BibitemShut {NoStop}%
\bibitem [{\citenamefont {Ismail}\ \emph {et~al.}(2003)\citenamefont {Ismail},
  \citenamefont {Seif},\ and\ \citenamefont {El-Gebaly}}]{ISMAIL200353}%
  \BibitemOpen
  \bibfield  {author} {\bibinfo {author} {\bibfnamefont {M.}~\bibnamefont
  {Ismail}}, \bibinfo {author} {\bibfnamefont {W.}~\bibnamefont {Seif}}, \ and\
  \bibinfo {author} {\bibfnamefont {H.}~\bibnamefont {El-Gebaly}},\ }\href
  {\doibase https://doi.org/10.1016/S0370-2693(03)00600-2} {\bibfield
  {journal} {\bibinfo  {journal} {Physics Letters B}\ }\textbf {\bibinfo
  {volume} {563}},\ \bibinfo {pages} {53} (\bibinfo {year} {2003})}\BibitemShut
  {NoStop}%
\bibitem [{\citenamefont {Gao-Long}\ \emph {et~al.}(2008)\citenamefont
  {Gao-Long}, \citenamefont {Xiao-Yun},\ and\ \citenamefont
  {Zu-Hua}}]{Gao_Long_2008}%
  \BibitemOpen
  \bibfield  {author} {\bibinfo {author} {\bibfnamefont {Z.}~\bibnamefont
  {Gao-Long}}, \bibinfo {author} {\bibfnamefont {L.}~\bibnamefont {Xiao-Yun}},
  \ and\ \bibinfo {author} {\bibfnamefont {L.}~\bibnamefont {Zu-Hua}},\ }\href
  {\doibase 10.1088/0256-307x/25/4/023} {\bibfield  {journal} {\bibinfo
  {journal} {Chinese Physics Letters}\ }\textbf {\bibinfo {volume} {25}},\
  \bibinfo {pages} {1247} (\bibinfo {year} {2008})}\BibitemShut {NoStop}%
\bibitem [{\citenamefont {Morehead}(1995)}]{doi:10.1063/1.531270}%
  \BibitemOpen
  \bibfield  {author} {\bibinfo {author} {\bibfnamefont {J.~J.}\ \bibnamefont
  {Morehead}},\ }\href {\doibase 10.1063/1.531270} {\bibfield  {journal}
  {\bibinfo  {journal} {Journal of Mathematical Physics}\ }\textbf {\bibinfo
  {volume} {36}},\ \bibinfo {pages} {5431} (\bibinfo {year} {1995})},\ \Eprint
  {http://arxiv.org/abs/https://doi.org/10.1063/1.531270}
  {https://doi.org/10.1063/1.531270} \BibitemShut {NoStop}%
\bibitem [{\citenamefont {Mao}\ \emph {et~al.}(2022)\citenamefont {Mao},
  \citenamefont {He}, \citenamefont {Gao}, \citenamefont {Zeng}, \citenamefont
  {Yun}, \citenamefont {Du}, \citenamefont {Lu}, \citenamefont {Sun},\ and\
  \citenamefont {Zhao}}]{9760631}%
  \BibitemOpen
  \bibfield  {author} {\bibinfo {author} {\bibfnamefont {D.}~\bibnamefont
  {Mao}}, \bibinfo {author} {\bibfnamefont {Z.}~\bibnamefont {He}}, \bibinfo
  {author} {\bibfnamefont {Q.}~\bibnamefont {Gao}}, \bibinfo {author}
  {\bibfnamefont {C.}~\bibnamefont {Zeng}}, \bibinfo {author} {\bibfnamefont
  {L.}~\bibnamefont {Yun}}, \bibinfo {author} {\bibfnamefont {Y.}~\bibnamefont
  {Du}}, \bibinfo {author} {\bibfnamefont {H.}~\bibnamefont {Lu}}, \bibinfo
  {author} {\bibfnamefont {Z.}~\bibnamefont {Sun}}, \ and\ \bibinfo {author}
  {\bibfnamefont {J.}~\bibnamefont {Zhao}},\ }\href {\doibase
  10.34133/2022/9760631} {\bibfield  {journal} {\bibinfo  {journal} {Ultrafast
  Science}\ }\textbf {\bibinfo {volume} {2022}},\ \bibinfo {pages} {9760631}
  (\bibinfo {year} {2022})}\BibitemShut {NoStop}%
\bibitem [{\citenamefont {Brabec}\ \emph {et~al.}(1996)\citenamefont {Brabec},
  \citenamefont {Ivanov},\ and\ \citenamefont {Corkum}}]{PhysRevA.54.R2551}%
  \BibitemOpen
  \bibfield  {author} {\bibinfo {author} {\bibfnamefont {T.}~\bibnamefont
  {Brabec}}, \bibinfo {author} {\bibfnamefont {M.~Y.}\ \bibnamefont {Ivanov}},
  \ and\ \bibinfo {author} {\bibfnamefont {P.~B.}\ \bibnamefont {Corkum}},\
  }\href {\doibase 10.1103/PhysRevA.54.R2551} {\bibfield  {journal} {\bibinfo
  {journal} {Phys. Rev. A}\ }\textbf {\bibinfo {volume} {54}},\ \bibinfo
  {pages} {R2551} (\bibinfo {year} {1996})}\BibitemShut {NoStop}%
\bibitem [{\citenamefont {Chen}\ \emph {et~al.}(2000)\citenamefont {Chen},
  \citenamefont {Liu}, \citenamefont {Fu},\ and\ \citenamefont
  {Zheng}}]{PhysRevA.63.011404}%
  \BibitemOpen
  \bibfield  {author} {\bibinfo {author} {\bibfnamefont {J.}~\bibnamefont
  {Chen}}, \bibinfo {author} {\bibfnamefont {J.}~\bibnamefont {Liu}}, \bibinfo
  {author} {\bibfnamefont {L.~B.}\ \bibnamefont {Fu}}, \ and\ \bibinfo {author}
  {\bibfnamefont {W.~M.}\ \bibnamefont {Zheng}},\ }\href {\doibase
  10.1103/PhysRevA.63.011404} {\bibfield  {journal} {\bibinfo  {journal} {Phys.
  Rev. A}\ }\textbf {\bibinfo {volume} {63}},\ \bibinfo {pages} {011404(R)}
  (\bibinfo {year} {2000})}\BibitemShut {NoStop}%
\bibitem [{\citenamefont {Kondev}\ \emph {et~al.}(2021)\citenamefont {Kondev},
  \citenamefont {Wang}, \citenamefont {Huang}, \citenamefont {Naimi},\ and\
  \citenamefont {Audi}}]{NUBASE2020}%
  \BibitemOpen
  \bibfield  {author} {\bibinfo {author} {\bibfnamefont {F.}~\bibnamefont
  {Kondev}}, \bibinfo {author} {\bibfnamefont {M.}~\bibnamefont {Wang}},
  \bibinfo {author} {\bibfnamefont {W.}~\bibnamefont {Huang}}, \bibinfo
  {author} {\bibfnamefont {S.}~\bibnamefont {Naimi}}, \ and\ \bibinfo {author}
  {\bibfnamefont {G.}~\bibnamefont {Audi}},\ }\href@noop {} {\bibfield
  {journal} {\bibinfo  {journal} {Chinese Physics C}\ }\textbf {\bibinfo
  {volume} {45}},\ \bibinfo {pages} {030001} (\bibinfo {year}
  {2021})}\BibitemShut {NoStop}%
\bibitem [{\citenamefont {Huang}\ \emph {et~al.}(2021)\citenamefont {Huang},
  \citenamefont {Wang}, \citenamefont {Kondev}, \citenamefont {Audi},\ and\
  \citenamefont {Naimi}}]{CPC-2021-0034}%
  \BibitemOpen
  \bibfield  {author} {\bibinfo {author} {\bibfnamefont {W.}~\bibnamefont
  {Huang}}, \bibinfo {author} {\bibfnamefont {M.}~\bibnamefont {Wang}},
  \bibinfo {author} {\bibfnamefont {F.}~\bibnamefont {Kondev}}, \bibinfo
  {author} {\bibfnamefont {G.}~\bibnamefont {Audi}}, \ and\ \bibinfo {author}
  {\bibfnamefont {S.}~\bibnamefont {Naimi}},\ }\href@noop {} {\bibfield
  {journal} {\bibinfo  {journal} {Chinese Physics C}\ }\textbf {\bibinfo
  {volume} {45}},\ \bibinfo {pages} {030002} (\bibinfo {year}
  {2021})}\BibitemShut {NoStop}%
\bibitem [{\citenamefont {Wang}\ \emph
  {et~al.}(2021{\natexlab{b}})\citenamefont {Wang}, \citenamefont {Huang},
  \citenamefont {Kondev}, \citenamefont {Audi},\ and\ \citenamefont
  {Naimi}}]{CPC-2020-0033}%
  \BibitemOpen
  \bibfield  {author} {\bibinfo {author} {\bibfnamefont {M.}~\bibnamefont
  {Wang}}, \bibinfo {author} {\bibfnamefont {W.}~\bibnamefont {Huang}},
  \bibinfo {author} {\bibfnamefont {F.}~\bibnamefont {Kondev}}, \bibinfo
  {author} {\bibfnamefont {G.}~\bibnamefont {Audi}}, \ and\ \bibinfo {author}
  {\bibfnamefont {S.}~\bibnamefont {Naimi}},\ }\href@noop {} {\bibfield
  {journal} {\bibinfo  {journal} {Chinese Physics C}\ }\textbf {\bibinfo
  {volume} {45}},\ \bibinfo {pages} {030003} (\bibinfo {year}
  {2021}{\natexlab{b}})}\BibitemShut {NoStop}%
\bibitem [{\citenamefont {Dong}\ \emph {et~al.}(2010)\citenamefont {Dong},
  \citenamefont {Zuo}, \citenamefont {Gu}, \citenamefont {Wang},\ and\
  \citenamefont {Peng}}]{PhysRevC.81.064309}%
  \BibitemOpen
  \bibfield  {author} {\bibinfo {author} {\bibfnamefont {J.}~\bibnamefont
  {Dong}}, \bibinfo {author} {\bibfnamefont {W.}~\bibnamefont {Zuo}}, \bibinfo
  {author} {\bibfnamefont {J.}~\bibnamefont {Gu}}, \bibinfo {author}
  {\bibfnamefont {Y.}~\bibnamefont {Wang}}, \ and\ \bibinfo {author}
  {\bibfnamefont {B.}~\bibnamefont {Peng}},\ }\href {\doibase
  10.1103/PhysRevC.81.064309} {\bibfield  {journal} {\bibinfo  {journal} {Phys.
  Rev. C}\ }\textbf {\bibinfo {volume} {81}},\ \bibinfo {pages} {064309}
  (\bibinfo {year} {2010})}\BibitemShut {NoStop}%
\end{thebibliography}


\end{document}